\documentclass[a4paper,10pt,twoside]{cpc-hepnp}
\usepackage[colorlinks,linkcolor=black,citecolor=blue]{hyperref}
\usepackage{multicol}
\usepackage{graphicx}
\usepackage{booktabs}
\usepackage{amssymb,bm,mathrsfs,bbm,amscd,color}
\usepackage[tbtags]{amsmath}
\usepackage{lastpage}

\usepackage[utf8]{inputenc}
\def\be{\begin{eqnarray}}
\def\ee{\end{eqnarray}}
\def\st{\begin{equation}}
\def\stp{\end{equation}}

\def\p{{\bf p}}

\def\ve{\varepsilon}
\def\P{{\mathcal P}}

\def\E{{\mathcal E}}

\def\Re{{\mbox {Re}}}
\def\bP{\mathbb{P}}

\def\bra{\langle}
\def\ket{\rangle}
\def\bbra{\langle\!\langle}
\def\kket{\rangle\!\rangle}

\def\ln{\mbox{ln}}

\def\Eq#1{Eq.~(\ref{#1})}
\def\Eqs#1{Eqs.~(\ref{#1})}

\def\Fig#1{Fig.~\ref{#1}}
\def\Sect#1{{Section~\ref{#1}}}

\def\Ref#1{Ref.~\cite{#1}}
\def\Tab#1{Table~\ref{#1}}

\def\ie{{\it i.e.}}

\begin{document}

\fancyhead[c]{}

\fancyfoot[C]{\thepage}

\footnotetext[0]{}

\title{A flow paradigm in heavy-ion collisions}

\author{%
Li Yan\email{li.yan@physics.mcgill.ca}%
}
\maketitle

\address{%
	Department of Physics,
	McGill University, 
	3600 rue University
	Montr\'eal, H3A 2T8, QC, Canada\\
}

\begin{abstract}
The success of hydrodynamics in high energy heavy-ion collisions leads to
a flow paradigm, to understand the observed features of harmonic flow
in terms of the medium collective expansion with respect to
initial state geometrical properties. 
In this review, we present
some essential ingredients in the flow paradigm, including the hydrodynamic modeling, 
the characterization of initial state geometry and the medium response relations.
The extension of the flow paradigm
to small colliding systems is also discussed.

\end{abstract}




\begin{multicols}{2}

\tableofcontents

\section{Introduction}

In the theoretical modeling of high energy heavy-ion collisions  at the
Relativistic Heavy-Ion Collider (RHIC) at the Brookhaven National Laboratory and at
the Large Hadron Collider (LHC) at CERN, one essential 
concept often used is the system collectivity, and especially the collectivity
of Quark-Gluon Plasma (QGP)~\cite{Shuryak:2008eq,Shuryak:2014zxa} 
-- a hot
and dense medium consisting of quarks and gluons. 

This concept
plays a key role in the understanding of 
heavy-ion collisions, in several respects.
First, the dynamical properties
of QGP during its collective evolution contain the information of QCD
phase structure. With respect to heavy-ion experiments, 
it is crucial to have an appropriate description of the QGP evolution 
in the investigation of the
phase transition from quarks and gluons to hadrons, 
and the searching of 
the QCD critical point in the beam energy scan program~\cite{Stephanov:1998dy}.
Second, 
there is mounting evidence from heavy-ion experiments supporting QGP 
as a perfect fluid in
nature, with very small dissipation. 
Phenomenological analyses have revealed that  in QGP the ratio 
of shear viscosity to entropy density, $\eta/s$, is very
close to the lower bound predicted from the theory of gauge-gravity 
duality for a strongly-coupled system~\cite{Kovtun:2004de}, 
\be
\frac{\eta}{s} = O\left(\frac{\hbar}{4\pi k_B}\right)\,.
\ee
Since the transport properties of a medium are determined by 
the underlying microscopic dynamics,  
estimating the $\eta/s$ of the QGP system in heavy-ion collisions 
provides an unambiguous probe of the dynamical properties of QCD.
Third, 
QGP collective evolution is responsible for all the observed 
patterns in the spectrum of soft particles (particles of small transverse momentum, mass, etc.). 
These soft particles amount to 
over 99\% of the total particle yields in heavy-ion experiments. 
Additionally, the collective evolution of the QGP also
determines the background for hard probes (particles with large transverse momentum, mass, etc.), 
including the 
jet-quenching phenomenon~\cite{Gyulassy:2003mc} and
heavy quarkonia dissociation and regeneration~\cite{Rapp:2008tf}. 
It is also the background for 
electromagnetic signatures in heavy-ion collisions, such as direct photon 
production~\cite{Peitzmann:2001mz,Paquet:2015lta}, 
and those related to the chiral magnetic effect~\cite{Kharzeev:2007jp}.

The concept of medium collectivity has been investigated extensively 
 in heavy-ion experiments, in which an unprecedented
level of precision in the experimental observables 
has been achieved. 
Different types of measurables from the correlations of
of soft particles have been devised and explored,
in nucleus-nucleus collisions at RHIC and the LHC energies. 
These include, especially, the so-called harmonic flow.
Recently,
measurements of particle correlations 
have been generalized to smaller colliding systems, such as
proton-lead, deuteron-gold, and even proton-proton collision events with
extremely high multiplicity production. Surprisingly, the observed particle correlation
patterns in the small colliding systems are compatible with the picture
of medium collective expansion. 
Regarding all the remarkable experimental progress, 
theoretical frameworks 
have been proposed on various grounds. Especially, the success of 
hydrodynamic modeling has allowed
a flow paradigm to emerge.
The purpose of this review is to show how the flow paradigm is established
based on theoretical calculations via hydrodynamic modelings of heavy-ion collisions, 
and the corresponding analyses of the experimental results. 

The flow paradigm for heavy-ion collisions is inspired by the very idea of 
the system collective expansion. It assumes 
the dominant evolution stages of the  
created system in nucleus-nucleus collisions or in small colliding systems, 
as a collectively expanding QGP medium or hadron gas, 
which is close to local thermal equilibrium. As a result, the dynamics of 
the medium evolution is dominated by long wave-length hydrodynamic modes.
In this way, the created partons or hadrons in each collision event evolve
coherently as a fluid medium, in response to the system's geometrical
structures at earlier times, so that their dynamical behaviors 
affect the observed correlations 
of soft particles in experiments. 

Accordingly, the flow paradigm commonly employs viscous hydrodynamics
for the description of medium evolution. 
However, note that the flow paradigm does not prevent the use of a kinetic approach 
with respect to individual partons or hadrons, as long as sytem collective evolution
in the kinetic description is well-established.
Viscous hydrodynamics is
an effective theory for systems close to local thermal equilibrium, where
the dominant degrees of freedom are long-wavelength hydrodynamic variables.
In the theoretical framework of hydrodynamics, the dissipative effect in a fluid system
is described in terms of a gradient expansion order-by-order. Each term in the 
gradient expansion is specified by a transport coefficient. The first order terms
correspond to the shear and the bulk viscosity. 
Application of viscous hydrodynamics to a system crucially depends on a separation
between the microscopic scale and the macroscopic scale which controls the convergence
of the gradient expansion. In the present analyses, a truncation at second order
viscous corrections is generally applied throughout all the collective evolution
stages in heavy-ion collisions. 
Solving viscous hydrodynamics requires initial conditions. 
With respect to the hydro modeling of heavy-ion collisions, an effective 
characterization of the
initial stage with fluctuations is employed, accounting for the fact 
that nucleus-nucleus collisions
fluctuate from event to event. 
After the system collective expansion stages in heavy-ion collisions, the 
observables should be calculable as hadrons emitted
independently towards detectors. 

Theoretical calculations always depend on parameters. For instance,
in hydrodynamic modeling, effective descriptions of initial state,
transport properties, etc, must be specified for
different colliding systems. 
For the purpose of extracting the medium transport properties, these parameterizations
bring in substantial uncertainties in analyses. Therefore, it is important in
the flow paradigm to capture some common features
of the observables in experiments, 
such as those in correlations and fluctuations of harmonic flow,
so as to minimize the uncertainties from effective parameterizations.
These common features of flow observables 
can be understood empirically following some quantitative relations, 
which are summarized based on hydrodynamic simulations with respect to all
existing observables in heavy-ion experiments. These relations between harmonic flow 
and the geometrical properties of initial state are expected to be
model independent, but contain essential information about the initial
state geometry,  event-by-event fluctuations, and the medium dissipations.

These are several types of average considered in this review.
Unless specified, we shall use double brackets $\bbra\ldots\kket$ to notate
the average of a quantity over events. In each single event, the average
over a density profile in the transverse plane is denoted by curly brackets,
$\{\ldots\}$. Considering the approximation of Bjorken boost invariance of 
the system generated in heavy-ion collisions, the Milne space-time
coordinates are often introduced in a theoretical framework, in which the proper
time $\tau=\sqrt{t^2-z^2}$ and space-time rapidity $\xi=\tanh^{-1}(z/t)$ are 
used instead of $t$ and $z$.

This review is organized as follows. 
Theoretical ingredients in the flow paradigm are described
in \Sect{sec:3}, including a brief description of viscous hydrodynamic modeling of
heavy-ion collisions in \Sect{sec:hydro}. \Sect{sec:ini_ecc} presents discussions on 
initial state geometrical
properties captured
in terms of initial eccentricities. 
Empirically, results from hydro modeling of heavy-ion
collisions lead to the medium response 
relations between harmonic flow and initial state eccentricities,
which are introduced in \Sect{sec:resp}. In \Sect{sec:Exp}, after an overview of the 
flow observables compatible with the flow paradigm, we present quantitative 
characterizations of flow observables
based on 
the medium response relations. Especially, 
the observed flow fluctuations and correlations are analyzed in \Sect{sec:vnfluc} 
and \Sect{sec:vncor}, respectively.
\Sect{sec:challenge} focuses on recent measurements involving flow observables in
small colliding systems, and the corresponding theoretical developments that generalize
the flow paradigm in small systems. A summary of the review is given in \Sect{sec:sum}.

\section{Ingredients of the flow paradigm}
\label{sec:3}

To understand the observed flow observables 
in heavy-ion collisions, in the flow 
paradigm 
a hydrodynamic model is applied which solves the 
hydrodynamic equations of motion
for the medium collective expansion. 
The information of the initial state, especially the geometry of the colliding systems,  
is incorporated accordingly 
in the initial condition of the coupled equations. 
In this way, the observed fluctuations and
correlations of flow harmonics are recognized as a consequence of 
the combined effects from the geometrical properties of
initial state and transport properties of the medium. 
In the flow paradigm, the geometry of initial state is decomposed and characterized by a set of 
eccentricities $\E_n$, while medium dynamical properties are contained in the proposed medium
response relations.
In this section, we present these three essential ingredients 
in the flow paradigm: hydrodynamic modeling, characterization 
of initial state geometry, and medium response relations. 

\subsection{Hydrodynamic modeling}
\label{sec:hydro}

In a hydrodynamic simulation of the system evolution in heavy-ion collisions, there are
three stages: initialization of the fluid, solving the hydrodynamic
equations of motion, and particle generation.

The hydrodynamic equations of motion are nothing but a set of conservation laws, 
in which dissipative properties of the medium are introduced
and captured by transport coefficients, such as shear viscosity
$\eta$ and bulk viscosity $\zeta$. These transport coefficients are 
solely determined by the underlying dynamics of the system, \ie,
QCD in heavy-ion collisions, so they should be treated as inputs.
However, the complexity of the medium system in heavy-ion collisions makes it difficult for
a first-principles calculation of the transport coefficients. For instance,
it is expected that the coupling constant varies from weakly-coupled
to strongly-coupled as the QGP cools down. As a result, 
in practical simulations, $\eta/s$ is
often considered as some parameterized form, with or without temperature
dependence (cf. the calculations in \Ref{Niemi:2015voa}). 
In a similar manner, in practical hydro simulations there
are also effective parameterizations for
the characterization of initial state of the colliding system, 
the equation of state, and the freeze-out prescription. Compared to
experimentally measured observables, the primary goal of hydro simulations 
then becomes  to constrain these parameterizations, which contain information
about initial state geometrical fluctuations~\cite{Retinskaya:2013gca}, 
medium dissipative properties, and
the equation of state of QCD~\cite{Moreland:2015dvc,Monnai:2017cbv}.

The conservation of energy and momentum is written in hydrodynamics as
\be
\partial_\mu T^{\mu\nu} = 0\,,
\ee 
where the energy-momentum tensor $T^{\mu\nu}$ is defined in terms of hydrodynamic variables:
energy density $e$, pressure $\P$ and flow four-velocity $u^\mu$,
\be
\label{eq:Tmn}
T^{\mu\nu}=eu^\mu u^\nu + \P\Delta^{\mu\nu} + \Pi^{\mu\nu}\,.
\ee
The operator $\Delta^{\mu\nu}=u^\mu u^\nu + g^{\mu\nu}$ is a projection operator, 
  which with the flow four-velocity $u^\mu$ can be used to put the formulation of
hydrodynamics in a covariant form. Namely, one may write the 
spatial gradient and temporal derivative covariantly as
$
\Delta_{\mu\nu}\partial^\nu = \nabla_\mu\,
$
and $ D = u^\mu \partial_\mu$, respectively. Note that in \Eq{eq:Tmn}, we have
taken the mostly plus matrix convention, $g^{\mu\nu}=(-,+,+,+)$, which leads to the
flow velocity normalization as $u^2=-1$. When an equation of 
state is introduced and coupled to the hydrodynamic equation of motion, these 
hydro variables are completely determined. In practical simulations with
respect to heavy-ion collisions, an equation of state from lattice QCD
calculations is generally incorporated~\cite{Huovinen:2009yb}.

Dissipative corrections to the energy-momentum tensor are reflected in the 
stress tensor $\Pi^{\mu\nu}$. The scale separation between
a microscopic scale related to system mean free path $l_{\mbox{\tiny mfp}}$, and
a macroscopic scale associated with the system size $L$, allows one to expand 
viscous hydrodynamics in a series of gradients. The ratio of these two scales is
recognized as a small quantity, the Knudsen number Kn$\sim l_{\mbox{\tiny mfp}}/L$. 
To the first order in the 
gradient expansion (the first order viscous correction), 
the stress tensor has the well-known Navier-Stokes form,
\be
\label{eq:NS}
\Pi^{\mu\nu}=-\eta \sigma^{\mu\nu} - \zeta \Delta^{\mu\nu}\nabla\cdot u
=\pi^{\mu\nu}+\Delta^{\mu\nu} \Pi\,,
\ee
where $\pi^{\mu\nu}$ and $\Pi$ are the viscous corrections associated with the shear
and the bulk channel. The structure of the first order in the gradient expansion
\be
\label{eq:sigma}
\sigma^{\mu\nu}= \nabla^\mu u^\nu + \nabla^\nu u^\mu - \frac{2}{3}\Delta^{\mu\nu}
\nabla\cdot u\equiv \nabla^{\bra\mu} u^{\nu\ket}
\ee
is a symmetric and traceless tensor, and it is transverse to $u^\nu$, $\sigma^{\mu\nu}u_\mu=0$. 
Note that in \Eq{eq:sigma}, and in the following discussion
of this section, the single brackets around
indices of a tensor denote that the tensor has been made symmetric, traceless
and transverse to $u^\mu$. 
In practical hydrodynamic simulations, 
to avoid acausal mode evolution, second order 
viscous corrections must be taken into account in the gradient expansion 
(cf. discussions in \Ref{Romatschke:2009im}).  

There are more terms which stem from more involved gradient structures in
higher order viscous corrections, and accordingly more transport coefficients.
In a simplified case at the second order, considering
a form of relaxing second order terms to their Navier-Stokes correspondence 
in
\Eq{eq:NS}, one finds the relativistic generalization of the Israel-Stewart
hydrodynamics, with new transport coefficients
$\tau_\pi$ and $\tau_\Pi$ associated with the 
relaxation time of the shear channel and bulk channel,
respectively. For instance, the shear channel has
\be
\label{eq:is}
(\tau_\pi D +1)\pi^{\mu\nu} = -\eta \sigma^{\mu\nu} + \ldots,
\ee
where the ellipsis implies structures generated from the second order of gradient expansion, e.g.,
$\sigma^{\bra \mu}_{\ \;\;\;\;\alpha}\sigma^{\alpha\nu\ket}$, $\sigma^{\mu\nu}\nabla\cdot u$,
and $\sigma^{\bra \mu}_{\ \;\;\;\;\alpha}\Omega^{\alpha\nu\ket}$ ($\Omega^{\mu\nu}=\frac{1}{2}(\nabla^\mu u^\nu-
\nabla^\nu u^\mu)$ being an anti-symmetric tensor).
Variants of definitions of the second order viscous
hydrodynamics have been achieved in a conformal fluid~\cite{Baier:2007ix}, 
and from the moment
expansion techniques~\cite{Denicol:2012cn}.

\Eq{eq:is} is the most commonly solved hydro equation of motion
for hydro modeling of heavy-ion collisions at RHIC and the 
LHC~\cite{Niemi:2014wta,McDonald:2016vlt,Qian:2016fpi,Noronha-Hostler:2015dbi,Chattopadhyay:2017bjs}. 
However, it is \emph{not}
a complete theoretical formulation, 
because thermal fluctuations are ignored. Thermal fluctuations in a 
fluid system, also known as the hydrodynamic fluctuations, are present in
all the evolution stages of the expanding medium in heavy-ion collisions.
The strength of hydrodynamic fluctuations is related to dissipative properties
of the fluid through the fluctuation-dissipation relations~\cite{Landau-sp2}. In the theoretical
framework, hydrodynamic
fluctuations can be introduced as an extra stochastic tensor $S^{\mu\nu}$ to the 
energy-momentum tensor~\cite{Sakai:2017rfi,Young:2014pka,Yan:2015lfa},
\be
T^{\mu\nu}=e u^\mu u^\nu + \P\Delta^{\mu\nu} +\Pi^{\mu\nu}
+S^{\mu\nu}\,,
\ee
whose two-point auto-correlation is related to the corresponding dissipations.
For the Navier-Stokes hydrodynamics~\cite{Kapusta:2011gt}\footnote{
The two-point auto-correlation of thermal fluctuations is defined according to
the average over thermal ensembles, for which in this review 
we use the same notation of double
brackets $\bbra \ldots\kket$, despite its subtle difference from average 
over events in heavy-ion collisions.
},
\begin{align}
\label{eq:ss}
\bbra S^{\mu\nu}(x)S^{\alpha\beta}(x')\kket
=2T\Big[&
\eta(\Delta^{\mu\alpha}\Delta^{\nu\beta}+\Delta^{\mu\beta}\Delta^{\nu\alpha})\cr
&+\left(\zeta-\frac{2}{3}\eta\right)\Delta^{\mu\nu}\Delta^{\alpha\beta}\Big]
\delta^{(4)}(x-x')\,,\cr
\end{align}
with the strength of the thermal fluctuations determined by the shear and
bulk viscosity. The Dirac delta function in \Eq{eq:ss} is practically recognized
in realistic simulations of a finite thermal system as the inverse of space-time
volume. Therefore, one naively expects significant contributions from thermal fluctuations
in small colliding systems, and systems close to the QCD critical point. In practical
simulations, solving the noisy viscous hydrodynamics is more challenging than most 
of the present hydro modelings.

To solve the hydrodynamic equations of motion, one needs inputs from 
an effective characterization of the system at the initial stages of 
heavy-ion collisions. That is to say, all hydro variables, as well
as the stress tensor $\pi^{\mu\nu}$, and bulk pressure $\Pi$, must be specified
at some proper time $\tau_0$, assuming at which the QGP system is sufficiently
close to local thermal equilibrium. 
Approaching local thermalization, or in a more relaxed term, the onset of
viscous hydrodynamics, is a crucial challenge to 
the application of viscous hydrodynamics
in heavy-ion collisions, despite all the success of hydro modelling 
has achieved regarding 
flow observables. In particular, in recent experiments 
with small colliding systems, the validity of viscous hydrodynamics needs to be re-examined.
The onset of hydrodynamics, although it is a topic beyond the scope of 
the present review, shall be briefly addressed in the context of 
applying hydrodynamics and the flow paradigm in small
colliding systems in \Sect{sec:thermalization}.
One may find more detailed discussions elsewhere in~\Ref{Fukushima:2016xgg}
and \Ref{Florkowski:2017olj}.
A value of $\tau_0\sim O(1)$ fm/c has been found necessary
to make reasonable predictions 
in nucleus-nucleus
collisions at RHIC and the LHC energies. For smaller colliding systems,
some hydro simulations suggest $\tau_0\sim O(0.1)$ fm/c~\cite{Habich:2015rtj}.

In hydrodynamic modeling, several effective models have been
developed to generate a density profile of the QGP system at 
$\tau_0$~\cite{Schenke:2012wb, Alver:2008aq, Kharzeev:2004if,Moreland:2014oya, Niemi:2015qia}, 
with fluctuations
implemented through Monte Carlo simulations. 
Inspired by a color-glass picture, in models such as 
IP-Glasma~\cite{Schenke:2012wb},
an initial state energy density profile is obtained by solving the gluon field
evolution. On the other hand, in the MC-Glauber model~\cite{Alver:2008aq}, energy is deposited
from nucleon-nucleon collisions in an eikonal approximation.
All of these models provide an event-by-event basis for hydrodynamic simulations,
and especially, a fluctuating initial state. 
\Fig{fig:ic} presents the distribution of energy density $e(x,y)$ 
in the transverse plane of
one typical event in heavy-ion collisions at initial time $\tau_0$,
generated from the MC-Glauber,
MC-KLN and IP-Glasma models. Bumpy structures seen in these distributions
reflect fluctuations originating from energy deposition during nucleus-nucleus
collisions. 
As one can see, the fluctuations are stronger in the IP-Glasma model than in MC-Glauber.
Simply speaking, one may recognize gradients of the distribution, due
to an overall shape or fluctuations,
as the driving force of system expansion in hydrodynamics. Accordingly, the fluctuating
geometry of the initial state is converted into asymmetries of the generated
particles in the momentum space. This is the intuitive interpretation of 
medium response to initial state geometry, which we shall detail in the next subsection.

\begin{center}
\includegraphics[width=0.5\textwidth] {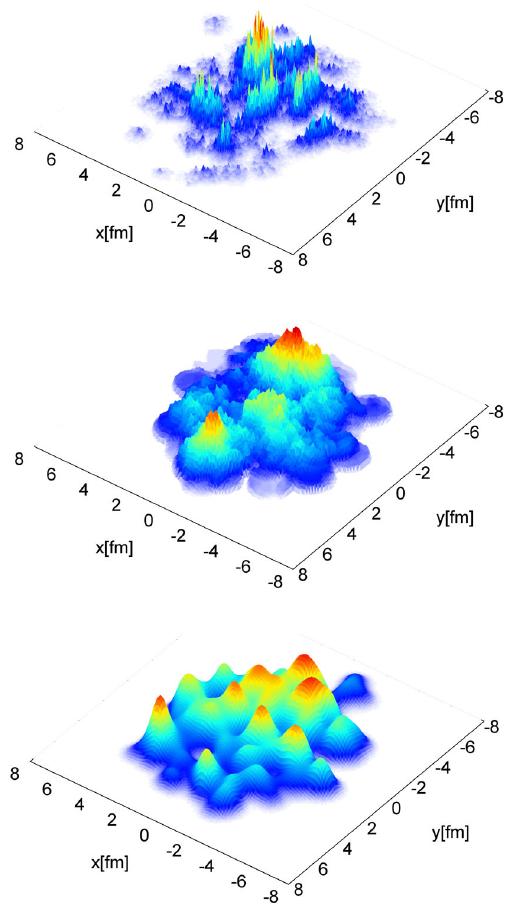}
\figcaption{Initial state distribution of energy density in the transverse
plane from 
IP-Glasma (top), MC-KLN (middle) and MC-Glauber (bottom) models.
Figures reproduced from
\Ref{Schenke:2012wb}, with permission.
\label{fig:ic}
}
\end{center}

The medium system created in heavy-ion collisions cools down as it expands.
Once the system becomes so locally dilute that the dominant degrees of
freedom are excited particles instead of hydro variables, the system starts
to freeze out from the fluid description. A common procedure of freeze-out was given by
Cooper and Frye~\cite{Cooper:1974mv}, 
in which hadrons are generated from a fluid on a 3D hyper-surface $\Sigma$
determined by the freeze-out condition,
\be
\label{eq:freeze}
E\frac{dN}{d^3 {\bf p}} = \int_\Sigma d\sigma\cdot p f(u^\mu, T, \mu)\,.
\ee
The phase-space distribution function $f(u^\mu, T, \mu)$ should be incorporated properly
corresponding to viscous hydrodynamics through Landau's matching
condition. More precisely, it is
\be
f(u^\mu, T, \mu)=n(u^\mu, T, \mu)+\delta f(u^\mu, T, \mu)\,,
\ee
with the local equilibrium distribution $n(u^\mu, T, \mu)$ corresponding to
ideal hydrodynamics and the viscous correction
\be
\int \frac{d^3 \p}{E} p^\mu p^\nu \delta f(u^\mu, T, \mu)=\Pi^{\mu\nu}\,.
\ee
For the shear
channel, it is conventionally taken as
$
\delta f(u^\mu, T, \mu) \propto p^\mu p^\nu \pi_{\mu\nu}
$~\cite{Teaney:2003kp}
while the bulk channel has
$
\delta f(u^\mu, T, \mu)\propto \Pi
$~\cite{Monnai:2009ad,Bozek:2009dw}. The canonical form of the viscous
correction to the phase-space distribution at second order has more involved
dependence in the second order gradients and momentum, but converges to
the first order viscous corrections with respect to small dissipations~\cite{Teaney:2013gca}. 
The chemical potential $\mu$ in \Eq{eq:freeze} is
specified with respect to the desired particle species. 

For more realistic
hydro simulations of heavy-ion collisions, the particle spectrum
receives further modifications from subsequent interactions among hadrons,
including hadrons collisions in kinetics and resonance decays, 
which can be described by effective models, 
such as UrQMD~\cite{Bass:1998ca}. 

Eventually, the  spectrum of particles from each collision event is determined
from numerical simulations, accounting for all the effects mentioned above. The
single-particle spectrum represents the hydro prediction of the particles 
observed in detectors in experiments. It should be emphasized that hydrodynamic
predictions of the particle spectrum are characteristic in long-range correlations
in rapidity, which are best quantified by  
the harmonic flow, $V_n$. 
Defined according to a Fourier decomposition of the emitted single-particle spectrum,
\be
\label{eq:vndef1}
E\frac{d N}{d^3 \p}
=\frac{1}{2\pi}\frac{d N}{p_T dp_T d\eta}\left[1+\sum_{n=1}^\infty \left(V_n(p_T,\eta) 
e^{-in\phi_p} + c.c.\right)\right]\,,
\ee
the harmonic flow $V_n$ characterizes the momentum anisotropy in azimuth of the 
particle spectrum, of order $n$. In \Eq{eq:vndef1}, 
$\eta=\tanh^{-1}(p_z/|{\bf p}|)$ is the pseudo-rapidity, which should be 
distinguished from the notation of shear viscosity. In \Eq{eq:vndef1}, 
$c.c.$ indicates a complex conjugate. 
Harmonic flow $V_n$'s are complex by definition,
\be
\label{eq:vndef2}
V_n(p_T,\eta) \equiv v_n e^{in\Psi_n}=\bra e^{in\phi_p}\ket
\ee
where the magnitude $v_n$ characterizes the magnitude of azimuthal anisotropies 
of the particle 
spectrum in the transverse directions, while the phase $\Psi_n$, also known as the
event-plane, determines orientation
of the anisotropy.
In the last equation of \Eq{eq:vndef2}, the single brackets denote
an average with respect to the single particle spectrum in one collision event. 

Fluctuations in the initial state density
profile from event to event result in fluctuations of the generated paricle spectrum,
and also flucutating harmonic flow $V_n$. Therefore, in event-by-event hydrodynamic
simulations, as in experiments, harmonic flow $V_n$ should be extracted via
multi-particle correlations.
In experiments,
it has been noticed that anisotropy of the particle spectrum depends on collision centrality, 
transverse momentum $p_T$,
and rapidity $y$ (implying the dependence on particle species), or pseudo-rapidity $\eta$, as does the
corresponding harmonic flow $V_n$.

Different orders of the harmonic flow have specified physical interpretations, in terms
of the azimuthal symmetry. 
The most dominant flow signature is $n=2$, the elliptic flow $V_2$, which 
characterizes the asymmetric distribution of particles
generated in- and out-of the event-plane $\Psi_2$ in one collision event~\cite{Ollitrault:1992bk}. 
For the case of $n=1$, which characterizes asymmetric particle yields from one side
of the system to the other, there is a rapidity-even component which
is often referred to as the dipolar flow~\cite{Teaney:2010vd,Retinskaya:2012ky},
and a rapidity-odd piece. 
In multi-particle correlations, although both the dipolar flow and the 
rapidity-odd $V_1$ receive contributions from
momentum conservation in multi-particle correlations~\cite{Borghini:2006yk}, 
the dipolar flow is generated
as a result of medium collective expansion, as in the flow paradigm. The rapidity-odd
$V_1$ is rooted in the properties of nucleus scatterings, and has a strong correlation with the 
reaction plane.
Regarding higher order azimuthal anisotropies, there are also 
trianglar flow $V_3$~\cite{Alver:2010gr}, quadranglar flow $V_4$, etc.  It should emphasized that 
the flow harmonics of order $n$ has a determined 
rotational symmetry in the azimuth, $\phi_p\rightarrow \phi_p+2\pi/n$.

In experiments, it has been found that the elliptic flow is more significant than others.
Especially in nucleus-nucleus collisions, such as the Au+Au at RHIC, the signature of ellitptic flow
is understood as a medium evolution with respect to 
an almond-shaped geometry determined by the initial overlap of the colliding system.
Relating flow observables to initial state geometry is generalizable to higher order
flow harmonics, which motivates the analyses of medium response relations in the flow paradigm.

Although hydrodynamic modeling of heavy-ion
collisions gives rise to results with quantitative agreement with the observed flow signatures, 
these calculations rely on several effective parameterizations. Especially, the
effective description of the initial state contributes to the greatest extent to 
the uncertainties of the analyses. These are the dominant source of uncertainties 
in the extraction of $\eta/s$ in hydro modeling. 
In addition, there are also fundamental issues related to the application of
dissipative fluid dynamics to heavy-ion collisions, which are more severe
for the recent experiments carried out with small colliding systems. 
Nevertheless, the success of hydro modeling in heavy-ion collisions allows one to
emperically correlate the observed flow signatures to the geometrical properties
of the initial state and the dissipative feature of the expanding medium, as in the 
flow paradigm. By doing so, what really matter in analysis are the common features of the 
observed flow harmonics in various colliding systems, in a way that not only can 
one understand the generation of 
harmonic flow, flow correlation and fluctuation, but also
provides quantitative constraints on the exatraction of $\eta/s$, with minimized dependence 
on the effective parameterizations of the model.

\subsection{Geometrical properties of the initial state}
\label{sec:ini_ecc}

The quantitative relations established in the flow paradigm to relate the observed
flow harmonics to initial state and medium dissipations requires appropriate characterizations
of the initial state geometry. This can be achieved by using the so-called initial state
eccentricity $\E_n$, which by definition is introduced according to the azimuthal 
symmetry of the initial state density profile. Fluctuations of the initial state density
profile lead to fluctuations of $\E_n$ from event to event. Additionally, on an event-by-event
basis, $\E_n$ are correlated owing to the background
geometry resulting from the colliding systems. Fluctuations and correlations of $\E_n$ are 
crucial in understanding the similar behaviors in $V_n$.

\subsubsection{Initial state anisotropy $\E_n$} 
The idea to relate harmonic flow and initial state geometry is inspired by
the fact that the linearised hydrodynamic response does not mix the evolution of modes,
once these modes are considered small perturbations. In terms of the expanding
medium systems in heavy-ion collisions, these are modes associated 
with azimuthal asymmetries which are responsible for the generation of
flow harmonics (cf. ~\Ref{Gubser:2010ui}). 
Also, as in the original observation of elliptic flow, $V_2$~\cite{Ollitrault:1992bk},
in a somewhat crude analysis ignoring fluctuations in the initial state, 
people realized that the observed 
elliptic flow is correlated to an almond shape of the initial
state density profile. 

In high energy heavy-ion 
collisions, such an almond-shaped distribution is simply expected from 
the overlap of two colliding nuclei when collisions are non-central, 
with the shorter dimension of the almond
shape aligned with the reaction plane. As a result, elliptic flow can be 
interpreted as a result of medium expansion 
driven 
by the asymmetric density profile. More precisely, gradients, which
play the role of force in hydrodynamics, 
are anisotropic in- and out-of reaction plane, leading correspondingly to
 an anisotropic expansion. This expansion
scenario has been justified 
by the similar expansion out of an anisotropic 
medium system realized in the cold atom experiments~\cite{OHara:2002pqs}.
The key
in this relation is the azimuthal structure of the initial distribution,
which is asymmetric between in- and out-of reaction plane,
analogous to that of elliptic flow.
The extent of 
asymmetry is characterized by a dimensionless quantity called ellipticity. 
With respect to the 
reaction-plane, it can be defined as
\be
\label{eq:ecc2}
\varepsilon_2^{RP} = \frac{\{x^2-y^2\}}{\{x^2+y^2\}}\,,
\ee
where the curly brackets denote the average with respect to the
initial state energy (or entropy) density distribution,
\be
\{\ldots\} = \frac{\int dx dy e(x,y) \ldots}{\int dx dy e(x,y)}\,.
\ee
The reaction-plane ellipticity $\varepsilon^{RP}_2$ is bounded by unity. 
It is clear that the elliptic asymmetry vanishes when $\varepsilon_2^{RP}=0$, 
corresponding to a density profile with absolute azimuthal symmetry.
Elliptic asymmetry maximizes when $\varepsilon_2^{RP}=1$. To a good 
approximation, a linear relation between the ellipticity and $v_2$ has been 
found~\cite{Song:2008si}.

When event-by-event fluctuations are taken into account, the initial state geometry 
in heavy-ion collisions depends not only on the background shape, but also
deformations induced by the extra fluctuations. 
A generalization of ellipticity to higher orders can be applied, which provides
a
mode decomposition with respect to the azimuthal asymmetry of the initial state geometry.
If one takes the complex expression
$z=x+iy=r e^{i\phi}$ for the transverse coordinates, 
a standard 
generalization of the $n$-th order eccentricity is
defined in terms of the n-th order moment of the density, as
\be
\label{eq:eccn}
\E_n\equiv \varepsilon_n e^{in\Phi_n}=-\frac{\{ z^n\}}{\{|z|^n\}}=-\frac{\{r^n e^{in\phi}\}}{\{r^n\}}\,,\qquad
n>1\,,
\ee
where $\{|z|^n\}$ in the denominator plays the role of normalization.
The minus sign is conventionally taken so that $\E_n$ is potentially aligned with
respect to $V_n$, although the alignment is often
broken due to the complexity induced from the medium response. 
It can also be understood as a Fourier decomposition of the energy density 
in terms of azimuthal angle $\phi$, with a $r^n$-weight corresponding to the
fluctuation modes along the radial direction. 
For the case of $n=1$, since $\{z\}$ vanishes by a re-centering of the density profile\footnote{
Re-centering of the initial density in a theoretical analysis is always allowed, since the 
physical observables in heavy-ion collisions are invariant under translations in
the transverse plane.},
the non-trivial leading 
contribution is
\be
\label{eq:ecc1}
\E_1 \equiv \varepsilon_1 e^{i\Phi_1}=-\frac{\{z^2 z^*\}}{\{|z|^3\}}=-\frac{\{r^3 e^{i\phi}\}}{\{r^3\}}\,,
\ee
which captures a dipolar structure in the initial energy density. 
$\E_1$ is the dipolar anisotropy, which is rapidity-even.
Note that in \Eq{eq:eccn}, $\E_n$ is complex with its module $\varepsilon_n$ characterizing
the magnitude of asymmetry, while its phase $\Phi_n$ defines the orientation. The phase
$\Phi_n$ is sometimes referred to as the participant plane of the initial state.
Both $\varepsilon_n$ and $\Phi_n$ fluctuate from event to event in heavy-ion collisions
as the density profile fluctuates. Again, by definition in \Eq{eq:eccn} and \Eq{eq:ecc1}, 
$\E_n$ is bounded by unity, with a vanishing $\E_n$ indicating a vanishing n-th order 
anisotropy, while a maximized anisotropy is achieved when $|\E_n|=\varepsilon_n=1$.  

\end{multicols}
\begin{center}
\includegraphics[width=1.\textwidth] {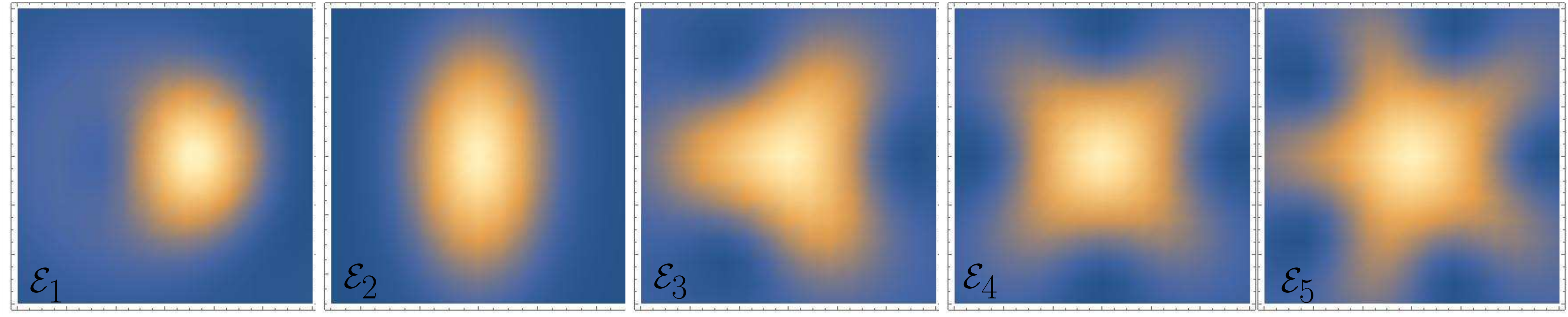}
\figcaption{
\label{fig:ecc_shape}
Characteristic shapes of the deformed initial state density profile, corresponding to anisotropies of $\E_1$, $\E_2$,
$\E_3$, $\E_4$ and $\E_5$ (from left to right).
}
\end{center}
\begin{multicols}{2}

By construction, $\E_n$ is invariant under rotation in azimuthal angle, $\phi\rightarrow \phi+2\pi/n$,
which implies that the azimuthal angle dependence of the
fluctuating density profile is decomposed into modes according to
$\E_n$. Each mode corresponds to
a characteristic shape, under the 
rotational symmetry. In addition to the elliptic shape associated with
$\E_2$, $\E_3$ defines the triangularity, $\E_4$ defines the quadrangularity, etc.
\Fig{fig:ecc_shape} demonstrates the eccentric shapes corresponding to $n=1$
to $n=5$. The phase $\Phi_n$'s in \Fig{fig:ecc_shape} are aligned with the x-axis.

Mode decomposition of the azimuthal angle dependence
in \Eq{eq:eccn} gives rise to correct definitions of dimensionless quantities,
with the desired symmetry dependence with respect to rotation. However, 
the definition of initial eccentricity is \emph{not} unique. 
Apparently, various types of weight function of the radial distance
$f(r)$
can be applied,
accounting for the dependence in the radial direction~\cite{Alver:2010gr,Floerchinger:2014fta,Floerchinger:2013rya}, 
so that one has in general
\be
\E_n[f(r)] = -\frac{\{f(r) e^{in\phi}\}}{\{|f(r)|\}}\,.
\ee
For instance, instead of the $r^2$ weighting in the definition of $\E_2$, 
$r^4$-weighted moments, $\{r^4 e^{i2\phi}\}=\{z^3 z^*\}$, also lead to the characterization ellipticity,
satisfying the same symmetry condition under rotation.
A systematic way of considering
the $r$-dependence of the initial state requires a generalized mode 
decomposition in $r$.
Analogous to the fact that different harmonic orders relate to different 
structures in the azimuth of the system, higher modes associated with $r$ in the definition of eccentricities 
correspond to different (and finer) structures in the radial direction. These higher modes are particularly
important regarding fluctuations of the initial state density profile.
With respect to the same harmonic order, these different modes
along $r$ contribute simultaneously to the corresponding harmonic flow, but they have different
dynamical properties responsible for the observed momentum dependence,
which motivates the proposed principal 
component analysis~\cite{Bhalerao:2014mua,Mazeliauskas:2015vea}.

In addition to the moment of the density profile, one may also consider using 
cumulants~\cite{Teaney:2010vd}. For 
instance, for the $4$-th order anisotropy, cumulants of the density profile gives 
\be
\E_4^c = -\frac{1}{\{|z|^4\}}\left[\{z^4\} - 3\{z^2\}^2 \right]\,,
\ee
where terms proportional to $\{z\}$ are suppressed as a result of re-centering
correction. Like $\E_4$, one may check that
$\E_4^c$ is also invariant under the rotation $\phi\rightarrow \phi+2\pi/4$. However,
unlike $\E_4$, $\E_4^c$ is \emph{not} bounded by unity, in particular in very
peripheral collisions where the ellipticity of the system is large. A cumulant definition
of the eccentricity coincides with the moment definition, when $n\le 3$. When $n\ge 4$, 
there are always subtractions or additions of nonlinear moment couplings 
in the cumulant definition. For the cumulant definition of the fifth order and the 
sixth order, one has
\begin{align}
\E_5^c=&-\frac{\{z^5\}-10\{z^2\}\{z^3\}}{\{|z|^5\}}\,,\cr
\E_6^c=&-\frac{\{z^6\}-15\{z^4\}\{z^2\}-10\{z^3\}^2+30\{z^2\}^3}{\{|z|^6\}}
\end{align}
Throughout discussions in this review, we shall take the momentum-based 
definition by default. 

\subsubsection{Fluctuations and correlations of $\E_n$}

In the flow paradigm, fluctuations and correlations of $\E_n$ play an essential
role in the understanding
of fluctuations and correlations of the harmonic flow $V_n$.
For the discussion of fluctuations, we shall focus on lower harmonic orders ($n\le 3$),
to avoid complexities due to the nonlinear mode couplings during the medium
evolution. Correlations amongst
$\E_n$ will be addressed with respect to the correlations of participant plane $\Phi_n$.  
In this subsection, we concentrate on
some of the analytical results from independent sources, with respect to
very fundamental assumptions. Deviations from the present assumptions
are expected to be sub-leading.

Fluctuations and correlations of initial state eccentricities are rooted in the 
event-by-event fluctuations of initial state density profile, and accordingly 
the induced fluctuating initial state geometry.
Although it can be realized through Monte Carlo simulations of effective models, the
fluctuating density profiles of various effective models behave differently, 
which gives rise to different evaluations of the induced anisotropies, the probability
distributions of anisotropies and correlations among them. 
As is clearly shown in \Fig{fig:ic}, the density profile 
from IP-Glasma is more spiky, implying stronger fluctuations of anisotropy.
Despite ambiguities 
in different effective models, 
there are two essential
concepts one has to take into account in a theoretical analysis of the initial state geometry:
background geometry and fluctuations. The background geometry of a fluctuating system 
reflects the event-averaged density profile. For instance, in collisions such as Au+Au at RHIC, 
and Pb+Pb at the LHC energies, 
the event-averaged density presents a determined almond shape in each centrality
class. Given a background density profile which captures the correct shape,
fluctuations are treated as extra sources. 

\subparagraph{Fluctuations of $\E_n$}
In the simplest scenario, one considers a fluctuating density profile 
in terms of N-independent point-like sources, distributed according to 
a two-dimensional Gaussian background geometry.
It has been found that with respect to such a configuration, the probability distribution of
initial state eccentricity 
can be solved analytically~\cite{Yan:2013laa,Yan:2014afa}, leading to the so-called
elliptic-power distribution. The two-dimensional elliptic-power distribution
for ellipticity $\E_n=\ve_{x}+i\ve_{y}$ is 
\be
\label{eq:ep_2d}
P(\ve_{x},\ve_{y})=\frac{\alpha}{\pi}(1-\ve_0^2)^{1+\frac{1}{2}}
\frac{(1-\ve_{x}^2-\ve_{y}^2)^{\alpha-1}}{(1-\ve_0\ve_{x})^{2\alpha+1}}\,,
\ee
where $\ve_0$ and $\alpha$ are parameters associated
with the background shape and number of
independent sources, respectively. Regarding the Gaussian background, $\ve_0$
characterizes ellipticity in the reaction plane. 
Note, however, that although $\ve_0$ is expected to be close to $\ve_2^{RP}$, it should be
distinguished from the real ellipticity in the reaction plane defined for a fluctuating
system, owing to the effects of fluctuations. 
The parameter $\alpha=\frac{N-1}{2}$ is related to the fluctuation strength,
which is roughly proportional to $1/\sqrt{N}$. One remarkable property of 
the elliptic-power distribution 
is its consistency with the condition that 
eccentricity is bounded by unity. This is implied in the normalization
\be
\int d\ve_{x} d\ve_{y}
P(\ve_{x},\ve_{y})=1\,, 
\ee
where integration runs over a unit disk, $|\E_n|\le 1$.  The upper bound of
the eccentricity is not generally satisfied in other distribution functions. For instance,
considering a simplification of the elliptic-power distribution in the limit $\alpha\gg1$,
corresponding to a system with small fluctuations, the distribution \Eq{eq:ep_2d}
reduces to a two-dimensional elliptic Gaussian,
\be
\label{eq:epGaus}
P(\ve_x,\ve_y)=\frac{1}{2\pi\sigma_x\sigma_y} \exp\left(
-\frac{(\ve_x-\ve_0)^2}{2\sigma_x^2}-\frac{\ve_y^2}{2\sigma_y^2}\right)\,,
\ee 
which is normalized in the entire plane of $(\ve_x,\ve_y)$.

In deriving the elliptic-power distribution, $\ve_0$ is introduced to characterize 
the intrinsic anisotropy induced from the 
background geometry. There are circumstances when $\ve_0=0$, 
so that the initial state eccentricity is entirely
generated by fluctuations. For instance,
in proton+Pb collisions, the created medium is rotationally symmetric
if the shape of the proton is isotropic. Also, as in collisions like Pb+Pb,
all odd-order anisotropies vanish from the background geometry.
In such cases, the elliptic-power reduces to the power distribution,
\be
\label{eq:2d_power}
P(\ve_{x},\ve_{y})
=\frac{\alpha}{\pi}(1-\ve_{x}^2-\ve_{y}^2)^{\alpha-1}\,.
\ee
One may check that the power distribution is also normalized on a 
unit disk.

The probability distribution of the magnitude can be found by integrating out the dependence 
of angle $\varphi_n = \arg(\E_n)$. The elliptic-power distribution gives,
\begin{eqnarray}
\label{eq:ep_1d}
P(\varepsilon_n)&=&2\varepsilon_n\alpha (1-\varepsilon_n^2)^{\alpha-1} 
(1-\varepsilon_n\varepsilon_0)^{-1-2 \alpha} 
(1-\varepsilon_0^2)^{\alpha+\frac{1}{2}}\times \cr
&&
{_2}F_1\left(\frac{1}{2},1+2\alpha;1;\frac{2\varepsilon_n\varepsilon_0}{\varepsilon_n\varepsilon_0-1}\right)\,,
\end{eqnarray}
where ${_2}F_1$ is the hypergeometric function. The power distribution gives,
\begin{equation}
P(\varepsilon_n)=2\alpha\varepsilon_n(1-\varepsilon_n^2)^{\alpha-1}\,.
\label{eq:power}
\end{equation}

\Fig{fig:his_fit_ecc} displays the probability ditribution of $\ve_2$, 
$\ve_3$ and $\ve_4$, generated from MC-Glauber simulations, 
with respect to Pb+Pb collision events in the 75\%-80\% centrality class
at the LHC energy $\sqrt{s_{NN}}=2.76$ TeV. This very peripheral centrality
class is purposely chosen since the effects of fluctuations are sufficiently strong,
so is the influence from the bound of $\ve_n$ by unity.

The fit using an elliptic-power
distribution  describes $\ve_2$ and $\ve_4$ (red solid lines in \Fig{fig:his_fit_ecc}~(a) and (c)) well,
while since $\ve_3$ in Pb+Pb collisions is solely fluctuation-driven, its fluctuating
feature is compatible with a power distribution, as shown in \Fig{fig:his_fit_ecc}~(b).
When $\sigma_x=\sigma_y$ in the two-dimensional elliptic Gaussian \Eq{eq:epGaus},
integration over angle $\varphi$ is applicable, which leads to the 
Bessel-Gaussion function~\cite{Voloshin:2007pc},
\be
\label{eq:BG_dis}
P(\ve_n) = \frac{\ve_n}{\sigma^2}\exp\left(-\frac{\ve_n^2+\ve_0^2}{2\sigma^2}\right)
I_0\left(\frac{\ve_0\ve_n}{\sigma^2}\right)
\ee
As a comparison, the fit using the Bessel-Gaussian function 
is shown in \Fig{fig:his_fit_ecc} as the 
green dashed lines. The Bessel-Gaussian results in a worse description of 
the eccentricity distribution. Especially, one notices the non-zero tails 
of the Bessel-Gaussian at $\ve_n=1$ in \Fig{fig:his_fit_ecc}, as anticipated,
since in the Bessel-Gaussian distribution the upper bound of eccentricities
is infinity.

As a simple summary, we notice that the probability distributions of 
initial state eccentricities are apparently non-Gaussian,
as one sees in \Fig{fig:his_fit_ecc} that the elliptic-power and power 
distributions give rise to much better fits 
than the Bessel-Gaussian function. 
Considering the derivation
of elliptic-power and power distribution functions, assuming only N-independent
point-like sources on top of a Gaussian background, the non-Gaussianity comes 
dominantly from the fact that $\E_n$ is bounded by unity.

Elliptic-power and
power distributions have been successfully applied to parameterize a general class
of the generated initial state eccentricity from effective models, where differences of
these effective models are quantitatively captured by the parameters $\ve_0$ and $\alpha$~\cite{Yan:2014afa}.
Note that in these models, there are many other sources that contribute to the
non-Gaussianity of the initial eccentricity fluctuations. For instance, there are
higher order corrections 
concerning a more sophisticated configuration of the initial
state density profile rather than a two-dimensional Gaussian,  and 
extra correlations among sources, etc.~\cite{Blaizot:2014nia,Blaizot:2014wba,Gronqvist:2016hym},
which have been taken into account in theoretical analyses.

\begin{center}
\includegraphics[width=0.45\textwidth] {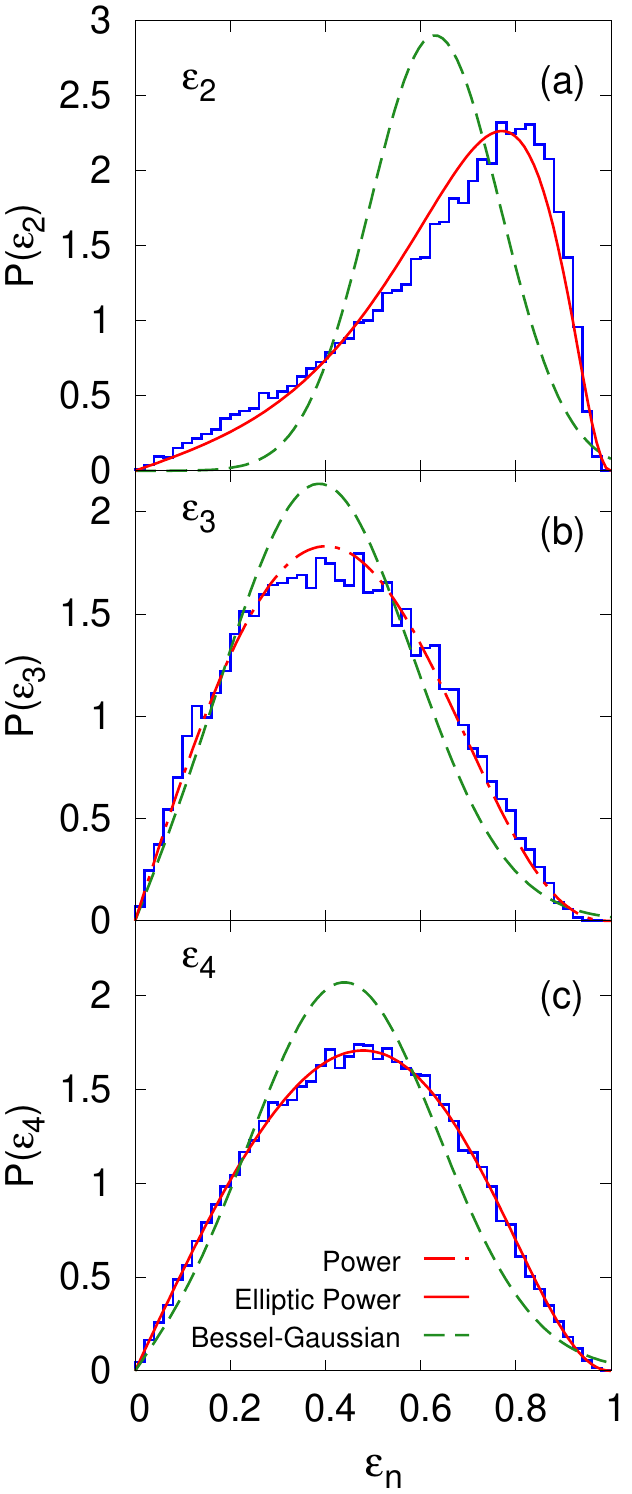}
\figcaption{Histogram of the initial state eccentricity (a) $\ve_2$,
(b) $\ve_3$ and (c) $\ve_4$ distributions, obtained from Monte-Carlo
simulations of the MC-Glauber model, with respect to the Pb+Pb collisions
in the 75-80\% centrality class, at $\sqrt{s_{NN}}=2.76$ TeV. Lines of different
styles and color are fit using elliptic-power (red solid), power (red dash-dot)
and Bessel-Gaussian (green dashed) functions. Figure reproduced from
\Ref{Yan:2014afa}, with permission.
\label{fig:his_fit_ecc}
}
\end{center}

\subparagraph{Correlations of $\E_n$}
Event-by-event fluctuations of the density profile also generate
correlations among initial state eccentricities~\cite{Teaney:2010vd,Jia:2012ju}.
These correlations are consequences of pure 
geometrical effects due to the averaged background geometry and fluctuations of
energy deposition from event to event~\cite{Yan:2015fva}. 
The similar idea of treating the density profile as independent sources
can be applied to the mixed correlations among $\E_n$~\cite{Bhalerao:2011bp}. 
In this way, one may consider introducing the event-by-event fluctuation of a
function $g$ as $\delta_g\equiv \{g\}-\bbra g\kket$. Note that $\{g\}$ is obtained
with respect to the density profile of one single event, while $\bbra g\kket$ 
is the averaged value over events. From the independent sources, one has the 
two-point correlation of fluctuations inversely proportional to the number of
independent sources~\cite{Alver:2008zza},
\be
\bbra \delta_f \delta_g \kket = \frac{\bbra fg\kket-\bbra f\kket \bbra g\kket}
{N}
\ee
Three-point and four-point correlations are found at the next leading-order,
in terms of $1/N^2$. 
Therefore, by 
taking account of the recentering corrections, 
fluctuating initial eccentricities $\E_2$ and $\E_3$ 
are written as
\begin{align}
\label{eq:id_e2e3}
\E_2 =& - \frac{\{(z-\delta_z)^2\}}{\{r^2\}}
\approx -\frac{\bbra z^2\kket+\E_0\delta_{r^2} +\delta_{z^2}}{\bbra r^2 \kket}+O(\delta^2)\cr
\E_3 =& - \frac{\{(z-\delta_z)^3\}}{\{r^3\}}
\approx -\frac{\delta_{z^3}-3\bbra z^2\kket \delta_z}{\bbra r^3 \kket}
+O(\delta^2)
\end{align} 
Note that we have assumed a non-zero averaged background ellipticity,
$\E_0=-\bbra z^2\kket/\bbra r^2\kket$ in the above equations, which corresponds to
the case of non-central nucleus-nucleus collisions. 
As a result, \Eq{eq:id_e2e3} implies the fact that $\ve_2\{2\}\sim \ve_0$,
while $\ve_3\{2\}\sim1/\sqrt{N}$.
One should be aware that in
colliding systems such as He$^3$+Au, the ultra-central collisions are expected to have
an averaged triangular shape, so that a non-zero background triangularity
should be assumed instead, $\E_0=-\bbra z^3\kket/\bbra r^3\kket$.

\Eq{eq:id_e2e3} can be generalized to other eccentricities, which
allows one to analytically derive the mixed correlations among $\E_n$, 
order by order with respect to $1/N$. For instance, at the leading-order,
the correlation 
between $\E_2$ and $\E_3$ has the following contribution,
\be
\label{eq:e2e3}
\bbra \E_2^3\E_3^{*2} \kket = 
-\frac{9|\E_0|^6}{N}\frac{ \bbra r^2\kket^3}{\bbra r^3\kket^2}\,.
\ee
Although the above estimate is only valid with respect to independent sources,
it generically captures  the feature of negative correlation. Besides, one also
expects from \Eq{eq:e2e3} that the strength of correlation grows when
$N$ decreases, corresponding to an increase of centrality percentile
in heavy-ion collisions. Both of these features are confirmed in model
simulations, as shown in \Fig{fig:ids_pc}.
\Fig{fig:ids_pc}
displays the 
mixed correlators involving $\E_1$, $\E_2$ and $\E_3$, from Monte-Carlo 
simulations of a CGC-typed model and the Glauber model.
Analytic solutions from independent sources are obtained in terms of the expansion
in inverse to the number of sources. One notices that the correlations are 
well-described by analytical results from independent sources.
The correlations of initial state eccentricities are observed quite generally
in various effective models, which inspire the measurement of the mixed
correlations in harmonic flow~\cite{Bhalerao:2013ina}.

\begin{center}
\includegraphics[width=0.5\textwidth] {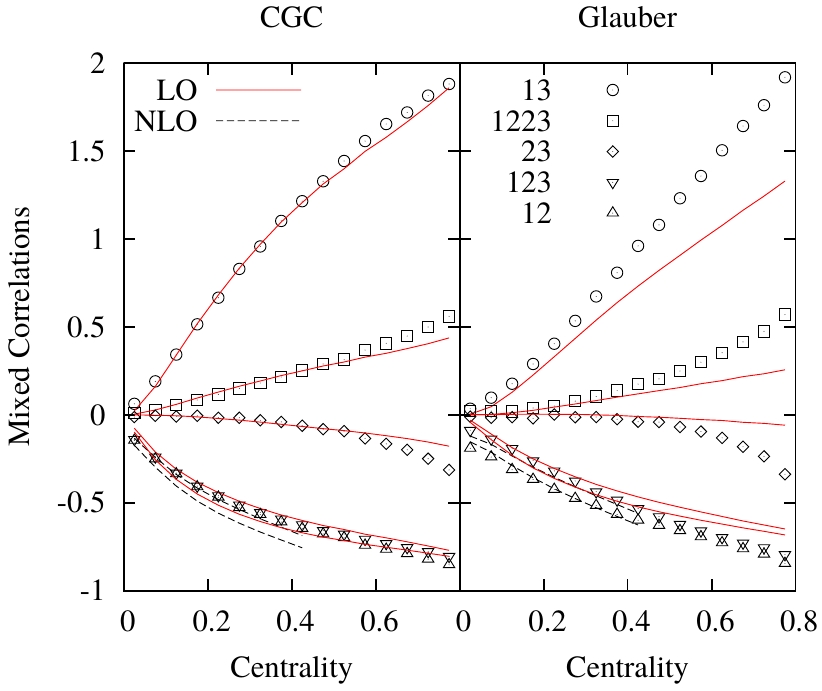}
\figcaption{ 
Mixed correlations among initial state eccentricities $\E_n$. Red solid and
black dashed lines are estimates from independent sources, to the leading-order
in $1/N$
and next leading-order, respectively. Symbols are Monte-Carlo simulations
from the CGC model (left panel) and the Glauber model (right panel).
The notations indicate to the mixed correlations of the corresponding 
harmonic orders, with a proper normalization of the magnitude of eccentricities.
Figure reproduced from \Ref{Bhalerao:2011bp}, with permission.
\label{fig:ids_pc}
}
\end{center}

\subsection{Medium response to $\E_n$}
\label{sec:resp}

Solutions to the hydrodynamic equations of motion are completely 
determined once an initial condition is specified. 
With respect to 
the fluctuating initial state characterized in
terms of 
initial state eccentricities, the hydrodynamic
predictions of harmonic flow are expected as a function of these eccentricities, 
\be
\label{eq:vn_form}
V_n = V_n(\E,\alpha)\,.
\ee
In \Eq{eq:vn_form}, $\E$ denotes a set of initial state eccentricities that are responsible
to $V_n$, while $\alpha$ contains parameters related to the medium dynamical properties,
such as the transport coefficient $\eta/s$. 
Although the explicit form of \Eq{eq:vn_form} is not known 
\emph{a priori} from first-principle calculations, 
there is mounting evidence from numerical hydrodynamic
simulations suggesting that one may expand 
$V_n(\E,\alpha)$ in terms of $\E$, 
\be
\label{eq:exp0}
V_n = \kappa(\alpha) \E 
+ O(\E^2) + \delta_n\,,
\ee
with respect to the fact that $|\E|<1$. The quantity 
$\delta_n$ is the residual introduced accounting for deviations 
due to additional fluctuations. 
By assumption, $\delta_n$ is
uncorrelated with initial state eccentricities, which we shall address later.
Note that $\delta_n$ is complex, since 
\Eq{eq:exp0} relates complex quantities on both sides of the equation.
More precisely, the magnitudes and the phases of both sides in the equation
are identical respectively.

In practice one would like to truncate the expansion at a finite order,
so that harmonic flow can be well approximated.
These terms in the expansion
correspond to the medium response to 
the initial geometry $\E$, from the linear order,
to nonlinear mixing of higher orders.
Comments are in order with respect to the expansion \Eq{eq:exp0}.
\begin{enumerate}

\item
The expansion relies on the fact that $\E_n$'s are small quantities. 
In collisions of large systems such as Pb+Pb and Au+Au,
this criterion is generally satisfied for 
harmonic orders $n\ne2$, because initial state eccentricities of 
order $n\ne2$ are generated entirely from fluctuations. One estimates the 
magnitude of eccentricity $\ve_n\sim 1/\sqrt{N}\ll 1$. 
Whereas when $n=2$,
ellipticity comes dominantly from the background geometry
in non-central collisions. Therefore, $\E_2$ is more significant
than other eccentricities in the expansion. Effective model simulations 
have shown that towards peripheral collisions, $\ve_2$
can grow above 0.5, which implies the role of nonlinear 
order terms involving $\E_2$ in the expansion.

\item 
By expanding in $\E_n$'s, the dependence on medium dynamical properties
is absorbed separately in the expansion coefficients, 
$\kappa(\alpha)$, etc.
We shall refer to the coefficient of the linear order $\kappa(\alpha)$ 
as the linear response coefficient.  Coefficients of higher
orders as nonlinear response coefficients. These coefficients
are remarkable in probing the medium dynamical properties, as their
dependence on the initial state is minimized. 
Although \Eq{eq:exp0} is written on an event-by-event basis, one would expect the
medium response coefficients not to fluctuate in one centrality class.
These coefficients are calculable in hydrodynamic simulations, and 
as will be discussed, some of them are accessible in experiments under
fairly general assumptions.

\item 
In each single event, with respect to the rotational symmetry of $V_n$, 
each term on the right-hand side of \Eq{eq:exp0} is invariant under the 
rotation $\phi\rightarrow \phi+2\pi/n$.
Since response coefficients are real quantities
according to parity conditions, 
at each order the allowed combinations of $\E_n$'s can be determined
by symmetry conditions. 
For the response of linear order, it is apparent that
$V_n^L$ (the linear part of flow $V_n$) is proportional to $\E_n$. 
It should be emphasized that there exist ambiguities in the definition of
$\E_n$. Symmetry constraints on the nonlinear order part of the 
flow $V_n^{NL}$ are more useful.
For instance, for the fourth order flow harmonics,
in addition to a linear response to $\E_4$,
the nonlinear contributions at quadratic order include the coupling of $\E_2$, i.e.,
$V_4^{NL} \propto  \E_2^2 $.
For $V_5$, rotational symmetry requires the nonlinear part to be
generated by quadratic coupling between $\E_2$ and
$\E_3$, 
$V_5^{NL} \propto \E_2 \E_3$.

\end{enumerate}

With respect to all these aspects, the following forms have been
found successful in practical applications of
the expansion \Eq{eq:exp0}. From
the elliptic flow $V_2$ to
harmonic order $n=6$~\cite{Yan:2015jma,Qian:2016fpi},
\begin{subequations}
\label{eq:vnform}
\begin{align}
\label{eq:v2form}
V_2=&\kappa_2 \E_2 + \kappa_2'\ve_2^2 \E_2 + \delta_2\,,\\
\label{eq:v3form}
V_3=&\kappa_3 \E_3 + \kappa_{23}'\ve_2^2 \E_3 + \delta_3\,,\\
\label{eq:v4form}
V_4=&\kappa_4 \E_4 + \kappa_{422}\E_2^2 + \delta_4\,,\\
\label{eq:v5form}
V_5=&\kappa_5 \E_5 + \kappa_{523}\E_2\E_3 + \delta_5\,,\\
\label{eq:v6form}
V_6=&\kappa_6 \E_6 + \kappa_{633}\E_3^2 + \kappa_{624}\E_2\E_4
+\kappa_{6222}\E_2^3+\delta_6\,.
\end{align}
\end{subequations}
One notices that higher order flow becomes more complicated since more
terms from the nonlinear mode mixings are required to achieve a good 
approximation.
There have also been some attempts in the similar analysis for the flow
$V_7$ and $V_8$, 
\begin{eqnarray}
V_7&=& \kappa_7\E_7+\kappa_{7223} \E_2^2\E_3 + \ldots+\delta_7\,\cr
V_8&=& \kappa_8\E_8+\kappa_{82222} \E_2^4+\ldots+\delta_8\,,
\end{eqnarray}
where even more involved terms up to the 
quartic order nonlinear couplings contribute. 
Although 
\Eqs{eq:vnform} are kind of idealistic, in the sense that they require
 explicit information of the $\E_n$'s, 
these relations are of great significance in the flow paradigm. 
These response
relations can be applied to quantitatively describe the final state flow observables
in terms of  the dynamical properties of the medium, and the initial state 
eccentricities $\E_n$. 
In the following subsections, we shall summarize
some important details in \Eqs{eq:vnform} based on solutions 
of viscous hydrodynamics.

\subsubsection{Viscous effects on the medium response }

The effect of fluid response to the initial state perturbations is suppressed
by viscous corrections in 
hydrodynamics. 
When the solution of the system evolution is decomposed into modes, it is further
expected that the 
higher order modes get stronger viscous corrections. 

In theory, the argument that the response of higher order mode is more damped
by viscous corrections can be verified 
in several cases with analytic solutions of viscous hydrodynamics. In the linear
response analysis with respect to a static fluid background, considering any
types of initial perturbations of hydrodynamic fields which are decomposed into
modes $k$ (wave-number), it is known that the evolution of these modes 
undergoes viscous corrections 
proportional to $-k^2 \eta/s$, as a result of
the diffusion feature of the Navier-Stokes equations (cf. \Ref{1963AnPhy..24..419K}). 

\begin{center}
\includegraphics[width=0.4\textwidth] {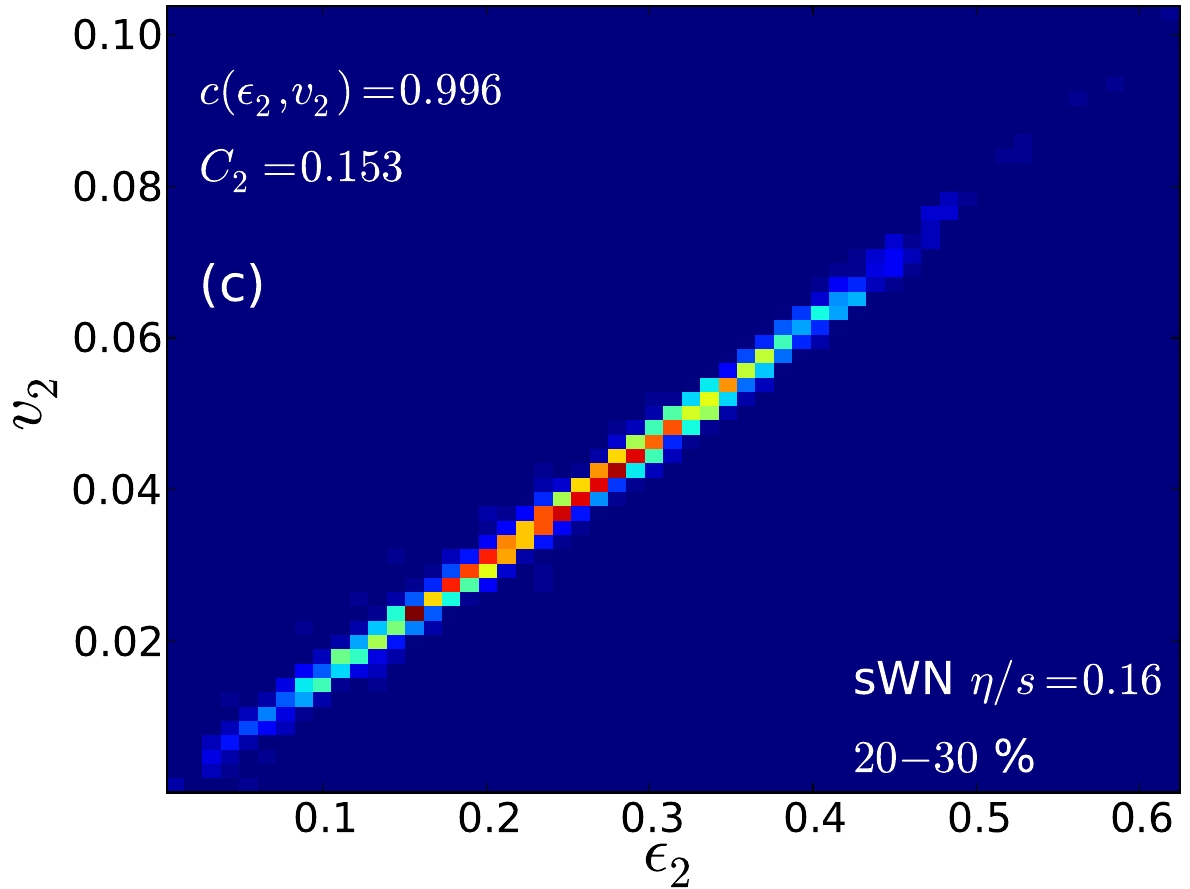}
\includegraphics[width=0.41\textwidth] {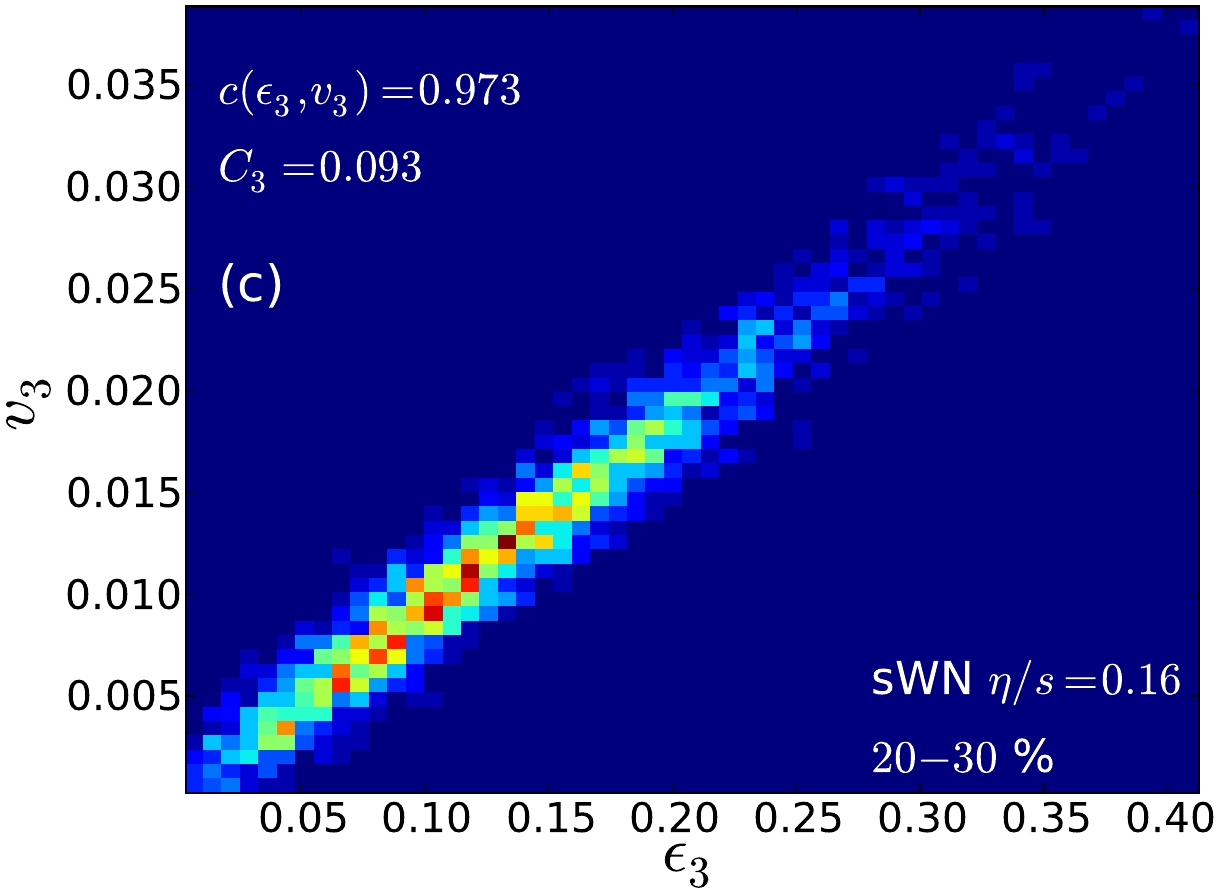}
\figcaption{ 
\label{fig:resp0}
Response of the flow magnitudes $v_2$ and $v_3$ to the magnitudes of initial ellipticity $\ve_2$
and triangularity $\ve_3$, obtained from event-by-event hydro simulations of the Au+Au collisions
in one centrality class, $20\%-30\%.$. Figures reproduced from \Ref{Niemi:2012aj}, with permission.
}
\end{center}

Similar behaviors of the fluid dynamics are also present in the analytic solutions 
of expanding systems: the 0+1 dimensional Bjorken flow~\cite{Bjorken:1982qr} and 1+1 dimensional 
Gubser flow~\cite{Gubser:2010ui}. 
Gubser flow is an analytical solution of a conformal fluid system, with
system expansion realized both in the longitudinal direction along space-time rapidity,
and in the radial direction along $r$. The analytical solution of the Gubser flow 
assumes a background which is rotationally symmetric. Deformation of the 
symmetric background results in mode decomposition in terms of spherical harmonics
$\E_n^{(l)}\sim Y_n^l$. The index $l$ captures the fluctuations along the radial direction, while
$n$ is associated with the azimuthal angle. It should be noted that the mode decomposition 
in terms of $n$ is
consistent with the Fourier decomposition used in the usual 
definition of initial state
eccentricities.
Corresponding to the initial state geometrical fluctuations in heavy-ion 
collisions, one may approximate the mode $\E_n^{(l)}$ in the Gubser flow,
as the ordinarily defined eccentricity $\E_n$ from moment $\{r^l e^{in\phi}\}$. 
For instance, in the Gubser flow the mode $\E_2^{(2)}$ can be recognized as the ellipticity $\E_2$.
For each $n$-th order initial eccentricity, the dominant one in the decomposition 
comes from the lowest order $l$, satisfying $n=l$,
which is suppressed in the Navier-Stokes hydrodynamics according to the factor
\be
\label{eq:gubser_sup}
\exp(-l^2 \eta/s)\sim
\exp(-n^2 \eta/s)\,.
\ee
\Eq{eq:gubser_sup} also reflects  the diffusion feature of the mode evolution in
hydrodynamics, in analogous to
the $k^2$ suppression in a static fluid. 

In realistic systems in heavy-ion collisions, which are not conformal, nor
rotationally symmetric,
there is no analytical relation solved between harmonic order $n$
and viscous corrections. However, as inspired in these analytic solutions,
especially \Eq{eq:gubser_sup},
one still expects a $n^2$-scaling of the viscous corrections~\cite{Staig:2011wj,Lacey:2013eia,Gubser:2010ui}. 
Indeed, numerical solutions
in the single-shot viscous hydrodynamics have found very similar trends
in the linear response to the moment (and cumulant) based initial eccentricities,
as viscosity increases~\cite{Teaney:2012ke}. 
Owing to the complexity in exacting linear response
coefficients, however, the $n^2$-scaling has not been validated 
in event-by-event
hydrodynamic simulations.

The $n^2$ dependence of the viscous suppression in the fluids has strong indications
in the flow paradigm. In particular, in addition to the simple fact that the linear response of 
higher order flow harmonics is more suppressed than
the lower order ones, nonlinear response coefficients are always less
suppressed than the corresponding linear response, for the same harmonic order\footnote{
This statement is true when the viscous correction of harmonic order $n$ is
proportional to 
$n^\alpha$, as long as $\alpha>1$, i.e., not necessarily $\alpha=2$.
}.
As a consequence, in higher order harmonic flow, with $n\ge4$, the
dominant contributions are from the nonlinear mode mixings from $\E_2$ and $\E_3$, etc., 
in the expansion
equation \Eq{eq:vnform}, when viscous effects are sufficiently strong.

\subsubsection{Medium response in $V_2$ and $V_3$}

\end{multicols}
\begin{center}
\includegraphics[width=0.45\textwidth] {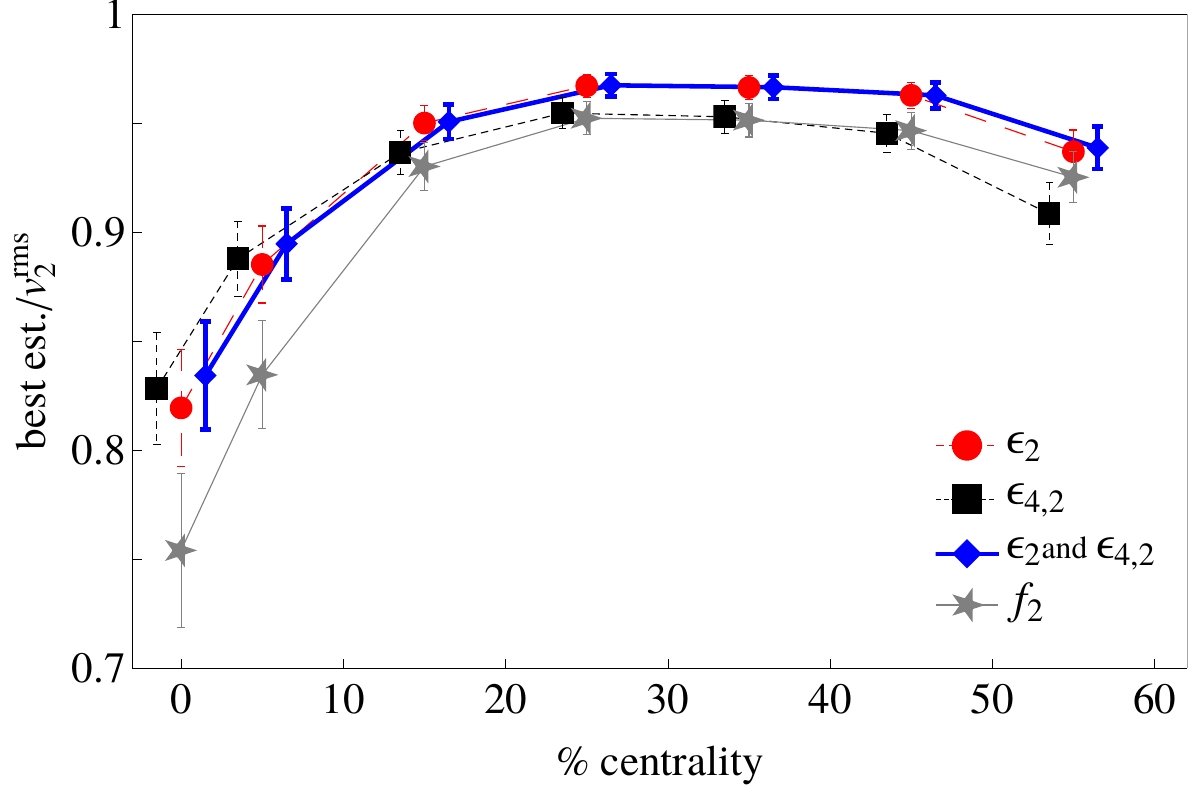}
\includegraphics[width=0.45\textwidth] {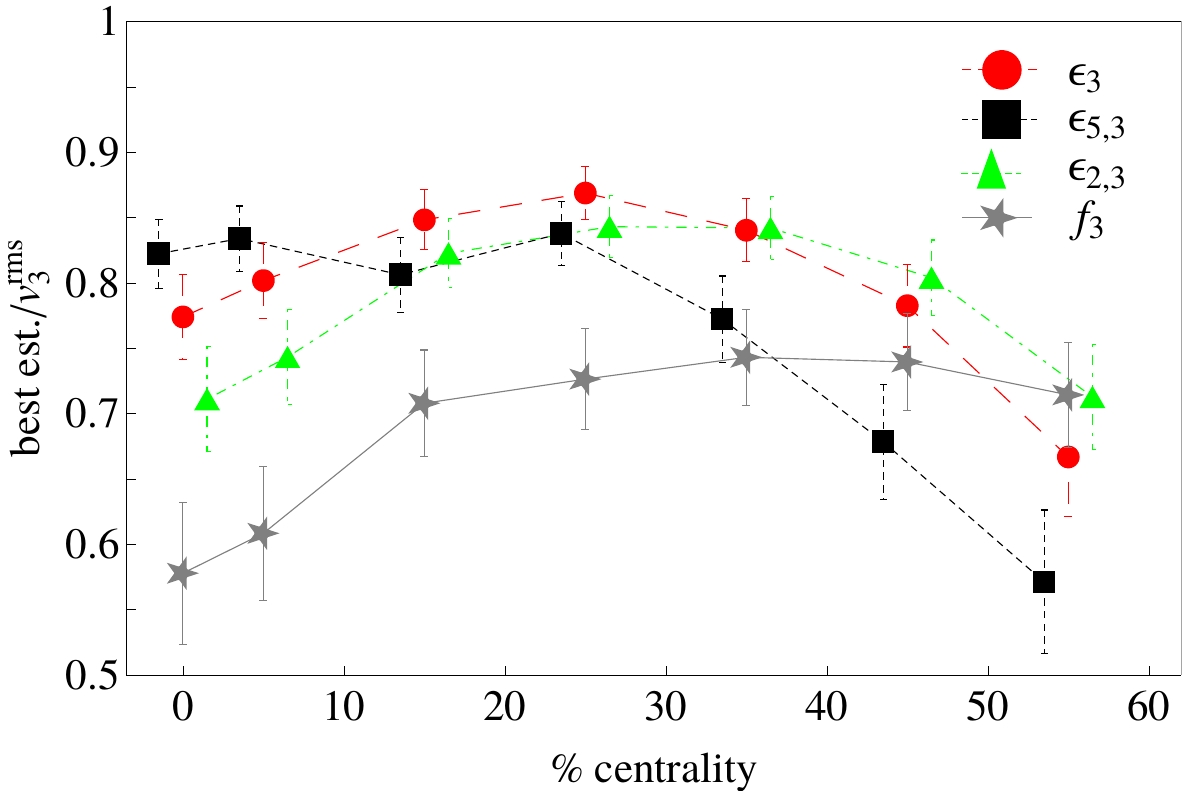}
\figcaption{ 
\label{fig:resp1}
Correlations of  the elliptic flow $v_2$ (left panel) and 
triangular flow $v_3$ (right panel) for different definitions of initial 
eccentricities, characterized in terms of Pearson correlation
coefficients as a function of centrality percentile. Figures taken
from \Ref{Gardim:2011xv}, with permission.
}
\end{center}
\begin{multicols}{2}

\subparagraph{Linear order response}
We start with the two lower order flow harmonics, elliptic flow $V_2$ and triangular
flow $V_3$. These two types of flow harmonics are different 
from the
higher order ones in the flow paradigm, as they are dominated by a linear medium 
response. Nonlinear mode mixings
appearing in \Eq{eq:v2form} and \Eq{eq:v3form} are of cubic order, whose
effects are minor expect in some particular situations.
Ignoring nonlinear terms, the linear relations 
\begin{align}
\label{eq:linear}
V_2=\kappa_2 \E_2 \,,\qquad
V_3=\kappa_3 \E_3\,, 
\end{align}
simplify the theoretical analysis significantly. From \Eq{eq:linear},
one expects the magnitudes of $V_2$ and $V_3$ to be linearly proportional to 
those of $\E_2$ and $\E_3$, respectively. Besides, the initial state participant
planes $\Phi_2$ and $\Phi_3$ are aligned respectively to the event-plane $\Psi_2$ and 
$\Psi_3$, which are determined from the observed particle spectrum. 
As a result, fluctuations and correlations
seen in flow harmonics are to a large extent understandable in terms of those in $\E_2$ and $\E_3$.
For instance, the probability
distribution of $v_2$ is then a rescaled distribution of $\ve_2$. In turn, 
features of the cumulants of initial $\ve_2$ are detectable once the 
$v_2$ cumulants are measured in experiments.

The linear relation can be examined simply in hydrodynamic
simulations~\cite{Qiu:2011iv,Niemi:2012aj}. Scatter plots in \Fig{fig:resp0}, for example, present the linear 
relation between the magnitude $v_2$
and magnitude $\ve_2$, and between $v_3$ and $\ve_3$, based on
simulations of viscous
hydrodynamics with respect to Au+Au collision events of $\sqrt{s_{NN}}=200$ MeV,
in the centrality class 20\%-30\%~\cite{Niemi:2012aj}. 
The linearity is obvious in the figures, with the slope corresponding to the
linear response coefficient $\kappa$. However, it should be emphasized that the slope 
is \emph{not} identical to the linear response coefficient $\kappa$. One notices a larger slope in
the scatter plot of $v_2$ than $v_3$  
in the same centrality class, signifying a stronger medium linear response to
the ellipticity than triangularity.
A stronger correlation between 
$v_2$ and $\ve_2$ is observed, 
as the dispersion in the scatter plot of triangularity is more 
pronounced.

To validate the linear response relations in a more quantitative way,
 one defines the Pearson correlation coefficient~\cite{Bhalerao:2011bp}. For $V_2$, it is written
as
\be
\P_{2}= \frac{\Re\bbra V_2 \E_2^* \kket}{\sqrt{\bbra v_2^2\kket}\sqrt{\bbra \ve_2^2\kket}}\,,
\ee
in terms of event-averaged quantities.
The Pearson coefficient $\P_2$ measures the linear correlation between $V_2$ and $\E_2$. 
It should be emphasized that $\P_2$ measures the linear relation simultaneously
between both magnitudes and phases.  
By construction, $\P_2$ vanishes
when $V_2$ is uncorrelated with $\E_2$, while
$\P_2=\pm 1$ indicates an absolute correlation or
anti-correlation. 
Similarly, for 
$V_3$ one has
\be
\P_{3}= \frac{\Re\bbra V_3 \E_3^* \kket}{\sqrt{\bbra v_3^2\kket}\sqrt{\bbra \ve_3^2\kket}}\,,
\ee
which measures the linear correlation between $V_3$ and $\E_3$.
It is worth mentioning that the linear response coefficient $\kappa$ can be
calculated analogously via these event-averaged quantities, i.e.,
\begin{align}
\label{eq:lkappa}
\kappa_{2}= \frac{\Re\bbra V_2 \E_2^* \kket}{{\bbra \ve_2^2\kket}}\,,\qquad
\kappa_{3}= \frac{\Re\bbra V_3 \E_3^* \kket}{{\bbra \ve_3^2\kket}}\,,
\end{align}
The linear response coefficients deviate from the slope of the scatter plot  in \Fig{fig:resp0},
which is given as $\bbra v_n\ve_n\kket/\bbra |\ve_n|^2\kket$.

In \Fig{fig:resp1}, the Pearson correlation coefficients are calculated in viscous
hydrodynamics, with various types of characterizations of the 
initial state ellipticity (left panel) and triangularity (right panel)
examined. On a general ground, the ellipticity defined in \Eq{eq:eccn} with
a $r^2$ weight (blue points)
is most responsible in the linear relation to $V_2$, leading to $\P_2$ above 0.95 for most
of the centrality bins. The correlation gets stronger towards non-central collisions, as 
one observes an increasing trend of $\P_2$. Ellipticity defined with a $r^4$ weighting
can be found in the cumulant expansion of the initial density profile, whose linear
correlation with $V_2$ is also found to be strong in \Fig{fig:resp1} (black points). Actually
the $r^4$-weighted ellipticity is more responsible for $V_2$ in central collisions,
which signifies the role of outer layer in the density profile in the generation
of elliptic flow in central collisions. The quantity $f_2$ in \Fig{fig:resp1} is adopted from
the mode decomposition with respect to the Gubser flow~\cite{Gubser:2010ui}, whose
correlations with $V_2$ are weaker in all centrality classes.

\end{multicols}
\begin{center}
\includegraphics[width=0.45\textwidth] {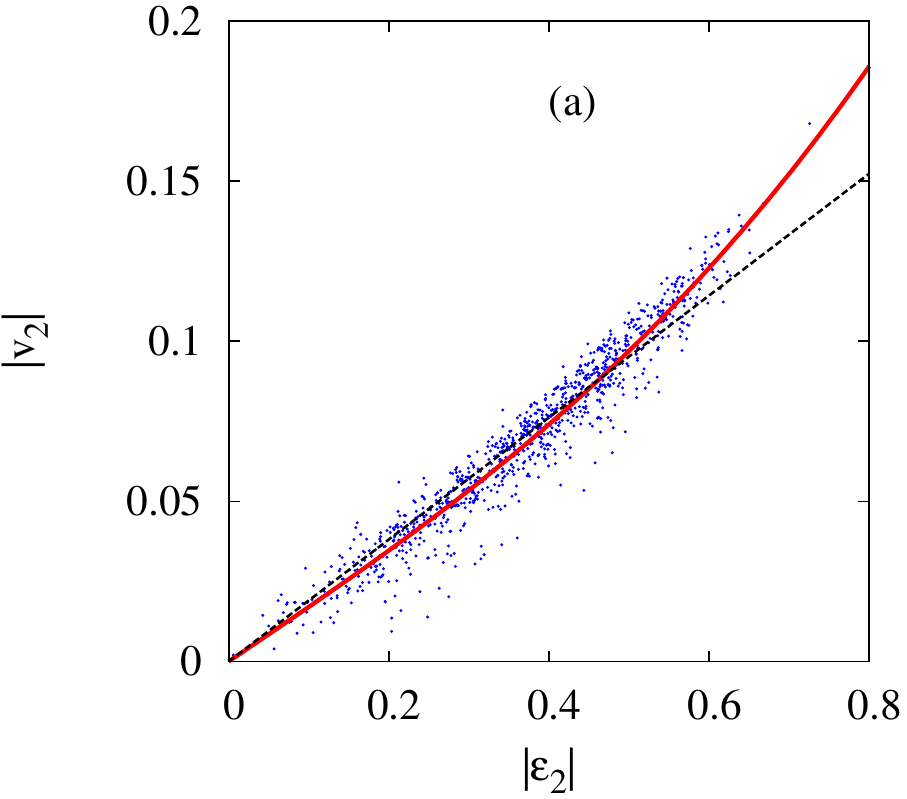}
\includegraphics[width=0.45\textwidth] {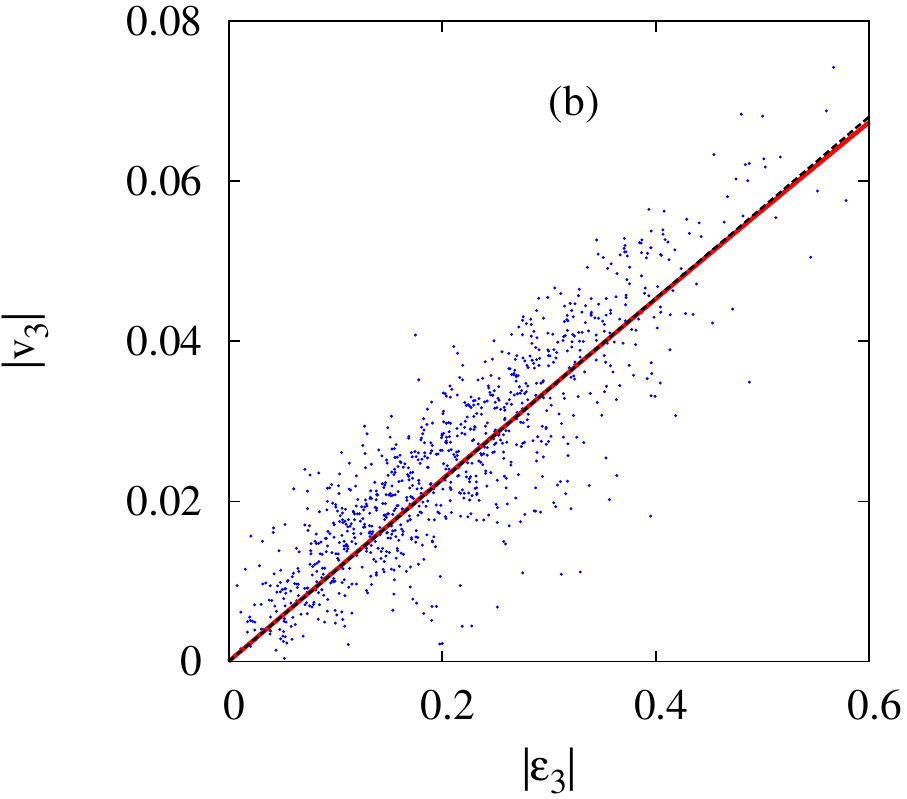}
\figcaption{ 
\label{fig:resp2}
Correlation between the magnitudes
of anisotropic flow $v_n$ and initial eccentricity $\ve_n$ for Pb+Pb
collisions at 2.76 TeV in the 45-50\% centrality range. Each
point corresponds to a different initial geometry. Dotted line:
linear estimator. Full line: cubic estimator.
(a) Elliptic flow. (b) Triangular flow. Figures reproduced from \Ref{Noronha-Hostler:2015dbi}, with permission.
}
\end{center}
\begin{multicols}{2}

Similar analyses for the triangular flow are shown in the right-hand panel of \Fig{fig:resp1}.
The correlation of the linear relation is generally weaker in $V_3$ than $V_2$, consistent
with the observation in \Fig{fig:resp0}. The linear correlation with 
the triangularity defined in \Eq{eq:eccn} with a $r^3$ weighting  
is reasonably good, though a $r^2$ weighting is more favorable in peripheral bins, which
implies the significance of inner layer in the generation of $V_3$ in systems of smaller
sizes. But in larger systems, the outer layer is more sensitive,
as one notices the stronger linear correlation with the $r^5$-weighted definition.
The $r^5$-weighted triangularity is again derived from the cumulant expansion 
of initial state density profile. Similarly, in the mode decomposition developed from the
Gubser flow, $f_3$ provides weaker correlations to $V_3$.

Viscous effects on the linear medium response are expected to suppress the response 
coefficient $\kappa$, as has been discussed in the previous section. Besides, event-by-event
hydro simulations have also shown that viscosity tends to enhance the linear correlation 
between initial
state eccentricity and flow~\cite{Niemi:2012aj}.

\subparagraph{Cubic order response}

The increasing trend of the Pearson correlation coefficients $\P_2$ and $\P_3$ seen in \Fig{fig:resp1} 
from central to non-central 
bins can be understood as
owing to the growth of ellipticity and triangularity, against residual fluctuations,
etc. When going towards very peripheral collisions, however, one notices a weak but clear 
drop of the correlation with respect to linearity, indicating the role of
nonlinear terms in \Eq{eq:v2form} and \Eq{eq:v3form}.

Deviation from the linear relation
in the peripheral collision bins for 
$V_2$ is visible in hydrodynamic simulations~\cite{Niemi:2015voa}.
Shown in \Fig{fig:resp2}(a) is a scatter plot of the correlation between
the magnitudes $v_2$ and $\ve_2$, obtained in the
centrality class 45\%-50\% of Pb+Pb collisions
at $\sqrt{s_{NN}}=2.76$ TeV~\cite{Noronha-Hostler:2015dbi}. Each point in the plot corresponds to
an event with a randomly specified initial geometry from the Monte Carlo
Glauber model.
A nonlinear trend in the plot is obvious for the events with 
large values of $\ve_2$.
In the same centrality class, deviation from the linear correlation between the
magnitude $v_3$ and the magnitude $\ve_3$ is  seen to be negligible
in \Fig{fig:resp2}(b). Black dashed lines correspond to the linear 
response relation, with the slope calculated according to \Eqs{eq:lkappa}. 
Red solid lines are results containing a cubic order response, with
the response coefficients determined as in \Eq{eq:kap2p}
and \Eq{eq:kap3p} below.

\end{multicols}
\begin{center}
\includegraphics[width=0.45\textwidth] {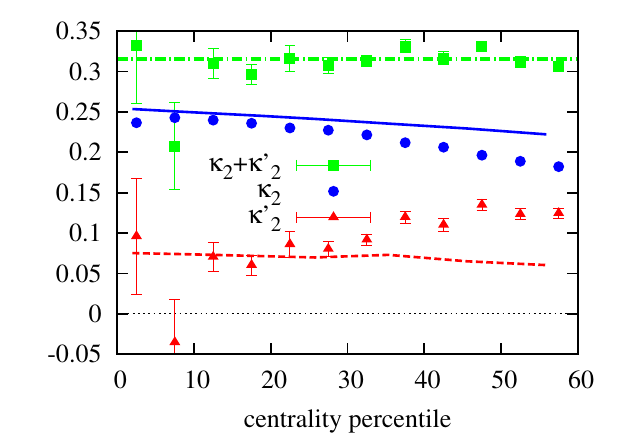}
\includegraphics[width=0.45\textwidth] {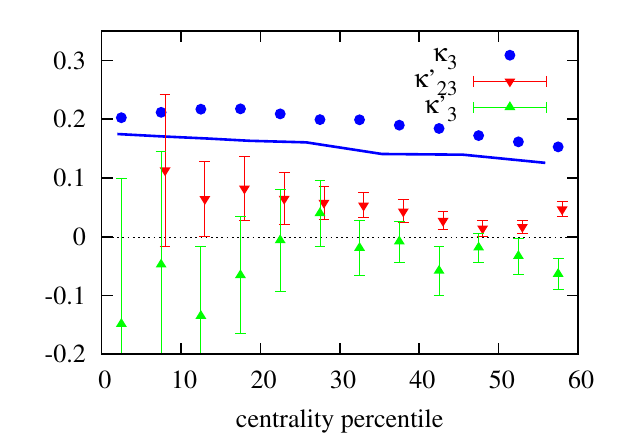}
\figcaption{ 
\label{fig:kappa}
Linear and cubic order response coefficients calculated from
event-by-event hydro simulations for the Pb+Pb collisions
of $\sqrt{s_{NN}}=2.76$ TeV. Upper panel: elliptic flow.
Lower panel: triangular flow. Figures reproduced from \Ref{Noronha-Hostler:2015dbi}, with permission.
}
\end{center}
\begin{multicols}{2}

To include nonlinear effects in the response 
function, for $V_2$ and $V_3$ the allowed terms by symmetry are
of cubic order. Following the rule that $\E_2$ is dominant in the expansion,
one arrives at the terms $\propto \ve_2^2 \E_2$ for $V_2$ and 
$\propto \ve_2^2 \E_3$ for $V_3$, respectively.
There are new response coefficients corresponding to the
cubic order terms. Determination of the linear response coefficient
$\kappa$ and the cubic order response coefficient $\kappa'$ 
can be approached 
by minimizing the effects of residual fluctuations $\bbra |\delta_n|^2\kket$.
Or equivalently, it is the maximization of the Pearson correlation coefficient
with respect to
the flow constructed with the cubic order response terms.
For the elliptic flow $V_2$, it leads to
\begin{subequations}
\label{eq:kap2p}
\begin{align}
\kappa_2 =&
\frac{\Re\left(\bbra \ve_2^6\kket\bbra V_2\E_2^*\kket-\bbra \ve_2^4\kket\bbra \ve_2^2 V_2\E_2^*\kket\right)}
{\bbra \ve_2^6\kket\bbra \ve_2^2\kket-\bbra \ve_2^4\kket^2}\,,\\
\kappa_2'=&
\frac{\Re\left(-\bbra \ve_2^4\kket\bbra V_2\E_2^*\kket+\bbra \ve_2^2\kket\bbra V_2\E_2^* \ve_2^2\kket\right)}
{\bbra \ve_2^6\kket \bbra \ve_2^2\kket-\bbra \ve_2^4\kket^2}\,.
\end{align} 
\end{subequations}
Similarly for the triangular flow $V_3$,
\begin{subequations}
\label{eq:kap3p}
\begin{align}
\kappa_3 =&\frac{\Re\left(\bbra V_3\E_3^*\kket\bbra \ve_2^4\ve_3^2\kket-\bbra V_3\E_3^*\ve_2^2\kket\bbra \ve_2^2\ve_3^2\kket\right)}
{\bbra \ve_3^2\kket\bbra \ve_2^4\ve_3^2\kket-\bbra \ve_2^2\ve_3^2\kket^2}\,,\\
\kappa_{23}'=&\frac{\Re\left(-\bbra V_3\E_3^*\kket\bbra \ve_2^2\ve_3^2\kket+\bbra V_3\E_3^*\ve_2^2\kket\bbra \ve_3^2\kket\right)}
{\bbra \ve_3^2\kket\bbra \ve_2^4\ve_3^2\kket-\bbra \ve_2^2\ve_3^2\kket^2}\,.
\end{align} 
\end{subequations}

In comparison with \Eq{eq:lkappa}, when the cubic order response terms contribute,
the resulting linear response coefficients $\kappa_2$ and
$\kappa_3$ get extra \emph{negative} 
corrections which scale as $\ve_2^2$. The corrections are
potentially significant at very peripheral collision bins.
With  the cubic order response, one may also plug in the estimated flow out of
initial eccentricities in the evaluations of Pearson correlation coefficients. The resulting
correlation is improved for both
$V_2$ and $V_3$, but the improvements are not sizable.

The calculated result of these coefficients from viscous hydrodynamic simulations
can be found in \Fig{fig:kappa}, as a function of centrality percentile~\cite{Noronha-Hostler:2015dbi}. 
The symbols are obtained with respect to 
\Eq{eq:kap2p} and \Eq{eq:kap3p} through 
event-by-event simulations of viscous hydrodynamics. 
Solid blue lines are the linear response coefficients from the single-shot
hydrodynamic calculations. In the single-shot hydrodynamic calculations, 
hydrodynamic equations of motion are solved using a smooth initial condition, 
which is normally a deformed two-dimensional Gaussian 
distribution. The deformations can be introduced properly so that there is only one
type of initial eccentricity involved in the initial state. Hence it involves single mode
evolution. 

\begin{center}
\includegraphics[width=0.5\textwidth] {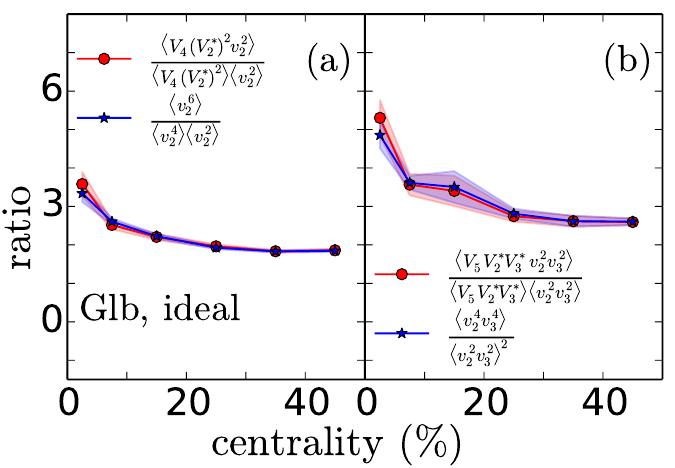}
\figcaption{ A test of the uncorrelation between the linear and nonlinear parts of $V_4$ (left panel)
and $V_5$ (right panel), based on event-by-event hydrodynamic simulations.
Figure reproduced from \Ref{Qian:2016fpi}, with permission.
\label{fig:vnvnl}
}
\end{center}

In either type of calculation, the obtained linear response coefficients
decrease as centrality percentile grows, in line with an increasing effect of 
viscous corrections in the medium system. The viscous effect is
stronger in the event-by-event simulations using MC-Glauber initial
conditions, although both calculations
took the same value of $\eta/s=0.08$\footnote{
An intuitive explanation can be given as follows:
in event-by-event simulations with respect to random initial distributions, 
the calculated response contain all higher modes. Hense if response of higher
modes are more sensitive to $\eta/s$, the viscous effects are expected 
stronger in the event-by-event calculations, 
comparing to the single-shot hydrodynamics where only the lowest
order mode is calculated.}.
It is consistent with the expectation in the linear response theory,
that the response behavior diffuses with the diffusion proporntional
to $\eta/s$. It is interesting to note that the viscous effect on 
cubic order response coefficient $\kappa_2'$ is opposite, as $\kappa_2'$
has a slight increase against the centrality percentile. 
Again, the increase is more evident in the results obtained
from event-by-event simulations. It is interesting to note that, for 
$v_2$, the sum of the linear
and the cubic order response coefficients remains constant, regardless
the change of centrality.
Although nonlinear deviations in the scatter plot in \Fig{fig:resp2} of $V_3$ are
not obvious, a finite cubic order response coefficient $\kappa_{23}'$ is
found for \Eq{eq:kap3p}.
 For comparison, the contribution from a cubic coupling of $\ve_3^2\E_3$ to
$V_3$ is examined in the calculations,
\be
V_3=\kappa_3 + \kappa_3' \ve_3^2 \E_3 + \delta_3\,,
\ee
with the corresponding cubic order response coefficient $\kappa_3'$ found compatible
with zero from the event-by-event hydrodynamic simulations.

\subsubsection{Medium response in higher harmonic orders}

When analyzing 
higher order flow harmonics in the flow paradigm, it is very useful to rewrite
the $\E_2$ and $\E_3$
in terms of $V_2$ and $V_3$ in \Eqs{eq:vnform}, respectively,
\begin{subequations}
\label{eq:vnform1}
\begin{align}
V_4 =&V_4^L + \chi_{422} V_2^2 + \delta_4\,\\
V_5 =&V_5^L + \chi_{523} V_2 V_3 + \delta_5\,\\
V_6 =&V_6^L + \chi_{633} V_3^2 + \chi_{624} V_2 V_4^L + \chi_{6222} V_2^3 + \delta_6\,,
\end{align}
\end{subequations}
where in a similar manner, the linear response part of higher order flow is left implicitly as 
$V_n^L$.
By doing so, it becomes possible to analyze directly the quantitative relations among
high order flow harmonics, and $V_2$ and $V_3$, so that ambiguities resulting from the 
definition of higher order eccentricities are avoided. Accordingly, 
uncertainties from the effective characterizations of initial state are reduced.

The re-expressions in terms of $V_2$ and $V_3$
are allowed in the expansion due to the fact that the 
linear response relation in the lower order flow harmonics 
is well approximated and has been tested in hydrodynamic simulations. Accordingly,
there are new nonlinear flow response coefficients
introduced. For the nonlinear contribution to $V_4$,
$\chi_{422}$ quantifies how much the contribution comes 
from the mixing of $V_2^2$. For $V_5$, $\chi_{523}$ quantifies how much the contribution to $V_5$ is
from the mixing of $V_2V_3$, etc. These new nonlinear response coefficients are related to
those $\kappa$'s  written in \Eqs{eq:vnform}, e.g., $\chi_{422}=\kappa_{422}/\kappa_2^2$ and
$\chi_{523}=\kappa_{523}/(\kappa_2\kappa_3)$.

Let us emphasize again that, unlike the lower order flow, 
there is subtlety in interpreting the linear response of higher order flow
harmonics as $V_n^L\propto \E_n$, due to the ambiguity 
in the $\E_n$ definitions. (Some recent hydrodynamic simulations suggest
stronger correlations between the linear part of $V_4$ and $V_5$ with
the corresponding initial state eccentricities from cumulant definition~\cite{Qian:2017ier}.)
Nevertheless, seen from \Eqs{eq:vnform1}, one
may 
recognize $V_n^L$ as the 
part 
uncorrelated to the nonlinear response contributions in the higher order
flow harmonics, on an event-by-event basis.  Namely, one has the event averages
$\bbra V_n^L V_n^{NL}\kket=0$. As indicated in the 
expansion formulae, these event averaged correlators are assumed to vanish,
\begin{subequations}
\label{eq:noncor}
\begin{align}
\label{eq:noncor4}
\bbra V_4^L(V_2^*)^2 \kket=&0\,,\\
\label{eq:noncor5}
\bbra V_5^L(V_2^*V_3^*)\kket=&0\,,\\
\label{eq:noncor6}
\bbra V_6^L(\chi_{633}V_3^{*2}+\chi_{624}V_2^*V_4^{L*}+\chi_{6222}V_2^{*3})\kket=&0\,.
\end{align} 
\end{subequations}

\Eqs{eq:noncor} are actually exact conditions on a general ground, 
as long as $V_n^L$ is identified as the part of flow projected out of 
the plane composited from nonlinear mode mixings. Indeed, these relations in
$V_4$ and $V_5$ have
been found to hold approximately in hydrodynamics~\cite{Qian:2016fpi} 
and AMPT calculations~\cite{Bhalerao:2014xra}.
Shown in \Fig{fig:vnvnl} are some specified ratios from
event-by-event hydrodynamic simulations,
in which the implied identities are equivalent to the uncorrelation of the linear and 
nonlinear parts in $v_4$ and $v_5$~\cite{Yan:2015jma},
\begin{subequations}
\begin{align}
&\frac{\bbra V_4(V_2^*)^2v_2^2\kket}{\bbra V_4(V_2^*)^2 \kket\bbra v_2^2\kket}
=\frac{\bbra v_2^6 \kket}{\bbra v_2^4\kket \bbra v_2^2 \kket}\,,\\
&\frac{\bbra V_5V_2^*V_3^* v_2^2v_3^2\kket}{\bbra V_5V_2^*V_3^* \kket\bbra v_2^2v_3^2\kket}
=\frac{\bbra v_2^4 v_3^4\kket}{\bbra v_2^2 v_3^3\kket^2}\,.
\end{align}
\end{subequations}
In experiments, measurements also support the conclusion that the linear
and nonlinear parts in $V_4$ and $V_5$ are uncorrelated, as long as one consider $V_n^L$
as the part of projection out of the plane composed from nonlinear mode mixings~\cite{Acharya:2017zfg}.
Given these results, further assumptions can be made 
for $V_6$, that all the nonlinear terms are not correlated with each other, i.e.,
\be
\bbra V_6^L (V_2^*)^3 \kket=\bbra V_6^L(V_3^*)^2 \kket=\bbra V_6^L (V_2^*V_4^{L*})\kket =0\,. 
\ee

\setcounter{footnote}{0}

The analysis of higher order flow is simplified once the linear and nonlinear parts are uncorelated.
In particular,  the linear and nonlinear response 
of higher order flow for events in a centain centrality class are separable regarding
the rms values.  
Taking $V_4$ as an example,  the square of $v_4\{{\rm rms}\}=v_4\{2\}$
\footnote{
Taking into account the effect of event-by-event fluctuations, the quantity 
$v_4\{2\}$ is determined from two-particle correlations in experiments (or 
hydro simulations), whose definition and details will be given later. 
} 
becomes the 
sum of the square of $v_4^L\{{\rm rms}\}=v_4^L\{2\}$ and the square of $v_4^{NL}\{{\rm rms}\}=v_4^{NL}\{2\}$,
\be
v_4\{2\}^2=v_4^L\{2\}^2+v_4^{NL}\{2\}^2\,.
\ee
It is worth mentioning that
the rms value of the nonlinear $v_4$ is identical to the projection of $V_4$ 
onto the plane determined by $V_2^2$, or more explicitly $2\Psi_2$. This is known as 
the $v_4\{\Psi_2\}$. 
Hence, by using the event-averaged quantities,  $v_4^{NL}\{2\}$, or
the projection $v_4\{\Psi_2\}$, 
is accessible in event-by-event
hydrodynamic calculations and in experiments,
\be
v_4^{NL}\{2\}=
\frac{\Re\bbra V_4 (V_2^*)^2\kket}{\bbra |V_2|^4\kket^{1/2}}
\equiv
v_4\{\Psi_2\}\,.
\ee
Note that the denominator is the fourth order moment of $V_2$, which
in experiments should be measured with a rapdity-gap so that non-flow effects 
are not important.
The linear part can be obtained accordingly,
\be
v_4^L\{2\}=\sqrt{\bbra |V_4|^2\kket-v_4\{\Psi_4\}^2}\,.
\ee
It is very similar for $V_5$, except that the plane from the nonlinear coupling of $V_2V_3$ 
is $\Psi_{23}=2\Psi_2+3\Psi_3$. 
The nonlinear part of $v_5$ is measured in the $\Psi_{23}$ plane, 
\be
v_5^{NL}\{2\}=\frac{\Re\bbra V_5 (V_2^*V_3^*)\kket}{\bbra |V_2|^2|V_3|^2\kket^{1/2}}
\equiv
v_5\{\Psi_{23}\}\,, 
\ee
which results in the linear part of $v_5$ 
\be
v_5^L\{2\}=\sqrt{\bbra |V_5|^2\kket-v_5\{\Psi_{23}\}^2}\,.
\ee
The sixth order harmonic flow has several types of nonlinear mode mixings.
Except the one that involves linear response of the quadrangular flow $V_4^L$, 
there are planes well-defined in terms of $V_3$ of the quadratic order ($2\Psi_3$),
and $V_2$ of cubic order ($3\Psi_2$). Measured in these specified planes respectively,
one has  $v_6$ projected onto $\Psi_2$ and $\Psi_3$ planes,
\begin{align}
v_6\{\Psi_2\}\equiv&
\frac{\Re\bbra V_6 (V_2^*)^3\kket}{\bbra |V_2|^6\kket^{1/2}}\,,\\
v_6\{\Psi_3\}\equiv&
\frac{\Re\bbra V_6 (V_3^*)^2\kket}{\bbra |V_3|^4\kket^{1/2}}\,.
\end{align}
To extract the linear part of $v_6$, one would have to substract also the
projected flow onto the $V_2V_4^L$ plane, which has not been done yet.
Apart from that, the strategy has been applied in experiments to disentangle 
the linear and nonlinear parts in higher order flow~\cite{Acharya:2017zfg,Tuo:2017ucz}, 
leading to results consistent with those obtained by using the
event-shape engineering method~\cite{Aad:2015lwa}.
It is worth mentioning that the separated nonlinear part of the higher 
order harmonic flow is related to the measured event-plane correlations~\cite{Yan:2015jma},
which we discuss in \Sect{sec:vncor}.

The projection of higher order flow, and event-plane correlation, are understood in
the flow paradigm  as a combined effect of initial state geometry and medium
dynamical expansion. Namely, they depends on the initial eccentricities and medium response coefficients
in \Eqs{eq:vnform1}, in a similar way to many other flow observables. However, 
a closer look at the formulae of the projections
of higher flow harmonics reveals the possibility of extracting  the nonlinear response 
coefficients $\chi$'s directly, by taking proper ratios with respect to the specified moments
\begin{eqnarray}
\label{eq:chi}
\chi_{422}&=&\frac{\bbra V_4 (V_2^*)^2\kket}{\bbra |V_2|^4\kket}
=\frac{v_4\{\Psi_2\}}{\sqrt{\bbra |V_2|^4\kket}}
\cr
\chi_{523}&=&\frac{\bbra V_5 V_2^*V_3^*\kket}{\bbra |V_2|^2 |V_3^2|\kket}
=\frac{v_5\{\Psi_{23}\}}{\sqrt{\bbra |V_2|^2 |V_3^2|\kket}}\cr
\chi_{6222}&=&\frac{\bbra V_6 (V_2^*)^3\kket}{\bbra |V_2|^6\kket}
=\frac{v_6\{\Psi_2\}}{\sqrt{\bbra |V_2|^6\kket}}\cr
\chi_{633}&=&\frac{\bbra V_6 (V_3^*)^2\kket}{\bbra |V_3|^4\kket}
=\frac{v_6\{\Psi_3\}}{\sqrt{\bbra |V_3|^4\kket}}
\end{eqnarray} 
These $\chi$'s are calculable in hydrodynamics, and measurable in experiments according
to the expressions above. 
Note that these moments in the denominators are supposed to be entirely due to
medium collectivity, thus in experiments non-flow contributions should be 
excluded carefully.
By design,  dependence on initial states eccentricities is cancelled in
these ratios, hence $\chi$'s are not sensitive to the details of initial state.
As a result, $\chi$'s are ideal probes reflecting directly the dynamical 
properties of the system evolution.  

\end{multicols}
\begin{center}
\includegraphics[width=.95\textwidth] {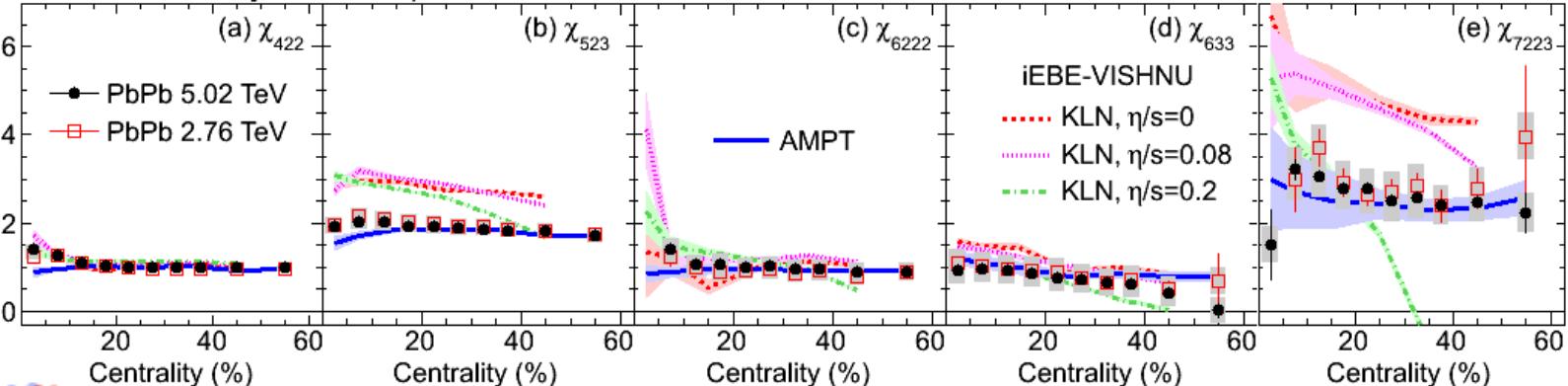}
\figcaption{
\label{fig:chi_exp}
Non-linear response coefficients measured by the ALICE collaboration with respect to
Pb+Pb collisions at $\sqrt{s_{NN}}=2.76$ TeV, as a function of 
centrality percentile. Hydro calculations are shown for comparisons. Figure reproduced from
\Ref{Tuo:2017ucz} (DOI: https://doi.org/10.1016/j.nuclphysa.2017.05.064), under the CC-BY 4.0 license (http://creativecommons.org/licenses/by/4.0/).
}
\end{center}
\begin{multicols}{2}

These nonlinear response coeffiicents have been measured in experiments at the LHC
energies~\cite{Acharya:2017zfg,Tuo:2017ucz}. 
Shown in \Fig{fig:chi_exp} are the measured $\chi$'s as a function of centrality percentile,
from the CMS collaboration. Compared to  the direct measurements of higher order flow,
the nonlinear response coefficients present weaker dependence on the collision centrality.
Considering the major effect on flow observables from central to peripheral collisions 
due to the 
increase of medium viscous corrections, $\chi$'s are less affected by viscosity. This could be
due to the cancellation of viscous effects in the ratios. Besides, there is also
cancellation of statistical errors in the ratios, so that the measured nonlinear response
coefficients are more accurate than their corresponding higher order flow.

In \Fig{fig:chi_exp}, the hydro calculations from VISH2+2~\cite{Qian:2016fpi}
and AMPT calculations~\cite{Yan:2015lwn} 
are shown as well, regarding different types of initial 
conditions and parameterizations of medium dissipative effects. 
The hydro results are compatible with experiments, in terms of overall magnitudes and centrality
dependence, etc., although $\chi_{523}$ is apparently over-estimated.
In contrast to the original proposal, 
the resulted nonlinear response coefficients do present dependence on initial conditions~\cite{Qian:2016fpi}. 
For instance, with the same
$\eta/s=0.08$, hydro gives larger $\chi_{422}$ with respect to the MC-Glauber initialization than MC-KLN.
Similarly, MC-Glauber model leads to a larger value of 
$\chi_{6222}$\footnote{
One possible explanation for the discrepancy is that the KLN model has a more elliptic averaged background,
which results in a larger prediction of $v_2$. Accordingly, higher order mode mixings involving
$V_2$ are required in the expansion form \Eqs{eq:vnform1}, to give a good description of higher
order flow. However, these new terms are not considered in the definitions 
of nonlinear response coefficients.
On the other hand, these calculations
indeed show that $\chi_{523}$
and $\chi_{633}$ have weak sensitivity to initial conditions, where the role of $V_2$ is less
significant.}. Nonetheless, the dependence on initial conditions is quite weak in $\chi_{523}$ and
$\chi_{633}$.
 
Although viscous corrections to the $\chi$'s are expected to be minor, detailed hydro calculations
have shown strong dependence of $\chi$'s on the freeze-out prescription. For instance, the
$\chi$'s increase with respect to a lower freeze-out temperature in hydrodynamic 
simulations. Most of the 
viscous effects on $\chi$'s come from the viscous corrections to the phase space distribution
function $\delta f$ at freeze-out~\cite{Qian:2016fpi}, 
which signifies the importance of $\chi$ as a probe for future studies on the physics
of freeze-out and particle generation.

It is very likely that the nonlinear mode mixings in hydrodynamic modeling are dominantly generated
during the freeze-out process, which can be explained in a quantitative way as follows.
In ideal hydrodynamics, the nonlinear mode mixings at 
freeze-out lead the prediction that, at
fixed and large transverse momentum $p_T$, $\chi_{422}=\frac{1}{2}$, $\chi_{523}=1$, $\chi_{6222}=
\frac{1}{6}$, and $\chi_{633}=\frac{1}{2}$~\cite{Borghini:2005kd}. The integrated value of 
$\chi$'s are further modified by a factor from 
an average over the $p_T$ spectrum. For the nonlinear response 
coefficients associated with quadratic order couplings, the factor 
is roughly $\bra v^2\ket/\bra v\ket^2>1$, where
single brackets indicate the integrated flow ($v_2$ or $v_3$) according to the $p_T$ spectrum in 
a single hydro event. For those from the cubic order couplings,  a larger factor is expected, 
$\bra v^3 \ket/\bra v \ket^3>\bra v^2\ket/\bra v\ket^2$.
Since $v_2$ and $v_3$ have roughly similar $p_T$ dependence, the ratios are comparable for $\chi_{422}$,
$\chi_{523}$ and $\chi_{633}$, and comparable for $\chi_{6222}$ and $\chi_{7223}$. As a consequence,
in the flow paradigm captured by hydrodynamics, one expects the approximate relations
$\chi_{422}\sim\chi_{633}\sim \frac{1}{2}\chi_{523}$ and $\chi_{6222}\sim\frac{1}{3}\chi_{7223}$.
These approximate relations are confirmed in experimental observations. Accordingly,
up to a semi-analytical level, these observed relations demonstrate 
the success of the flow paradigm, and especially, the proposed medium response  
relations.

\subsubsection{Fluctuations in the medium response}

\begin{center}
\includegraphics[width=0.5\textwidth] {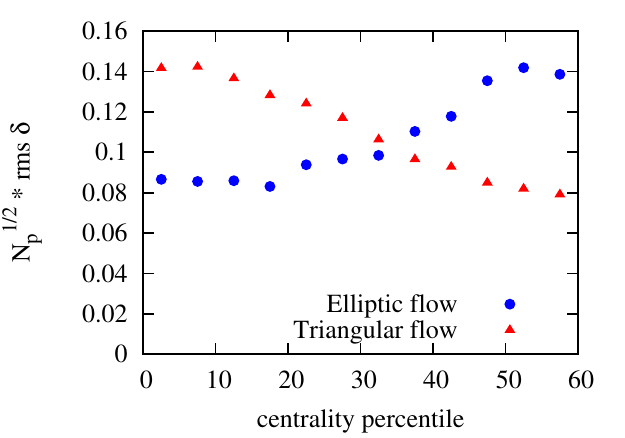}
\figcaption{ Rms values of the residual $\delta_n$ scaled by $\sqrt{N_p}$,
from event-by-event hydro simulations with respect to Pb+Pb collisions at 
$\sqrt{s_{NN}}=2.76$ TeV, 
as a function of centrality. Figure reproduced from \Ref{Noronha-Hostler:2015dbi}, with permission.
\label{fig:delta}
}
\end{center}

Even though all the linear and nonlinear responses in the medium are well understood,
or have been quantitatively characterized in a hydro simulation, the predicted flow
observables are not deterministic, owing to random 
fluctuations. As well as fluctuations of 
initial state eccentricities,
there are also fluctuations associated with the dynamics of medium response, which
in the expansion formulae \Eqs{eq:vnform} are captured by the residual $\delta_n$'s.

In writing residuals in the expansion form, there are two assumptions:
1) It is assumed that $\delta_n$ has the same rotational symmetry in the azimuth 
as $V_n$;  2) The residual $\delta_n$ should 
be statistically uncorrelated with the rest of the terms in the expansion 
in  \Eqs{eq:vnform}, 
i.e., one expects  vanishing correlations
of $\delta_n$ and initial eccentricities,  $\bbra \delta_n \E_m^*\kket = 0$. 
There are two possible origins of the residual $\delta_n$ in heavy-ion collisions. 
It can be understood as an inadequate characterization of the flow 
generation on an event-by-event basis, so that additional effects related to 
initial state fluctuations contribute, $\delta^{ini}$. On the other hand,
it can be a consequence of thermal fluctuations, $\delta^{th}$. The former
can be studied in the present framework of event-by-event hydro simulations, without considering
thermal fluctuation effects. 

As has been shown in \Fig{fig:resp0} and \Fig{fig:resp2}, the response of $v_2$ and $v_3$ 
disperse around the prediction of linear and/or linear+cubic response, indicating the effects
of $\delta_2$ and $\delta_3$ on an event-by-event basis. One would expect these residuals
to be generated from the hydro response to short-scale structures in the initial state density
profile. Indeed, the dispersion, or the rms value of $\delta_n$, is suppressed as viscosity applied
in these calculations is increased. 
It is also consistent with the observation that dispersion in
triangular flow is stronger, where the role of short-scale structures is relatively more important than
that in the elliptic flow. Therefore, one may associate the effect of $\delta_n$ with initial 
state fluctuations, which scale as $1/\sqrt{N_p}$, with $N_p$ being the number of participants.
Shown in \Fig{fig:delta} are the corresponding results from event-by-event hydro simulations, 
with respect to Pb+Pb collisions at $\sqrt{s_{NN}}=2.76$ TeV.  The rescaled rms values of
the residual vary within a factor of two when centrality increases, considering the fact 
that $\sqrt{N_p}$ changes by a factor 10.

The presence of thermal fluctuations can be decomposed into similar mode structures.
That is to say, for each of the harmonic orders, in the medium evolution there is a fluctuating
mode correspondingly originated from thermal fluctuations~\cite{Yan:2015lfa}, 
in addition to the averaged response coefficients. 
Although the effects of thermal fluctuations can be absorbed formally in the residuals of 
the response relations \Eqs{eq:vnform} as $\delta_n^{th}$'s, the origins of thermal fluctuations are 
distinct from those 
related to the initial state, $\delta^{ini}_n$'s. 

Several remarks are in order 
with respect to the effects of thermal fluctuations in the generation of harmonic flow.
First, a thermodynamical origin
guarantees the condition $\bbra \delta^{th}_n\E_m\kket=0$ and  $\bbra \delta^{th}_n\delta^{ini}\kket=0$. 
Therefore,
in the two-particle (or multi-particle) correlations, contributions from 
a thermal origin and a quantum origin are separable.
Second, the effects of fluctuations 
are controlled in principle 
by the fluctuation strength, in terms of the two-point auto-correlations. 
For the effects associated with initial state fluctuations, one has parameterically 
\be
\label{eq:qq_cor}
\bbra \delta^{ini}\delta^{ini}\kket\sim 1/N_{p}\quad\mbox{or}\quad
\bbra \ve_n^2\kket\sim 1/N_p\,.
\ee
On the other hand, two-point auto-correlations of thermal fluctuations are determined
by the dissipative corrections of the medium system through the  flutucation-dissipation
relation~\cite{Yan:2015lfa},
\be
\label{eq:tt_cor}
\bbra \delta^{th}\delta^{th}\kket\sim \frac{\eta}{s}\Big/\Big\langle\!\!\!\Big\langle 
\frac{dS}{dy}\Big\rangle\!\!\!\Big\rangle\,.
\ee
with an extra factor quantified by the inverse of the total entropy per rapidity.  
From \Eq{eq:qq_cor} and \Eq{eq:tt_cor}, one realizes that the contributions from
thermal fluctuations to the two-particle correlation 
are suppressed parametrically by a factor of $(\eta/s K_s)^{1/2}$,
where the constant $K_s\gg1$ is associated with the entropy production from
nucleon-nucleon collisions. Therefore, even though in higher order flow harmonics thermal fluctuations
are expected stronger~\cite{Yan:2015lfa,Sakai:2017rfi}, 
the overall effects of thermal fluctuations in the observed flow are \emph{not}
significant compared to those from the initial state, 
unless on the occasions that the system is sufficiently close to the QCD critical
point. Third, the previous estimates of the thermal fluctuations are obtained with an
inclusion of the thermal fluctuations \emph{linearly} 
in hydrodynamics in a canonical way, corresponding to 
the gradient expansion. Recently, however, it was found that the nonlinear feature
of hydrodynamics results in non-linear couplings of thermal fluctuations, and
a distinct behavior 
is achieved with the appearance
of a fractional order term in the gradient expansion $O(\nabla^{3/2})$. 
This is the generic property of long-time tails in hydrodynamics~\cite{PhysRevA.16.732,Kovtun:2003vj},
which has conceptual influences in the theoretical framework. In particular,
nonlinear couplings of the thermal fluctuations effectively modify
transport coefficients, such as $\eta$~\cite{Kovtun:2011np,Akamatsu:2016llw} 
and $\zeta$~\cite{Akamatsu:2017rdu,Martinez:2017jjf}.

\section{Experimental observables in nucleus-nucleus collisions}
\label{sec:Exp}

In this section, we summarize some of the observables in heavy-ion experiments that
are compatible with the flow paradigm. We shall start with an overview
of the measured harmonic flow from multi-particle correlations.
For comparison, corresponding predictions from hydrodynamics are shown
with properly parameterized transport coefficients, such as $\eta/s$, $\zeta/s$.
These are concrete examples showing how these coefficients are estimated
in the analyses in the flow paradigm. 

\begin{center}
\includegraphics[width=0.45\textwidth] {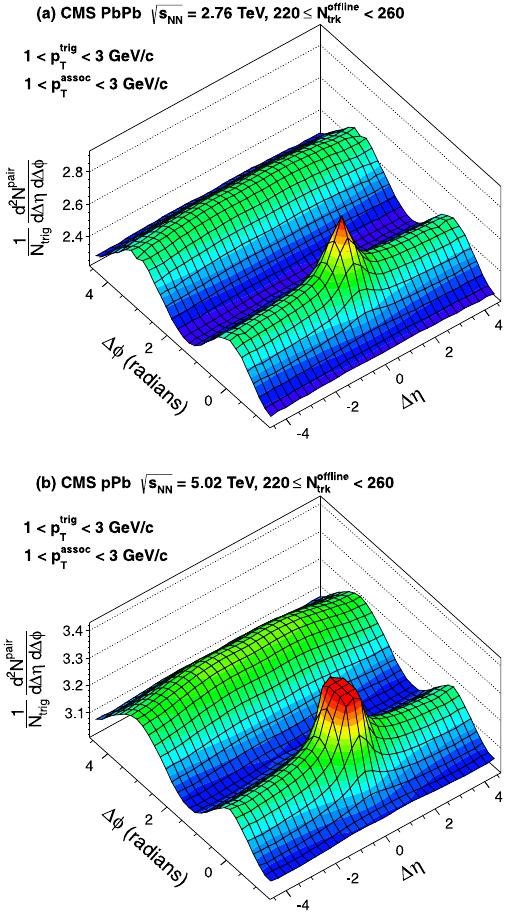}
\figcaption{
Two-particle correlation functions measured by the CMS collaboration, for
the Pb+Pb collisions at $\sqrt{s_{NN}}=2.76$ TeV (up)
and p+Pb collisions at $\sqrt{s_{NN}}=5.02$ TeV (down). 
Collision events in p+Pb and Pb+Pb are selected with comparable and 
sufficiently large
multiplicity productions. In these events, long-range correlation patterns can be identified as
the ``ridge'' structure:  the bump span along relative pesudo-rapidity at $\Delta \phi=0$. 
Trigger and associated particles are chosen in the
low $p_T$ range. 
Figure adapted from 
\Ref{Chatrchyan:2013nka} (DOI: https://doi.org/10.1016/j.physletb.2013.06.028, under the CC-BY-NC-ND license, with permission).
\label{fig:2pc_cms}
}
\end{center}

The concept of the flow paradigm depends essentially on the system collective expansion.
The most convincing evidence of the system collective expansion in heavy-ion collisions
comes from the observed structures in long-range multi-particle correlations. This is because, 
once a medium
system expands collectively, the short-scale structures at initial times 
are amplified in space as time
evolves, leading to long-range correlations. 
Normally, these long-range correlations are beyond the prediction of first principle
calculations based on particle
scatterings or classical fields at early evolution stages, if late-time medium evolution
is ignored.
In heavy-ion collisions, this long-range correlation is 
particularly recognized in the long-range correlations in rapidity, or pseudo-rapidity
$\eta$ 
(it should be distinguished from the notation of shear
viscosity)\footnote{
For the convenience of later discussions, we assume
the configuration of a colliding system as being to align the beam-axis with the $z$-axis, while
the plane perpendicular to the beam-axis is recognized as the transverse plane. 
When the nucleus-nucleus collisions are not head-on collisions, there is a non-zero
impact which defines 
the $x$-axis. The direction of the impact, together with the beam-axis,
 determine the reaction-plane $\Psi_R$.}.

One typical example can be found in 
\Fig{fig:2pc_cms}, which depicts the measured two-particle correlation functions in relative
azimuthal angle $\Delta \phi$ and relative pseudo-rapidity $\Delta \eta$, from the
CMS collaboration with respect to the Pb+Pb collisions at $\sqrt{s_{NN}}=2.76$ TeV, 
and p+Pb collisions at $\sqrt{s_{NN}}=5.02$ TeV. 
The collision events are purposely selected for illustration, 
in a way that the multiplicity productions
in Pb+Pb and p+Pb are comparable and sufficiently large.
Despite some fine details, one notices the very similar 
long-range
correlation structures at large $\Delta \eta$. Especially, at $\Delta\phi=0$ (near-side), a
bump 
appears in the Pb+Pb system, 
which is sometimes referred to as the ``ridge''. The ``ridge'' indicates a 
long-range correlation pattern which
is expected from the medium collective expansion scenario. 
Note that these Pb+Pb events correspond to peripheral collisions.
In more central collisions of larger multiplicity productions, 
the ridge, and thus the system collective
expansion, is more obvious.
On the contrary, as one might expect,
the ``ridge'' is absent in the two-particle correlation functions 
when the corresponding multiplicity
production of the collision event 
is small (see for instance \Ref{Khachatryan:2015lva} for p+p collisions),
or the trigger or associated particles are not soft.

Given the single-particle spectrum written in terms of harmonic flow $V_n$ in \Eq{eq:vndef1}, 
the two-particle 
correlation function 
can be described accordingly, 
\be
\label{eq:2pcvn}
\Big\langle\!\!\Big\langle\frac{d N^{pair}}{d\Delta \phi }
\Big\rangle\!\!\Big\rangle= 
\Big\langle\!\!\Big\langle \frac{dN_a}{d\phi_a}\frac{dN_b}{d\phi_b}
\Big\rangle\!\!\Big\rangle
\sim
1 + \sum_{n=1}^\infty \bbra V_n V_n^* \kket e^{in \Delta \phi}\,,
\ee
which explains the bump in \Fig{fig:2pc_cms} at $\Delta\phi=0$ as a consequence of the non-zero
correlation structure, 
$
\bbra V_n^a V_n^{b*}\kket \equiv V_{n\Delta}\,. 
$
Normally, a fit using \Eq{eq:2pcvn} for the two-particle correlation function
allows one to estimate the magnitude of harmonic
flow $v_n$ in experiments. More precisely, it defines the 
the measured flow using the scalar-product method, or the two-particle cumulant
$v_n\{2\}$,
\be
\label{eq:vn2}
v_n\{2\}^2
=\bbra v_n^2\kket\equiv\bbra e^{in(\phi_p^a-\phi_p^b)}\kket=\bbra V_n^a V_n^{b*}\kket\,.
\ee
The quantities $v_n\{2\}$ are the most generally measured harmonic flow signatures that
contain the information of the magnitude of $V_n$. They must be 
measured in experiments in the correlated particle
spectra as a consequence of event-by-event fluctuations of $V_n$.

\begin{center}
\includegraphics[width=0.45\textwidth] {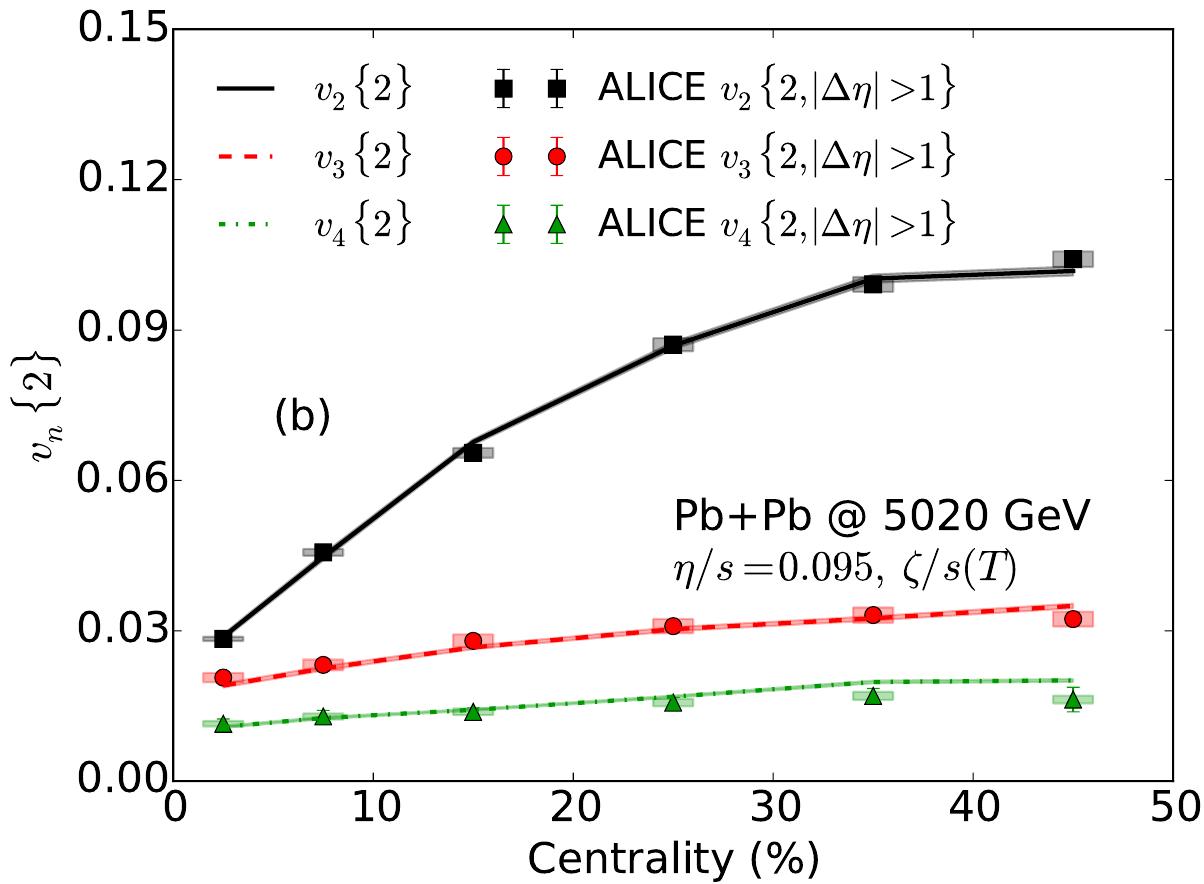}
\figcaption{ Harmonic flow measured from two-particle 
correlations from the ALICE collaboration for the Pb+Pb at
$\sqrt{s_{NN}}=5.02$ TeV. Lines are corresponding results from
hydrodynamic simulations, with a constant value of $\eta/s=0.095$ and a 
temperature dependent parameterization of $\zeta/s$.
Figure reproduced from \Ref{McDonald:2016vlt}, with permission.
\label{fig:vn2}
}
\end{center}

As an example, 
\Fig{fig:vn2} presents the recently measured flow harmonics, using the scalar-product method
for the correlated two-particle spectra, as a function of collision centrality.
The symbols correspond to elliptic flow $v_2\{2\}$ (black), the
triangular flow $v_3\{2\}$ (red) and the quadrangular
flow $v_4\{2\}$ (green), in the Pb+Pb system of $\sqrt{s_{NN}}=5.02$ TeV, 
by the ALICE collaboration. Hydro calculations from IP-Glasma+MUSICS with an input $\eta/s=0.095$
and a parameterized temperature dependent bulk viscosity over entropy 
density ratio $\zeta/s$ are shown in solid lines. One notices an increasing trend
of the flow harmonics along with the increase of collision centrality percentile\footnote{
Centrality percentile is used to identify collision events with respect to multiplicity production.
A centrality percentile approaching
to zero implies the collision with the largest multiplicity production of the whole events, namely, approximately
head-on collisions in nucleus-nucleus collisions. On the other hand, centrality percentile close
to 100\% corresponds to very peripheral collisions with very small particle yields.
}.
The increasing trend of the flow harmonics and 
the magnitude of each flow harmonics, are both captured quantitatively by hydro calculations.

As we mentioned before, harmonic flow $v_n\{2\}$ can be also measured 
differentially in experiments
as a function of the transverse momentum $p_T$.
These differential spectra of harmonic flow are obtained from two-particle 
correlations, assuming a factorization of the correlation~\cite{Aamodt:2011by}
\be
\label{eq:fact}
\bbra V^n(p_T^a)V_n^{*}(p_T^b)\kket \Rightarrow v_n\{2\}(p_T^a)\times v_n\{2\}(p_T^b)\,.
\ee
However, the above factorization can be broken. In the flow paradigm where a 
single-particle spectrum is well-established from hydrodynamic calculations, 
one has generally~\cite{Gardim:2012im} 
\be
\label{eq:cachy}
\bbra V^n(p_T^a)V_n^{*}(p_T^b)\kket \le v_n\{2\}(p_T^a)\times v_n\{2\}(p_T^b)\,,
\ee
with the equality satisfied only when the complex variables
$V_n(p_T^a)$ and $V_n(p_T^b)$ are linearly dependent.
That is to say, there exists a $p_T$-dependent function that does not fluctuate from
event to event in one centrality class. When the effects of initial state fluctuations are sizable, 
the linear dependent relation is broken, 
which in turn result in the breaking of the factorization in \Eq{eq:fact}. 
In particular, the breaking of 
factorization is expected to be stronger when the difference $p_T^a-p_T^b$ increases, 
owing to the effect of initial state fluctuations. Breaking of the factorization also
implies a $p_T$-dependent flow angle $\Psi_n(p_T)$~\cite{Heinz:2013bua}.
\Eq{eq:cachy} is a Cauchy-Schwarz inequality derived by 
correlating single-particle spectra. Therefore, violation of \Eq{eq:cachy} can be seen
as an indication of non-flow contribution. A quantity $r_n$ is introduced 
accordingly~\cite{Gardim:2012im},
\be
r_n\equiv \frac{V_{n\Delta}(p_T^a,p_T^b)}{\sqrt{V_{n\Delta}(p_T^a,p_T^a)}
\sqrt{V_{n\Delta}(p_T^b,p_T^b)}}
\ee
to quantify these effects. Indeed, experiments have found violation of the 
inequality \Eq{eq:cachy} with respect to large transverse momentum, that $r_n>1$. Besides,
a decreasing trend of $r_n$ is confirmed regarding the growth of the relative
difference transverse momentum $p_T^a-p_T^b$~\cite{Aamodt:2011by,Khachatryan:2015oea},
which are compatible with hydrodynamics~\cite{Heinz:2013bua,Kozlov:2014fqa}

 In addition to
the two-particle correlation function, 
in realistic analyses in experiments, the measurements of flow harmonics
also generalize to multi-particle correlations. 
Systematic generalization to four-, six- and eight-particle spectra leads to
the measured cumulants of harmonic flow $v_n\{4\}$, $v_n\{6\}$ and $v_n\{8\}$~\cite{Borghini:2001vi,Voloshin:2007pc}
\begin{subequations}
\label{eq:vn_cumu}
\begin{align}
v_n\{4\}^4 =& 2\bbra v_n^2\kket^2 - \bbra v_n^4\kket\\
v_n\{6\}^6 =& \frac{1}{4} \left[\bbra v_n^6\kket - 9\bbra v_n^4\kket \bbra v_n^2\kket
+12\bbra v_n^2\kket^3\right]\\
v_n\{8\}^8=&\frac{1}{33}\Big[-\bbra v_n^8\kket + 16\bbra v_n^6\kket\bbra v_n^2\kket
+18\bbra v_n^4\kket^2\nonumber\\
&\quad -144 \bbra v_n^4\kket\bbra v_n^2\kket^2+ 144\bbra v_n^2\kket^4\Big]\,,
\end{align}
\end{subequations}
where self-correlations are subtracted by definition. With respect to correlations of all particle yields in the collision events, 
there exists the Lee-Yang zero method and correspondingly harmonic flow $v_n\{\mbox{LY}\}$~\cite{Borghini:2000sa}.

\begin{center}
\includegraphics[width=0.45\textwidth] {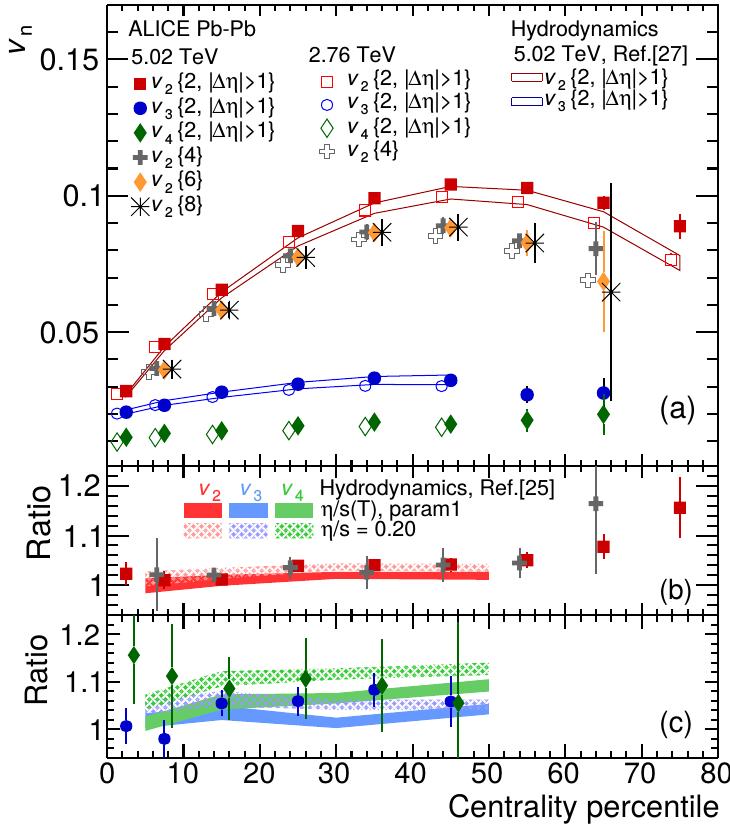}
\figcaption{ Harmonic flow measured from multi-particle 
correlations from the ALICE collaboration for the Pb+Pb at
$\sqrt{s_{NN}}=5.02$ TeV, in comparison to that from the Pb+Pb
at $\sqrt{s_{NN}}=2.76$ TeV. Lines are corresponding results from
hydrodynamic simulations. Figure reproduced from \Ref{Adam:2016izf} (DOI: https://doi.org/10.1103/PhysRevLett.116.132302, ``Anisotropic flow of charged particles in Pb-Pb collisions at $\sqrt{s_{\rm NN}}=5.02$ TeV''), under the CC-BY-3.0 license.
\label{fig:v2_cumu}
}
\end{center}

In heavy-ion experiments at RHIC and the LHC,
flow cumulants have been measured up to $v_2\{8\}$ in nucleus-nucleus collisions 
for the elliptic flow, as summarized 
in \Tab{tab:vn}. 
 \Fig{fig:v2_cumu} displays the recent results of the cumulants of
elliptic flow from multi-particle correlations from the ALICE collaboration~\cite{Adam:2016izf},
in the Pb+Pb collisions at $\sqrt{s_{NN}}=2.76$ TeV and $\sqrt{s_{NN}}=5.02$ TeV.
Higher order cumulants of $v_2$ tend to collapse, although they present similar 
centrality dependence as $v_2\{2\}$.
Note that there is a gap between $v_2\{2\}$ and these higher order cumulants, which
grows as centrality increases. As will be discussed later, the gap and the tiny splitting among
higher order flow cumulants signify the 
non-Gaussian behavior of $v_2$ flucutations.

These cumulants of $V_n$ are of particular significance with respect to the fluctuating 
nature of heavy-ion collisions, since they provide a quantitative characterization of the flow
event-by-event fluctuations. For instance, $v_n\{2\}$ captures the variance of flow
distribution on an event-by-event basis~\cite{Aad:2013xma}, while $v_n\{4\}$ and
$v_n\{6\}$ can be used to measure the skewness~\cite{Giacalone:2016eyu}.
Additionally,  flow measurements from
multi-particle correlations have the advantage of suppressing
non-flow contributions~\cite{Abelev:2014mda}. 
Note that in \Fig{fig:vn2} and \Fig{fig:v2_cumu}, a rapidity gap of $|\Delta \eta|>1$ 
has been applied to the measured two-particle spectra to
suppress non-flow contributions, assuming that non-flow effects are dominantly 
short-ranged.

Besides the measurements involving the same harmonic orders, one is allowed
to extract the correlations among mixed harmonic orders. In this way, the informaiton
of the phase $\Psi_n$ becomes detectable. Unlike the cumulants of harmonic flow which
reflect the fluctuation porperties of the colliding systems, to a large extent, 
flow correlations measure 
the correlation properties of flow angles on an event-by-event basis. There are several types
of correlators of the mixed harmonics that have been investigated in experiments, 
including event-plane correlations 
first measured by the ATLAS collaboration $\rho_{mn}$~\cite{Aad:2014fla},
the symmetric cumulants $SC(m,n)$ proposed by the ALICE collaborations~\cite{ALICE:2016kpq},
and the three plane event-plane correlators recently measured by the STAR 
collaboration at the RHIC energy~\cite{Adamczyk:2017hdl,Adamczyk:2017byf}. 

\Fig{fig:epcorrelator} displays a set of event-plane correlations measured by the 
ATLAS collaboration with respect to the Pb+Pb collisions at $\sqrt{s_{NN}}=2.76$ TeV.
Hydro simulations with the EKRT initial condition 
using different types of parameterization of $\eta/s$ are shown as lines. Hydro predictions
generally capture
the right trend and sign of correlations, while the correlation strength is found to be further
affected by the dissipative properties of the medium~\cite{Niemi:2015voa}. 
Note that an overall agreement is best achieved when a constant $\eta/s=0.2$ ($\zeta/s=0$) is
taken in the simulations. Since all other parameterizations involves a temperature dependent
$\eta/s$, which on average introduce effectively smaller dissipations to hydro simulaitons,
one realizes that the event-plane correlations are stronger in a more dissipative fluid, which
property we shall detail later in the context of the flow paradigm. 

\Fig{fig:epcorrelator} is again a typical example demonstrating the
extraction of $\eta/s$ from the hydro modeling of heavy-ion collisions. These flow
observables are by far the best probes for the extraction of $\eta/s$ of the QGP
system in heavy-ion collisions. It has been found from hydro simulations 
that a quite small value of $\eta/s\approx 0.08$ is realized in the system created from Au+Au
at RHIC~\cite{Romatschke:2007mq}, while at the LHC energies one would expect 
a larger value.

\end{multicols}
\begin{center}
\includegraphics[width=1.\textwidth] {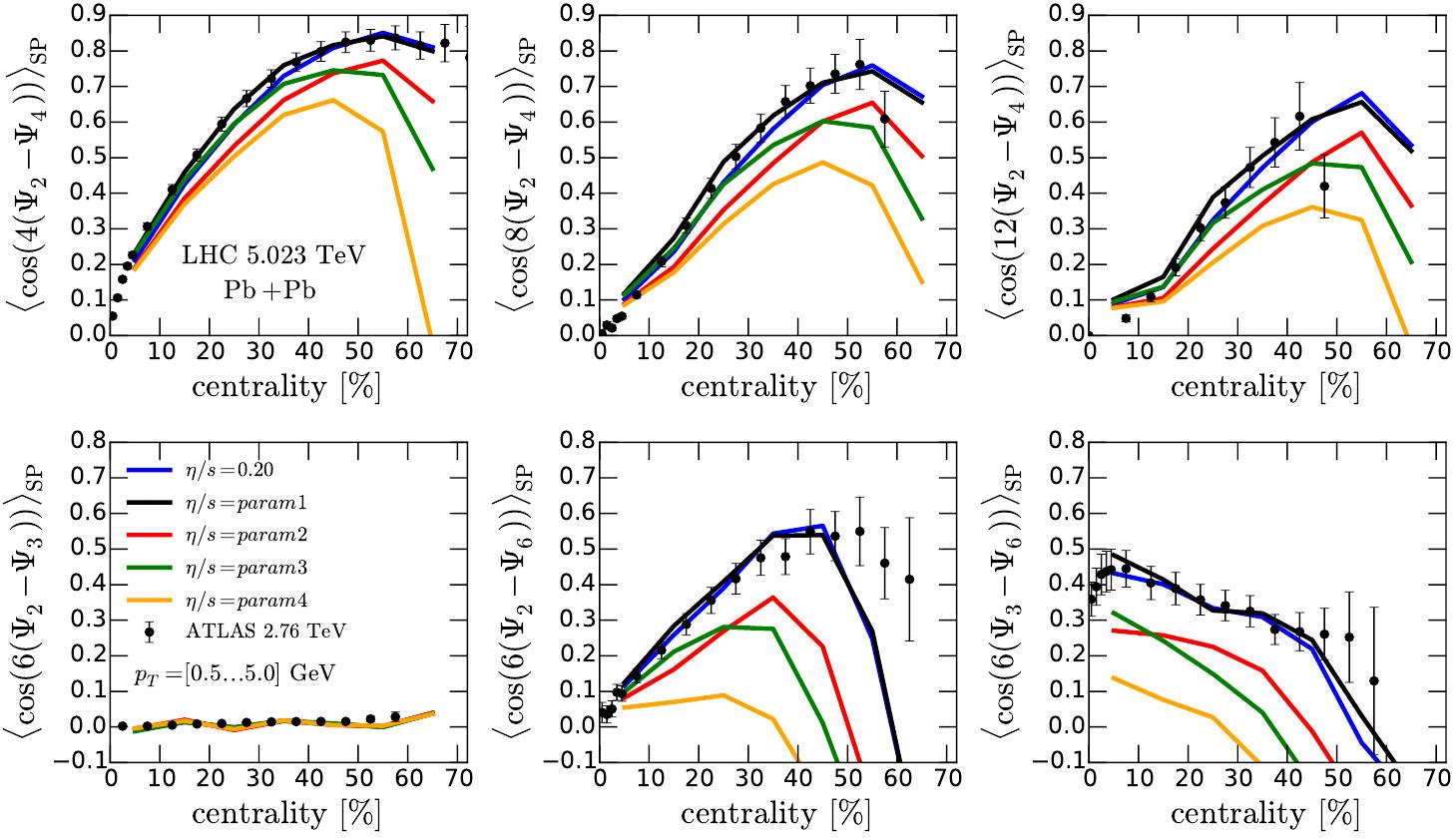}
\figcaption{
\label{fig:epcorrelator}
Event-plane correlations measured with respect to the Pb+Pb collisions at $\sqrt{s_{NN}}=2.76$ TeV
by the ATLAS collaboration~\cite{Aad:2014fla}, using the scalar-product method. Lines of different colors
correspond to  hydro simulations with respect to the Pb+Pb collisions at $\sqrt{s_{NN}}=5.02$ TeV, with
various parameterizations of $\eta/s$, and with
EKRT initial condition. Figure reproduced from \Ref{Niemi:2015voa}, with permission.
}
\end{center}
\begin{multicols}{2}

In addition to these aforementioned flow observables, there are many other types of
measurements associated with the obvserved long-range multi-particle correlations,
and flow harmonics, supporting the concept of system collective expansion in nucleus-nucleus collisions.
Similar measurements have also been generalized recently to  small colliding systems, e.g., p+Pb
at the LHC energies, where medium collectivity is observed in collision events of sufficiently 
high multiplicity.
We summarize them in \Tab{tab:vn}.
Similar to the results shown in \Fig{fig:vn2} and \Fig{fig:epcorrelator}, all of these observables 
agree with hydrodynamic simulations, to a quantitative level,
upon specified parameterization of 
initial state, equation of state, and transport coefficients. Thereby, a flow paradigm based on hydro modelings of
heavy-ion experiments is very well established.

\end{multicols}
\begin{center}
\tabcaption{ \label{tab:vn}  A summary of flow observables measured in experiments}
\footnotesize
\begin{tabular*}{180mm}{@{\extracolsep{\fill}}c|c|c|c}
\toprule   flow observables & harmonic order involved & colliding systems  & dependence \\
\hline
$v_n\{2\} $ & $n=$ 1,2,3,4,5,6 & PbPb, pPb, dAu, He$^3$Au & centrality, $p_T$, particle species, \\ 
&&AuAu&pseudo-rapidity  \\ 
\hline
$v_n\{4\} $ &  $n=$ 2,3 & PbPb & centrality \\
\hline
$v_n\{6\} $ &  $n=2$ & PbPb & centrality \\
\hline
$v_n\{8\} $ &  $n=2$ & PbPb & centrality \\
\hline
$r_n $ &  $n=2,3$ & PbPb, pPb & centrality,$p_T$,pseudo-rapidity \\
\hline
Event-by-event flow distribution $\bP(v_n)$ & $n=$ 2,3,4 & PbPb & centrality \\
\hline
Event-plane correlation &  $n\le 6$ & PbPb, AuAu & centrality, pseudo-rapidity \\
\hline
Projection of $V_n$ onto lower haromincs & $v_4\{\Psi_2\}$, $v_6\{\Psi_3\}$, $v_7\{\Psi_{23}\}$ & PbPb, AuAu & centrality, $p_T$ \\
\hline
Nonlinear medium response coefficients & $ n=$ 4,5,6,7 & PbPb & centrality \\
\hline
Symmetric cumulants & $n\le 5$ & PbPb, pPb & centrality \\
\bottomrule
\end{tabular*}
\end{center}

\begin{multicols}{2}

In the following discussions of the section, we select several flow observables in experiments.
These observables are categorized into two sets. One is related to the fluctuations of harmonic
flow, and the other contains information about flow correlations. It is our purpose to demonstrate, 
in the framework of flow paradigm,
in particular, in terms of the analytic response relations we have shown in the previous sections,
that these observables can be understood to a quantitative level.  Various constraints from these
observables, and also schemes developed correspondingly to reduce uncertainties from inital state modelings, help to 
improve the precision in the extraction of the medium transport coefficients. 

\subsection{Event-by-Event fluctuations of $v_n$}
\label{sec:vnfluc}

As discussed, event-by-event fluctuations in the observed flow signatures are mostly from
initial state eccentricities. 
Although there is no direct measurement of initial stages in heavy-ion collisions available,
fluctuating behavior of the flow harmonics 
provides the best probes of the fluctuations among 
the initial state of nucleon-nucleon collisions.
Fluctuations of flow harmonics can be studied in terms of the probability distribution $\bP(v_n)$ 
of the flow magnitude $v_n$,
or in a more quantitatively manner, the cumulants of harmonic flow. 
The quantitative analysis in this section depends on the linear response relation, $v_n=\kappa_n \ve_n$, 
which is approximately valid
for the lower order flow harmonics, $v_2$ and $v_3$, although effects due to nonlinear mode couplings 
are strong in some particular observables. 

\subsubsection{Fitting event-by-event flow fluctuations}

The ATLAS collaboration managed to measure the event-by-event distribution of harmonic flow in Pb+Pb
collisions at $\sqrt{s_{NN}}=2.76$ TeV~\cite{ALICE:2016kpq},
using an unfolding procedure to subtract non-flow effects~\cite{Jia:2013tja,Aad:2013xma}. 
The strategy has been applied recently
by the CMS collaboration, with respect to the 
updated Pb+Pb collisions at $\sqrt{s_{NN}}=5.02$ TeV~\cite{Castle:2017ywq}. 
Shown in \Fig{fig:pv} are the measured results of 
event-by-event distribution of $v_2$ (left panel), $v_3$ (middle panel) and 
$v_4$ (right panel), by the ATLAS collaboration,
for different centrality classes. 

The first attempt of fitting (and also interpreting) the probability distribution
 involves a pure Gaussian parameterization, which has been applied in \Fig{fig:pv} to
fit the event-by-event flow distribution (indicated as the solid curves),
\be
\bP(v_n) = \frac{v_n}{\sigma_n^2} e^{-v_n^2/2\sigma_n^2}\,.
\ee
A pure Gaussian corresponds to the Bessel-Gaussian function with a 
vanishing mean anisotropy $v_0=0$,
which gives rise to better description of $v_3$ than 
$v_2$, as expected. Note that in \Fig{fig:pv} the Gaussia fit is only available for
$v_2$ of the ultra-central collisions. 
However, even for $v_3$, 
non-Gaussian behavior is observed in collisions of larger centralities.
It should be noted that 
the fitting procedure using a Gaussian 
function reveals very little information about the
initial state fluctuations, nor about the dissipative properties of the medium
evolution. It should also be noted that the probability distribution 
of $v_4$ is more complicated in its physical origin, accounting for nonlinear medium
response, which we shall address later.

With the help of the formulae \Eqs{eq:vnform} to relate magnitudes $v_n$ and $\ve_n$, 
for the elliptic flow and triangular flow, one finds a change of variable
\be
\bP(v_n) = \frac{d \ve_n}{d v_n} P(\ve_n^{-1}(v_n))\,,
\ee 
can be applied to rewrite the distribution function of flow harmonics 
in terms of the probability distribution of initial state eccentricities.
For the linear response dominated scenario, in $v_2$ and $v_3$, the above relation is simply a rescaling
of the initial state eccentricity distribution,
\begin{align}
\label{eq:ebev2v3}
\bP(v_2)=\frac{1}{\kappa_2} P(v_2/\kappa_2)\,,\qquad
\bP(v_3)=\frac{1}{\kappa_3} P(v_3/\kappa_3)\,.
\end{align}
For the initial state eccentricity distribution characterized by the elliptic-power \Eq{eq:ep_1d} 
or power distribution \Eq{eq:power},
the rescaling introduces one extra parameter, the linear response coefficient $\kappa_n$. 
Therefore, it becomes possible to extract directly the 
linear response coefficient from the probability distribution of flow. 
In addition to $\kappa_n$,
fitting to flow fluctuations also identifies the parameters $\alpha$ and/or $\ve_0$, which
captures in the initial state the fluctuation strength and mean eccentricity.  

\end{multicols}
\begin{center}
\includegraphics[width=0.95\textwidth] {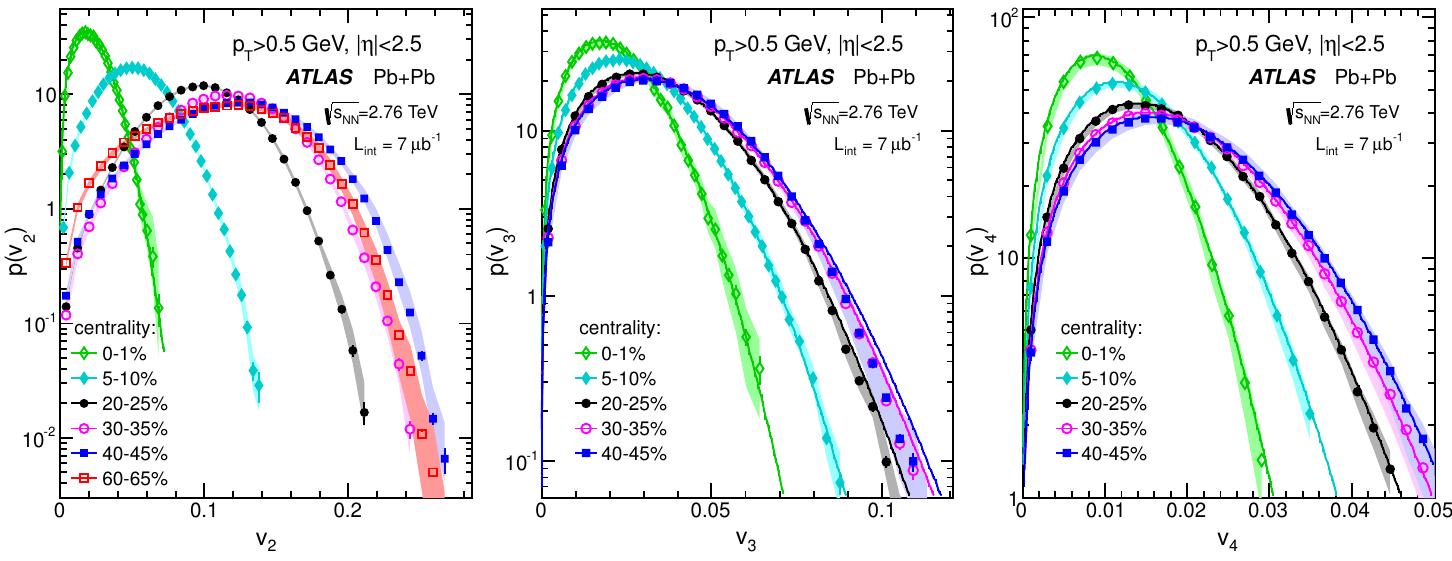}
\figcaption{Probability distribution of flow magnitude $\bP(v_n)$, measured 
by the ATLAS collaboration for the Pb+Pb collisions at $\sqrt{s_{NN}}=2.76$ TeV.
Left panel: elliptic flow $v_2$. Middle panel: triangular flow $v_3$. 
Right panel: quadrangular flow $v_4$. Figure adapted from
\Ref{Aad:2013xma} (DOI: https://doi.org/10.1007/JHEP11(2013)183), under the CC-BY-NC-ND license, with permission.
\label{fig:pv}
}
\end{center}
\ruledown
\begin{multicols}{2}

In a similar manner, one may consider applying the rescaled 
Bessel-Gaussian function for the parameterization
of initial state eccentricity fluctuations, so that the linear response coefficient $\kappa_n$ can be
determined simultaneously with $\sigma$ and $\ve_0$, the parameters
in the original Bessel-Gaussian distribution function. However, because 
the Bessel-Gausian function (or Gaussian) is scale invariant,
for which a rescaling the distribution is equivalent to $\sigma\rightarrow \kappa_n\sigma$ and
 $\ve_0\rightarrow\kappa_n\ve_0$, the fitting procedure cannot constrain the value of $\kappa_n$,
 nor $\sigma$ or $\ve_0$ individually.  
Actually, the ability to disentangle initial state parameters and the linear response coefficients, 
relies on the non-Gaussianity of the flow fluctuations.

The left-hand panel of \Fig{fig:pv2} presents the extracted values of $\kappa_n$, from the 
 fit of ATLAS measured $v_2$  and $v_3$ distributions, 
with Elliptic-Power and Power parameterizations.  The linear response coefficients $\kappa_2$ and 
$\kappa_3$ decrease monotonically as centrality percentile increases. This is expected, since 
viscous effects get stronger in smaller collision systems, leading to a suppressed medium
response. In comparison with hydro predictions, the centrality dependence of linear response coefficients 
allows one to estimate the value of $\eta/s$. In the left-hand panel of \Fig{fig:pv2}, 
hydro calculations with $\eta/s\simeq 0.19$ (green lines) 
are found to give rise to the best description. It should be emphasized that uncertainties normally
induced from initial state effective modelings are reduced 
in the present procedure of estimating
$\eta/s$. 

Simultaneously, the fit of flow event-by-event distribution results in the extracted 
values of $\alpha$ and $\ve_0$, which are shown in the middle panel and 
the right-hand panel of \Fig{fig:pv2}, respectively. Recalling that the parameter $\alpha$
characterizes the fluctuation strength in the initial state, the decrease of $\alpha$
from central to peripheral collisions,
indicates an increasing strength of fluctuations. On the other hand,
the background shape of the system gets more elliptic in non-central Pb+Pb collisions, which
is reflected in 
the observed growth of $\ve_0$. Both trends are 
expected. For comparisons, the fluctuation strength
and background shape in MC-Glauber and IP-Glasma are shown in \Fig{fig:pv2} as shaded bands.
Note that the extracted values of $\alpha$ and $\ve_0$ from the experiementally measured
$v_2$ distributions are not quite compatible with these effective models, upon the
linear medium response.

\end{multicols}
\begin{center}
\includegraphics[width=0.32\textwidth] {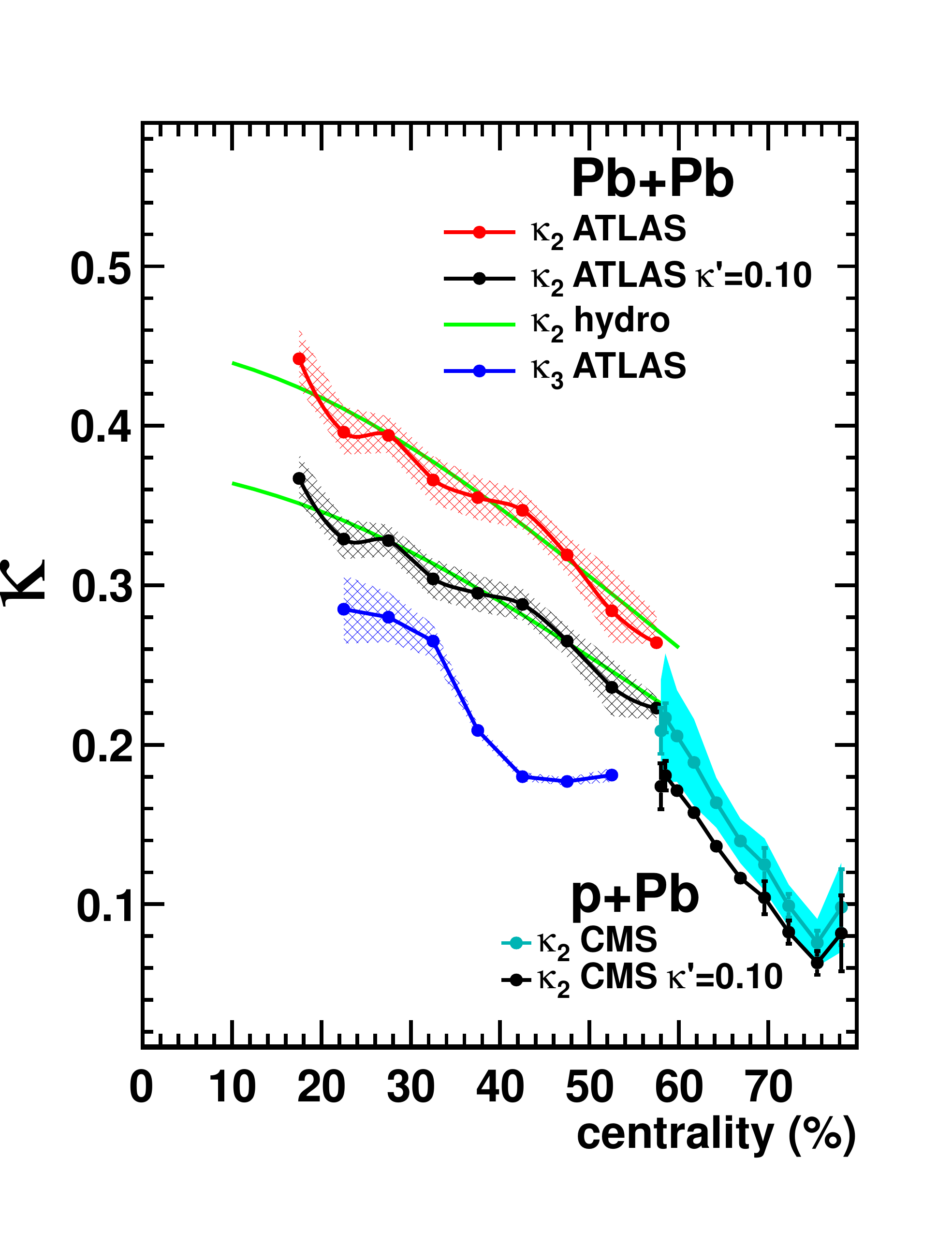}
\includegraphics[width=0.32\textwidth] {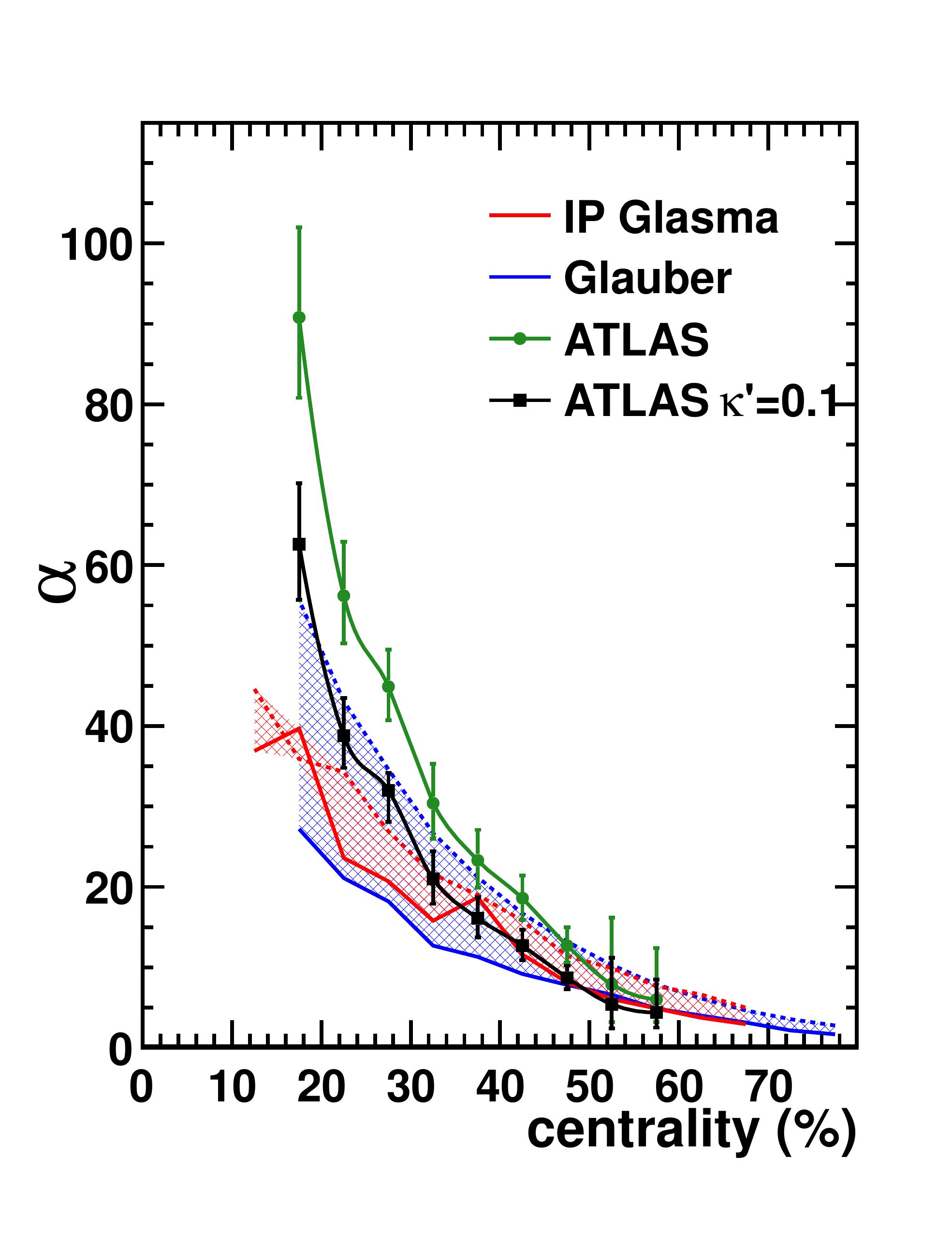}
\includegraphics[width=0.32\textwidth] {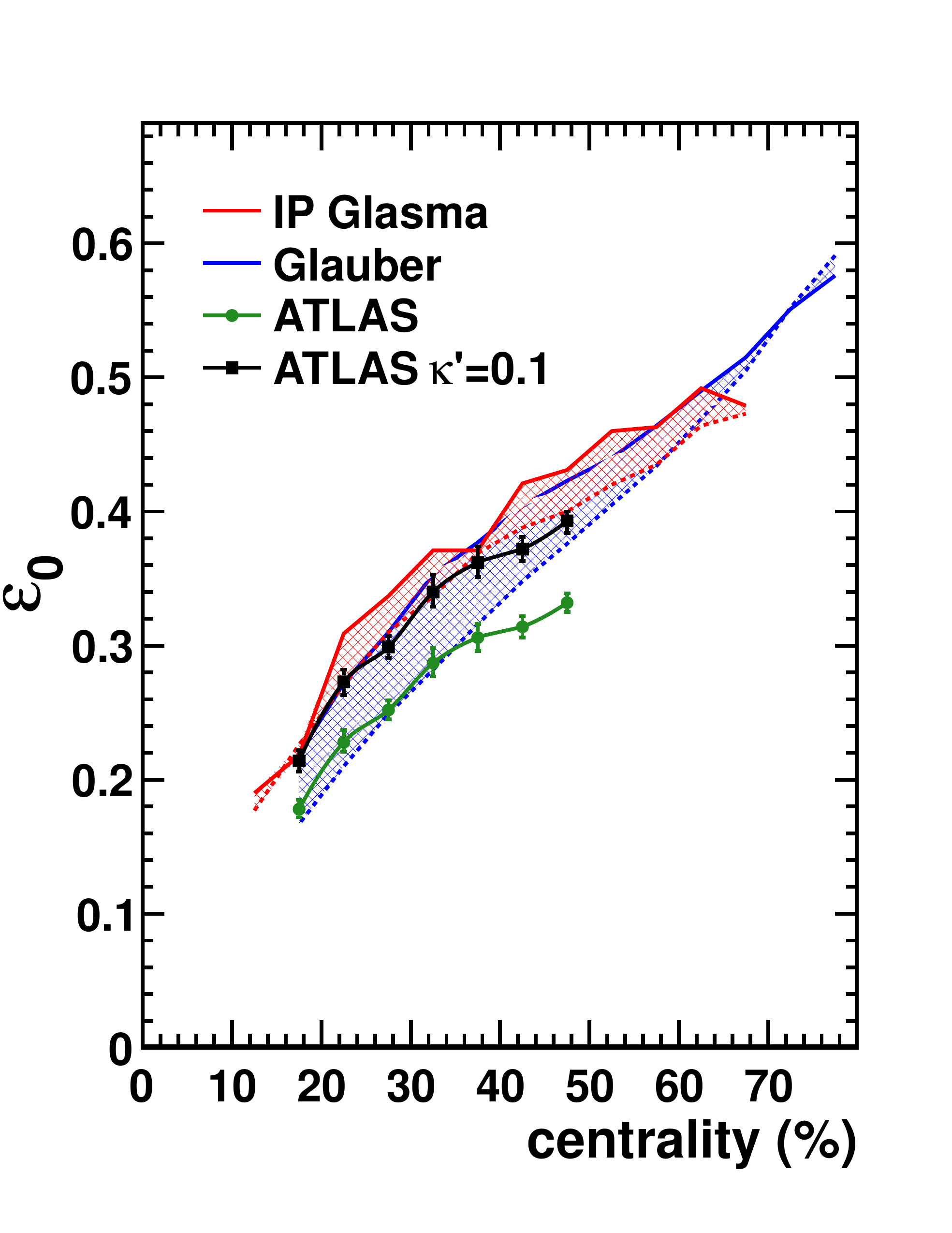}
\figcaption{
\label{fig:pv2}
Left panel: Extracted linear response coefficient $\kappa_n$ from the fit of flow
event-by-event distributions, with respect to the Pb+Pb collisions at $\sqrt{s_{NN}}=2.76$ TeV,
and p+Pb at $\sqrt{s_{NN}}=5.02$ TeV. A corrected estimate of $\kappa_2$ is obtained
accounting for an extra cubic order response in the relations (black points). Solid green lines 
are viscous hydro calculations of the linear response coefficient for Pb+Pb collisions, 
corresponding to $\eta/s=0.19$. Middle panel and the right panel:
Extracted values of $\alpha$ and $\ve_0$ from the fit of flow event-by-event distributions (green points).
Corrected
estimates accounting for a cubit order response are shown as well as black points. 
The extracted values are compared with initial state effective models, MC-Glauber (blue bands)
and IP-Glasma (red bands). Figures adapted from \Ref{Yan:2014nsa} (DOI: https://doi.org/10.1016/j.physletb.2015.01.039), under the CC-BY-4.0 license (http://creativecommons.org/licenses/by/4.0/).
}
\end{center}
\begin{multicols}{2}

The cubic order corrections to the linear response relations 
are normally tiny, but their effects are not negligible in the flow
event-by-event fluctuations.  
Apparently, it changes the expected probability distribution of
flow in \Eq{eq:ebev2v3}, which introduces the cubic order response coefficient accordingly as 
an extra parameter. It is particularly important in the elliptic flow, for which during analysis 
we shall assume a constant ratio 
between cubic order response coefficient and the linear response coefficient, 
$\kappa'=\kappa_2'/\kappa_2\simeq 0.1$, for all centralities. 
This constant ratio is an empirical assumption, but it is compatible with event-by-event
hydro simulations, where $\kappa'$ is roughly a constant around 0.2.
We shall ignore effects from the cubic corrections in $v_3$. As expected,
shown in \Fig{fig:pv2}, a cubic order correction reduces the value of the linear response coefficient.
However, it does not significantly alter  the centrality dependence of $\kappa_2$, nor the estimated
value of $\eta/s$. A reduced linear response coefficient requires the system to have a larger 
eccentricity, and eccentricity fluctuations, to maintain the generated flow, which explains
the observed decrease of $\alpha$ and increase of $\ve_0$. With the additional
cubic order corrections, extracted values of $\alpha$ and $\ve_0$ from experiments
are compatible with MC-Glauber and IP-Glasma.

The ATLAS collaboration has also measured event-by-event distributions of the 
quadrangular flow magnitude $v_4$. One would expect the linear part
of the flow, $v_4^L$, to be captured by a power law distribution, as it  originates from pure initial
state fluctuations, similar to $v_3$. However, since quadrangular flow contains strong nonlinear couplings
of $V_2$, the probability distribution function of the whole magnitude, $v_4$, is supposed to be
a particular convolution of the power and elliptic-power distributions.

Alternatively, the 
probability distribution $\bP(v_n)$ can be quantitatively characterized
by moments, or cumulants, of the corresponding harmonic flow $v_n$. 
Given the 
probability distribution
function measured in experiments, the moment of harmonic flow $\bbra v_n^m \kket$ 
can be calculated through the following integral 
\be
\label{eq:vn_mom}
\bbra v_n^m\kket = \int d v_n \bP(v_n) v_n^m\,.
\ee
Note that in realistic experiments, $\bbra v_n^m\kket$ is often measured
through an event average with a rapidity-gap to suppress non-flow contributions. 
Similarly, the cumulants of harmonic flow can be obtained 
in both methods in experiments, from the probability distribution function, 
or an average over collision events. 

In either way, the obtained moments (or cumulants) of the harmonic flow can be
used to analyze  the underlying probability distribution function, according to 
\Eq{eq:vn_mom}. For instance, 
the mean $\bbra v_n\kket$ and the second order moment, $\bbra v_n^2\kket$ (or the
second order cumulant $v_n\{2\}$), determine the
variance of the distribution function. The third order and fourth order moments, $\bbra v_n^3\kket$
and $\bbra v_n^4\kket$, are related to the skewness and kurtosis of the distribution function, respectively.
It should be emphasized that moments, or cumulants, of the flow, are very sensitive probes
of the detailed structure of 
the distribution function. As a result, moments of flow harmonics
lead to more accurate characterization than a direct fit of
the event-by-event flow distribution.

\subsubsection{Cumulants of harmonic flow}

\end{multicols}
\begin{center}
\includegraphics[width=0.9\textwidth] {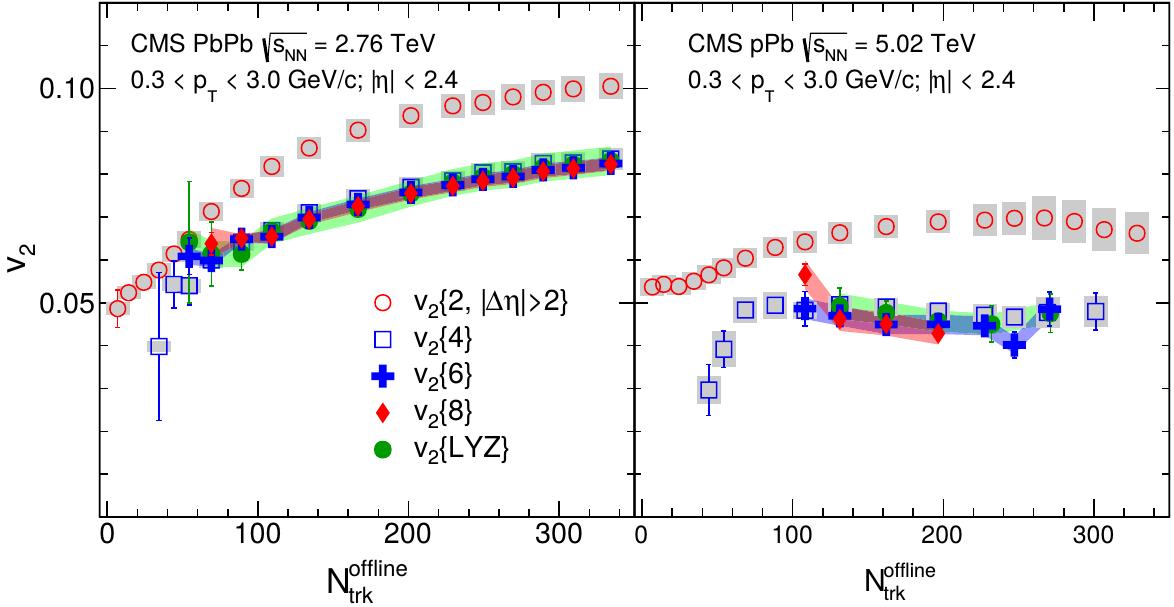}
\figcaption{
\label{fig:v2_cumulant}
Cumulants of elliptic flow $v_2$ are measured up to the 8th order in Pb+Pb collisions
at $\sqrt{s_{NN}}=2.76$ TeV and p+Pb collisions at $\sqrt{s_{NN}}=5.02$ TeV in the same multiplicity
range (in terms of number of offline tracks), from the CMS
collaboration. Figure reproduced from \Ref{Khachatryan:2015waa} (DOI: https://doi.org/10.1103/PhysRevLett.115.012301, ``Evidence for Collective Multiparticle Correlations in p-Pb Collisions''), under the CC-BY-3.0 license.
}
\end{center}
\begin{multicols}{2}

There are certain patterns of the flow cumulants observed in experiments, corresponding to the
underlying properties of the probability distribution. These patterns quantitatively
constrain the flow fluctuation behavior. 
For the convenience of discussion,
we limit ourselves to the case of linear medium response, $v_n=\kappa_n\ve_n$ (as in
$v_2$ and $v_3$),
so that the patterns in cumulants of flow harmonics are associated with those in the 
fluctuations of initial state eccentricities, 
through the scaled relation $\bP(v_n)=\bP(\kappa_n \ve_n)/\kappa_n$. 
With respect to a specified probability distribution function, a cumulant of order $m$ is identified
in the expansion of the cumulant generating function,
\begin{align}
G(k_x,k_y)\equiv&\ln\bbra \exp(ik_x v_x + ik_yv_y) \kket \cr
=&\ln\int dv_x dv_y \bP(v_x,v_y)\exp(ik_x v_x + ik_yv_y)\,,\cr
\end{align}
with respect to the $k^m$ term.

 A simple example is a Gaussian distribution, from which, except for the $v_n\{2\}$ 
being equal to a 
hydro linear response to the variance $v_n\{2\}=\kappa_n\sigma$,  
all higher order cumulants vanish,
\be
v_n\{4\}=v_n\{6\}=v_n\{8\}=\ldots=0\,.
\ee
This is understandable, since the generating function of the cumulant contains only the second order term,
\be
G(k_x,k_y) =G(k)=-\frac{1}{2}k^2 (\kappa_n\sigma)^2\,.
\ee
In a similar way, for a Bessel-Gaussian distribution, the cumulant generating function is (after
an integral over the relative angle in $\bf k$),
\be
G(k) = -\frac{1}{2}k^2 (\kappa_n\sigma)^2+\ln J_0 (k\kappa_n \ve_0)
\ee
The expected second order cumulant is 
$v_n\{2\}=\kappa_n\sqrt{\sigma^2+\ve_0^2}$, while all higher order cumulants become degenerate,
\be
v_n\{4\}=v_n\{6\}=v_n\{8\}=\ldots=\kappa_n\ve_0\,.
\ee
The Bessel-Gaussian distribution is a good demonstration that the second order cumulant is separated from
higher order ones, as long as fluctuations (described in terms of $\sigma$) play a significant role. The
experimentally measured cumulants of elliptic flow from the CMS 
collaboration~\cite{Khachatryan:2015waa} are shown in \Fig{fig:v2_cumulant}, for
the Pb+Pb collisions at $\sqrt{s_{NN}}=2.76$ TeV (the left-hand panel), where one indeed observes the feature that
$v_2\{2\}$ is separable from higher order ones, while higher order cumulants collapse.
However, the pattern is not uniquely associated with a Bessel-Gausian 
distribution. In particular, the pattern can be expected  
in the power distribution and elliptic-power distribution.
Moreover, one notices that the experimental measurements suggest 
an ordering in higher cumulants,
although the differences in higher order cumulants are tiny,
\be
\label{eq:vnorder}
v_2\{2\}>v_2\{4\}\gtrsim v_2\{6\}\gtrsim v_2\{8\}\,.
\ee

For the power distribution, all the values of cumulants are determined analytically
via the single parameter $\alpha$. With the help of the cumulant generating function, 
the second order cumulant is found to be 
\be 
\label{eq:v22_power}
v_n\{2\}=\kappa_n/\sqrt{1+\alpha}\,,
\ee
while higher order ones are~\cite{Yan:2013laa}
\begin{align}
\label{eq:vncumu}
v_n\{4\}= &\kappa_n\left[\frac{2}{(1+\alpha)^2(2+\alpha)}\right]^{1/4}\,,\cr
v_n\{6\}= &\kappa_n\left[\frac{6}{(1+\alpha)^3(2+\alpha)(3+\alpha)}\right]^{1/6}\,,\cr
v_n\{8\}= &\kappa_n\left[\frac{48\left(1+5\alpha/11\right)}{(1+\alpha)^4(2+\alpha)^2(3+\alpha)(4+\alpha)}\right]^{1/8}\,.
\end{align}
Note that the power distribution is derived assuming $N$ independent sources on top of a two-dimensional
Gaussian desnity profile with $\alpha=(N-1)/2$. Although in the large $N$ limit, $N\gg1$, one expects
a strong ordering $v_n\{8\}\ll v_n\{6\}\ll v_n\{4\} \ll v_n\{2\}$. In the opposite limit of 
$N\rightarrow 1$, 
(limit that corresponds to the case of strong flow fluctuations), 
these cumulants converge: $v_n\{8\}\approx v_n\{6\}\approx v_n\{4\} \approx v_n\{2\}$. However,
for practical N values corresponding to
p+Pb collisions at the LHC energies, the gap between second order cumulant and higher order ones
becomes sizable, and one finds similar relations $v_n\{8\}\lesssim v_n\{6\}\lesssim v_n\{4\}<v_n\{2\}$,
 in agreement with
what was found in the p+Pb collisions (right-hand panel in \Fig{fig:v2_cumulant}). 
It is worth mentioning that, given these flow cumulants up to
$v_2\{8\}$, the linear flow response 
coefficients $\kappa_2$ and the parameter $\alpha$, can be solved
according to \Eq{eq:v22_power} and \Eq{eq:vncumu}. The corresponding results are shown in
\Fig{fig:pv2}.

In the elliptic-power distribution,  the additional parameter
$\ve_0$ that describes a mean eccentricity in the reaction-plane breaks azimuthal symmetry,
and the derivation of cumulant generating function is complicated. Nevertheless, following the
calculation of moments,  the elliptic-power distribution leads to the analytical expression~\cite{Yan:2014afa}
\begin{align}
f_m\equiv \langle (1-v_n^2)^m\rangle
=&\frac{\alpha}{\alpha+m}
\left(1-v_0^2\right)^m\times\cr
&
{_2}F_1\left(m+\frac{1}{2},m;\alpha+m+1;v_0^2\right),,
\end{align} 
where we assume $v_0=\kappa_n \ve_0$ as the linear response to the mean eccentricity.
Using this equation, 
one obtains from the standard definition of cumulants from moments,
\begin{align}
\label{cumulantsEP}
v_n\{2\}=&(1-f_1)^{1/2} \cr
v_n\{4\}=&(1 - 2 f_1 + 2 f_1^2  - f_2)^{1/4} \cr
v_n\{6\}=&\left(1+\frac{9}{2}
f_1^2-3f_1^3+3f_1(\frac{3}{4}f_2-1)-\frac{3}{2} f_2
 -\frac{1}{4}f_3\right)^{1/6}.\cr
\end{align}
These are analytical expressions of cumulants obtained with respect to the elliptic-power 
distribution, as a function of parameters $(\ve_0, \alpha)$. To demonstrate the ordering
of cumulants, let us consider the limit
of large $\alpha$, $\alpha\gg1$, but keeping $\alpha \ve_0^2$ a constant (so that $\ve_0^2\ll 1$ is a small
variable). This is 
a plausible assumption regarding nucleus-nucleus collision events with small centrality percentile,
because larger $\alpha$ corresponds to systems of a large number of independent sources,
thus there are more central collisions. Similarly, 
treating $\ve_0$ as a small quantity from the constant constraint of $\alpha \ve_0^2$, implies
a background of the collision system with a small ellipticity.  Given these conditions, one is 
allowed to expand the expressions in \Eqs{cumulantsEP} in series of $1/\alpha$ and/or $\ve_0^2$.
Note that the expansion must be carried out simultaneously regarding $O(1/\alpha)\sim O(\ve_0^2)$.
To the leading order, one finds the ratios between cumulants,
\begin{eqnarray}
\frac{v\{4\}}{v\{2\}}&=&\sqrt{\frac{\alpha\varepsilon_0^2}{1+\alpha\varepsilon_0^2}}+{\cal
  O}\left(\frac{1}{\alpha}\right)\cr
\frac{v\{6\}}{v\{4\}}&=&1-\frac{1+\alpha\varepsilon_0^2}{2(\alpha\varepsilon_0^2)^2\alpha}+{\cal
  O}\left(\frac{1}{\alpha^2}\right)\cr
\frac{v\{8\}}{v\{6\}}&=&1-\frac{1}{22(\alpha\varepsilon_0^2)\alpha}+{\cal
  O}\left(\frac{1}{\alpha^2}\right)\,,
  \label{phobos}
\end{eqnarray}
which reflect a similar ordering pattern as in \Eq{eq:vnorder} in this limit.

 Actually, the tiny splitting in higher order cumulants predicted in the power and the 
 elliptic-power distribution functions, and 
 observed in experiments, is a generic feature of non-Gaussianity. 
 It can be realized  in effective models, such as MC-Glauber~\cite{Bzdak:2013rya}, where non-Gaussianity is introduced
 from the background shape, extra correlations among sources, etc. 
 
 Shown in \Fig{fig:glb_ratio}
 are the results of cumulant ratios of initial ellipticity from event-by-event simulations of 
 MC-Glauber model (symbols),
 which, upon a linear hydro response, is identical to the cumulant ratios of elliptic flow $v_2$,
 e.g.,
 \be
 \frac{v_2\{4\}}{v_2\{2\}}=\frac{\ve_2\{4\}}{\ve_2\{2\}}\,,\quad
  \frac{v_2\{6\}}{v_2\{4\}}=\frac{\ve_2\{6\}}{\ve_2\{4\}}
 \ee
 The curves in \Fig{fig:glb_ratio} are parameterizations with an elliptic-power distribution. While the
 elliptic-power distribution 
 gives a consistent estimate of the splitting between $v_2\{2\}$ and $v_2\{4\}$, it overestimates
 the gaps among $v_2\{4\}$, $v_2\{6\}$ and $v_2\{8\}$. 
 
\begin{center}
\includegraphics[width=0.45\textwidth] {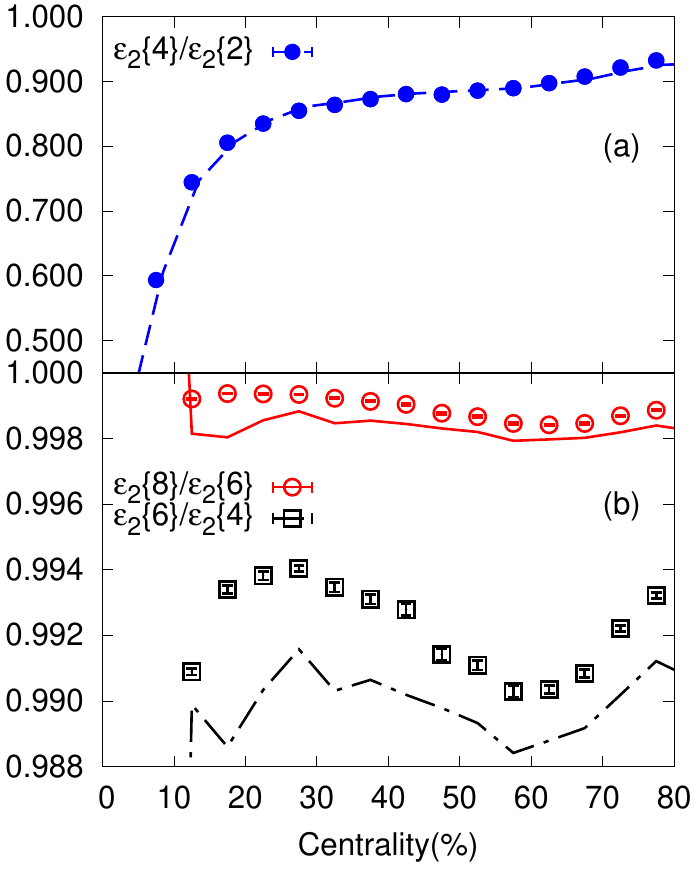}
\figcaption{
\label{fig:glb_ratio}
Cumulant ratios of the elliptic anisotropy obtained from event-by-event simulations with
MC-Glauber model, for the Pb+Pb collisions at $\sqrt{s_{NN}}=2.76$ TeV. Lines of the same
order correspond to paramterizations with elliptic-power distribution function.
Figure reproduced from \Ref{Yan:2014afa}, with permission.
}
\end{center}

 The success of the elliptic-power distribution in describing the cumulant ratios (at least up to the fourth
 order cumulant), comparing to the Bessel-Gaussian, is to large extent 
due to the fact that the initial state eccentricity is bounded by unity. 
The upper bound on the initial eccentricity induces a major source of 
the non-Gaussianity in the initial eccentricity fluctuations, and also in the event-by-event
flow fluctuations. In particular, the upper bound on the ellipticity
induces skewness of the $v_2$ fluctuations.

Using event-by-event hydro simulations of the Pb+Pb collisions in centrality
class 50\%-55\%, the distribution of elliptic flow
in- ($v_x$) and out-of- ($v_y$) the reaction-plane are shown as the histogram in \Fig{fig:v2skew1}. The magnitude
of the flow is $v_2=\sqrt{v_x^2+v_y^2}$. 
Correspondingly, initial state event-by-event distributions of $\ve_x$ and $\ve_y$ 
scaled by a linear hydro response coefficient $\kappa=0.21$ are shown for comparisons. Note this
linear response coefficient is a real quantity, which applies for both $v_x$ and $v_y$.
There are two characteristic properties of the 
non-Gaussian fluctuations associated with the $v_2$ distribution, as
one finds in \Fig{fig:v2skew1}. The first is reflected in the skewness.
The two-dimensional distribution of $\E_2$ is symmetric under $v_y\leftrightarrow -v_y$, as required
by parity. Thus one observes in \Fig{fig:v2skew1} (a) the symmetric distribution of $\ve_y$, 
leading to a symmetric distribution of $v_y$
via linear hydro response. On the other hand, the distribution of $\ve_x$ is mostly found as positive,
reflecting a non-zero mean $\bbra \ve_x\kket=\ve_0$, and $\bbra v_x\kket=v_0$. Together with the upper bound condition $\ve_2\le1$,
the corresponding x-component of initial eccentricity $\ve_x$ is negatively skewed\footnote{
The convention of negative skewness of a distribution is reflected as the distribution being concentrated
on the right side, while positive skewness corresponds to a left-side concentrated distribution. 
}  
Upon a linear hydro response, negative skewness is achieved in the event-by-event distribution of
$v_x$, seen as the comparison of histograms in \Fig{fig:v2skew1} (b). 

\begin{center}
\includegraphics[width=0.45\textwidth] {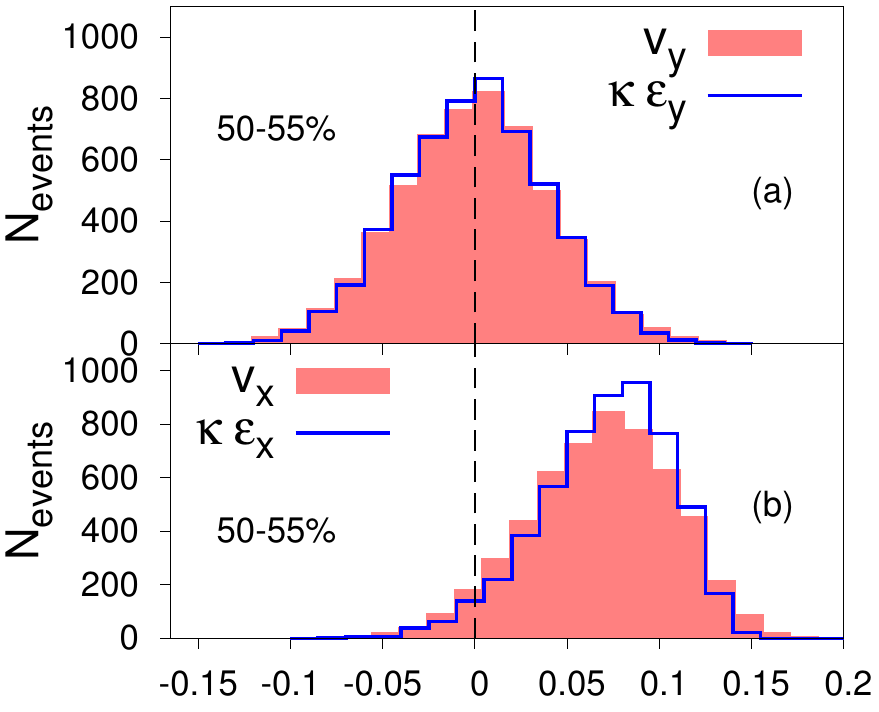}
\figcaption{
\label{fig:v2skew1}
Shaded area: Histograms of $v_x$ (b) and $v_y$ (a) event-by-event distributions from hydro simulations. 
Lines: Histograms of $\ve_x$ and $\ve_y$ distributions scaled by linear repsonse coefficient $\kappa=0.21$.
Figure reproduced from \Ref{Giacalone:2016eyu}, with permission.}
\end{center}

In addition to the skewness, one also 
observes from the simulations that the variance in the $v_y$ distribution and the variance in the $v_x$ distribution, 
\be
\sigma_y^2 \equiv \bbra v_y^2 \kket\,,\qquad
\sigma_x^2 \equiv \bbra (v_x  - v_0)^2\kket  
\ee
are not equal: $\sigma_y>\sigma_x$, 
another signature of
non-Gaussian fluctuations.

To quantify skewness of the distribution via the non-zero
third order moments, one needs the following non-zero third order moments
\be
s_1=\bbra (v_x-v_0)^3 \kket\,,\qquad
s_2 =\bbra (v_x-v_0) v_y^2 \kket\,.
\ee 
where $s_1<0$ describes the negatively skewed distribution of $v_x$. The skewness $s_1$ defines
the standardized skewness 
\be
\label{eq:gamma1_real}
\gamma_1 \equiv \frac{s_1}{\sigma_x^3}\,.
\ee

Therefore, in terms of the non-Gaussian 
fluctuations characterized by $\sigma_y^2-\sigma_x^2$, $s_1$ and $s_2$, the cumulants of flow are
\begin{eqnarray}
\label{asymskew}
v_2\{2\}&=&\sqrt{v_0^2+\sigma_x^2+\sigma_y^2} ,\cr
v_2\{4\}&\simeq& v_0+\frac{\sigma_y^2-\sigma_x^2}{2
  v_0}-\frac{s_1+s_2}{v_0^2} ,\cr
v_2\{6\}&\simeq& v_0+\frac{\sigma_y^2-\sigma_x^2}{2
  v_0}-\frac{\frac{2}{3}s_1+s_2}{v_0^2} ,\cr
v_2\{8\}&\simeq& v_0+\frac{\sigma_y^2-\sigma_x^2}{2
  v_0}-\frac{\frac{7}{11}s_1+s_2}{v_0^2} ,
\end{eqnarray}
where higher order cumulants are expanded in powers of fluctuations. The present analysis
focuses on the dominant source of non-Gaussian fluctuations in the elliptic
anistotropy, assuming a linear hydro response. Accounting for these fundamental concepts
in the flow paradigm, the derived results in \Eqs{asymskew} provide a very simple interpretation
of the fine splitting structures in higher order cumulants of $v_2$:
the splitting in higher order cumulants is solely generated by skewness $s_1$. It then allows
one to get some interesting findings. Especially, with respect to the fact that
 the splittings in cumulants are analytically related to the 
skewness  up to the leading order in the expansion of fluctuations, one expects
the differences
\begin{align}
&v_2\{4\}-v_2\{6\}=-\frac{s_1}{3 v_0^2}\,,\cr
&v_2\{6\}-v_2\{8\}=-\frac{s_1}{33 v_0^2}\,,
\end{align}
or equivalently~\cite{Jia:2014pza,Giacalone:2016eyu},
\begin{equation}
\label{v8}
v_2\{6\}-v_2\{8\}=\frac{1}{11}(v_2\{4\}-v_2\{6\}).
\end{equation}

\end{multicols}
\begin{center}
\includegraphics[width=0.45\textwidth] {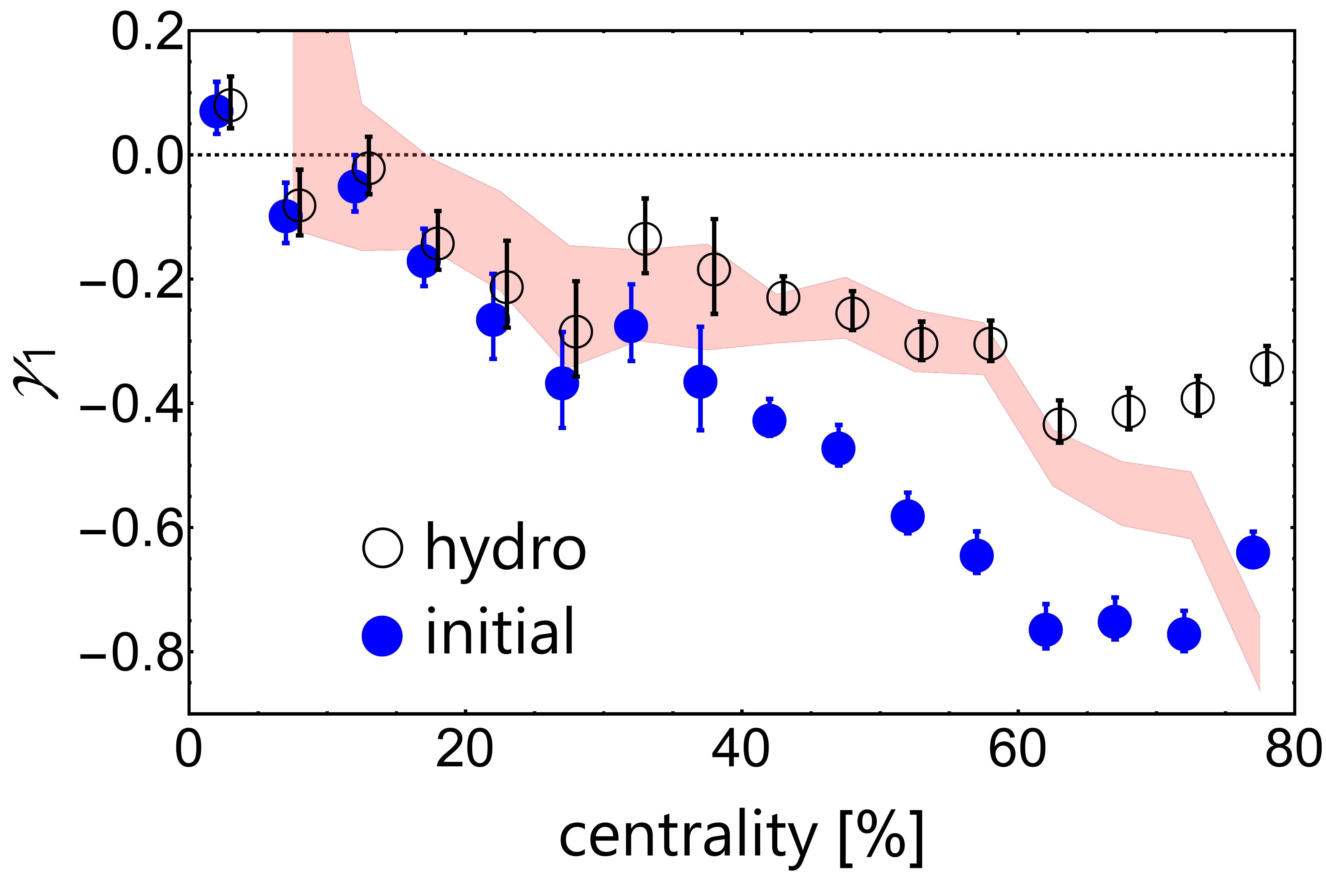}
\includegraphics[width=0.45\textwidth] {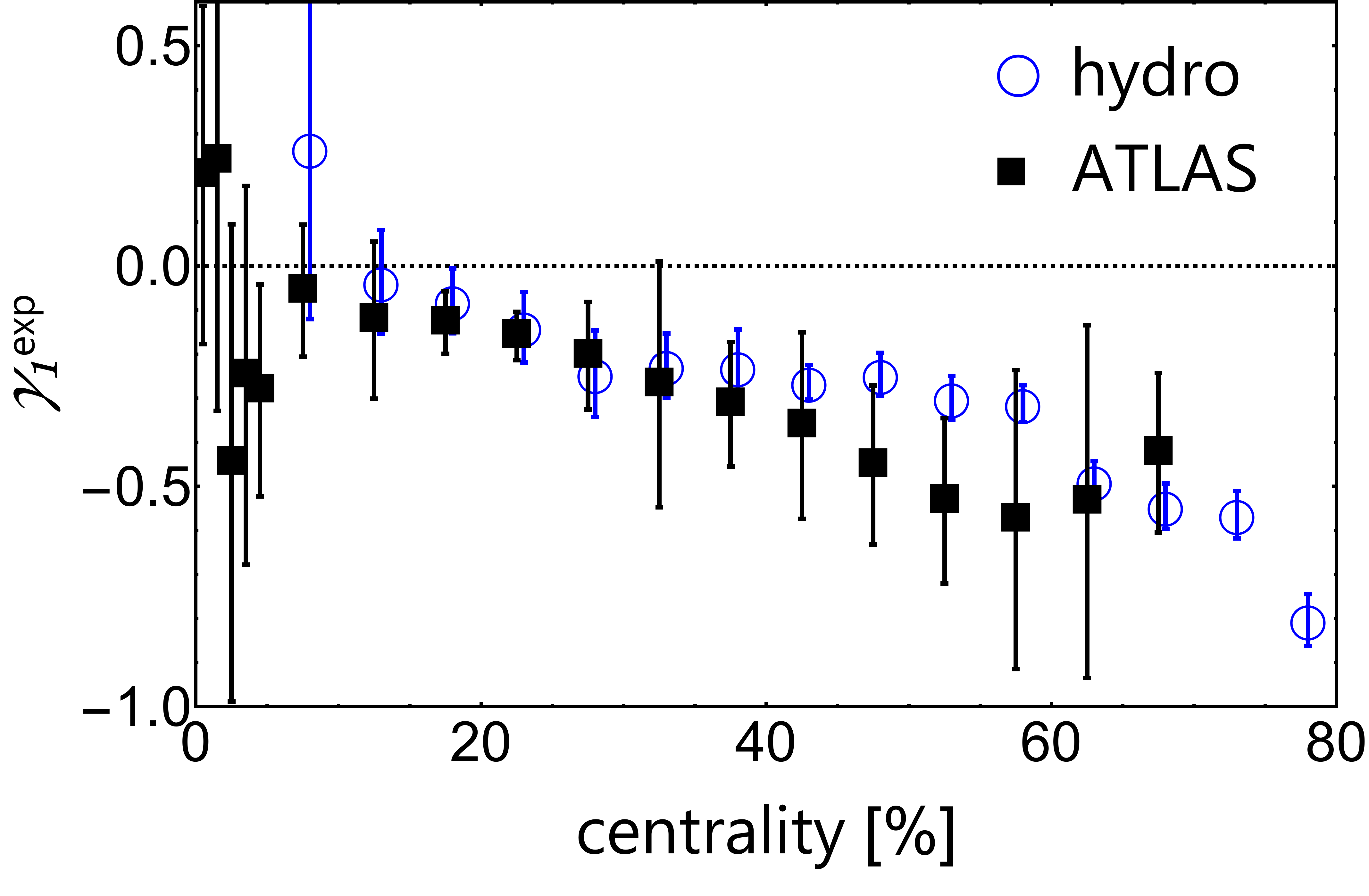}
\figcaption{
\label{fig:v2skew2} Left:
The standardized skewness of elliptic anisotropy from event-by-event hydro simulations
with respect to Pb+Pb collisions of $\sqrt{s_{NN}}=2.76$ TeV. Symbols correspond to
the results calculated according to \Eq{eq:gamma1_real} for the initial ellipticity (blue
points) and $v_2$ (open symbols). The shaded band corresponds to $\gamma_1^{exp}$, obtained
following \Eq{gamma1exp}. Right:
Comparison of hydro predictions on $\gamma_1^{exp}$ to experimental results from the
ATLAS collaboraion. Figure reproduced from \Ref{Noronha-Hostler:2015dbi}, with permission.
}
\end{center}
\begin{multicols}{2}

The skewness of $v_2$ is present only in the distribution of $v_x$, which is not approachable
in experiment on an event-by-event basis. However, since $s_1$ is related to the splitting 
of higher order cumulants of $v_2$, under the assumption that fluctuations are sub-dominant, 
one finds the standardized skewness
\Eq{eq:gamma1_real} approximates to the following quantity $\gamma_1^{exp}$,
\begin{equation}
\label{gamma1exp}
\gamma_1\approx
\gamma_{1}^{\rm exp}\equiv -6\sqrt{2}\, v_2\{4\}^2\frac{v_2\{4\}-v_2\{6\}}{(v_2\{2\}^2-v_2\{4\}^2)^{3/2}}.
\end{equation}
\Eq{gamma1exp} can be verified in event-by-event hydro simulations. Shown in the left-hand panel of
\Fig{fig:v2skew1} are the skewness results calculated from hydrodynamics with respect to  
Pb+Pb collisions at $\sqrt{s_{NN}}=2.76$ TeV, given a constant $\eta/s=0.08$, 
as a function of centrality. 
The simulations are initialized
via the MC-Glauber model, from which the standardized skewness of initial eccenticity $\ve_x$,
$\gamma_1^\ve$ is obtained (blue points), according to 
\Eq{eq:gamma1_real}  but
with a replacement of $v_x$ by $\ve_x$ in calculating $s_1$ and
$\sigma_x$. The open symbols in the left-hand panel of 
\Fig{fig:v2skew2} correspond to the exact predictions
of the standardized skewness from hydrodynamics following \Eq{eq:gamma1_real}, while the shaded band
is obtained with respect to the approximated estimate from \Eq{gamma1exp}, in terms of flow cumulants. 
The agreement between $\gamma_1$ and $\gamma_1^{exp}$ is remarkable, up to 60\% of centrality,
where flow fluctuations are significant that the assumption applied in \Eq{gamma1exp} breaks 
down. Hydro predicted $\gamma_1^{exp}$ of $v_2$ is compatible with the experimental results 
from the ATLAS collaboration, as can be seen
in the comparison in the right-hand panel of \Fig{fig:v2skew2}. Recently the CMS collaboration 
has extended the measurements of the standard skewness to $\sqrt{s_{NN}}=5.02$ TeV~\cite{Castle:2017ywq}.

In the case of the absolute linear medium response, the effects of medium dynamics drop out of the definition
of the standardized skewness, hence one may identify $\gamma_1$ with $\gamma_1^\ve$.  However,
\Fig{fig:v2skew1} displays sizable deviation between $\gamma_1$ and $\gamma_1^\ve$ in 
collisions of centrality greater than 30\%,  indicating the role of the nonlinear medium response.
Actually, one observes $|\gamma_1|<|\gamma_1^\ve|$, which means that 
medium dynamics from hydro washes out the skewness of initial eccentricity
fluctuations, so that the generated $v_2$ distribution is less skewed. This effect is mostly
due to the cubic order response in $v_2$, which gets stronger as centrality grows, in line with
the trend observed in \Fig{fig:v2skew1}. 

In addition to the corrections from cubic order response to the fluctuations of $v_2$, 
the effects beyond the linear medium response are more pronounced 
in the higher flow harmonics.  For instance, the quadrangular flow $v_4$ has a negative
value of $v_4\{4\}^4$ found in non-central collisions in experiments~\cite{Aad:2014vba}, which
has a natural origin from the nonlinear couplings of $v_2^2$, as
$2\bbra v_2^4\kket^2-\bbra v_2^8\kket$~\cite{Giacalone:2016mdr}.

\subsection{Correlations among harmonic flow}
\label{sec:vncor}

In experiments, the flow fluctuations measure the fluctuating properties of flow
magnitude $v_n$, by correlating flow of the same harmonic orders.
Correlations of harmonic flow involving mixed harmonic orders, on the other hand, reflect the 
correlation nature of flow magnitude $v_n$ and phase $\Psi_n$. 
In experiments, there have been several types of correlators investigated so far,
which we discuss within the context of the flow paradigm.

\subparagraph{The event-plane correlators.} At first, the event-plane correlators, 
measured by the ATLAS collaboration,
were designed to detect the correlations among phases of different flow 
harmonics, on an event-by-event basis. It was later 
realized that event plane correlators 
also involve contributions through the correlated flow magnitudes.
In practice, there exists subtle difference in the definitions of event-plane correlators 
owing to the issues of detector resolution~\cite{Luzum:2012da}. In the present discussion, we 
take the definition of event-plane correlators using the Pearson correlation coefficient, although
the original results obtained by the ATLAS collaboration using the scalar-product method
are slightly different, under some circumstances.
For instance, for the correlation $V_4$ and $V_2$ 
one defines,
\be
\label{eq:ep24}
\rho_{24}=\frac{\Re \bbra V_4^* V_2^2 \kket }{\sqrt{\bbra v_4^2\kket \bbra v_2^4\kket}}
=\bbra \cos4(\Psi_2-\Psi_4)\kket_w\,,
\ee
which is identical to the ATLAS measured event-plane correlator between $\Psi_2$ and $\Psi_4$.
The notation $\bbra \kket_w$, following \Ref{Aad:2014fla}, denotes the measured event-plane 
correlations with the scalar-product method.  The Pearson correlation coefficient has a clear
physical interpretation, quantifying the correlation strength between the \emph{complex} quantities $V_2^2$
and $V_4$. It is thus clear that correlation between magnitudes also contributes to $\rho_{24}$. 

For mixings among more than two 
harmonic orders,
the Pearson correlation coefficient definition differs slightly from event-plane correlations
measured by the scalar product method. For instance, the correlation among $V_2$, $V_3$ and $V_5$ is
\begin{subequations}
\label{eq:ep235}
\begin{align}
&\rho_{235} = \frac{\Re\bbra V_5^* V_2 V_3\kket}{\sqrt{\bbra v_5^2\kket\bbra v_2^2 v_3^2\kket}}\,,
\\
&\bbra \cos(2\Psi_2+3\Psi_3-5\Psi_5)\kket_w=\frac{\Re\bbra V_5^* V_2 V_3\kket}
{\sqrt{\bbra v_5^2\kket\bbra v_2^2 \kket \bbra v_3^2\kket}}\,.
\end{align}
\end{subequations}
Note that the deviation comes from correlation between $v_2$ and $v_3$, 
$\bbra v_2^2 v_3^2\kket\ne \bbra v_2^2 \kket \bbra v_3^2\kket$.
The absolute value of the Pearson correlation coefficient is strictly constrained between 0 and 1,
with 1 corresponding to absolute (anti)-correlation, while 0 indicates the case with no correlation at all.
Event-plane correlators among flow harmonics have been measured at the LHC energy involving
$V_2$, $V_3$, $V_4$, $V_5$ and $V_6$ (see \Fig{fig:epcorrelator}).

With the characterization of flow harmonics in terms of the medium response to 
initial geometrical properties, one
can understand the event-plane correlations in the flow paradigm. 
For the convenience of discussion, we take the correlation between $V_2$ and $V_4$ 
as an example. Recalling that $V_4=\kappa_4 \E_4 + \chi_{422} V_2^2 + \delta_4$, it is not difficult to recognize that
the Pearson coefficient reaches unity (absolute correlation) once  the linear medium response 
$\kappa_4\E_4$ and residual $\delta_4$ in $V_4$ are neglected, namely, 
$V_4$ is linearly dependent on $V_2^2$. On the other hand, if one considers only the
generation of $V_4$ from linear flow response to an initial eccentricity $\E_4$, the resulting
correlation is
\be
\rho_{24}=\frac{\Re \bbra \E_4^* \E_2^2 \kket }{\sqrt{\bbra \ve_4^2\kket \bbra \ve_2^4\kket}}
=\bbra \cos4(\Phi_2-\Phi_4)\kket_w\,.
\ee 
Namely, the event-plane correlation is identical to the participant-plane correlation.
 Since the realistic flow generation interpolates between
linear response and nonlinear mode couplings, the expected event-plane correlation would be
between these two extreme scenarios. 

\Fig{fig:pc24_show} illustrates the expected
event-plane correlation $\rho_{24}$, between the linear and nonlinear dominated extreme scenarios. 
There are two lines corresponding to the case generated 
by linear flow response, considering initial state fourth order eccentricity characterized
in terms of moments or cumulants. Since the relative contribution of linear flow response to
$V_4$ decreases as centrality grows, the correlation $\rho_{24}$ gets stronger towards peripheral
collisions. Similarly, it is understandable that when there are larger dissipations in the medium, 
the linear medium response gets more damped than the nonlinear mode couplings, 
and the induced correlation $\rho_{24}$ should be stronger.
A similar strategy has been generalized to all the observed event-plane correlators from the 
ATLAS collaboration, which leads to successful estimates, except 
$\bbra \cos(2\Psi_2+4\Psi_4-6\Psi_6)\kket$~\cite{Teaney:2013dta}. 

\begin{center}
\includegraphics[width=0.45\textwidth] {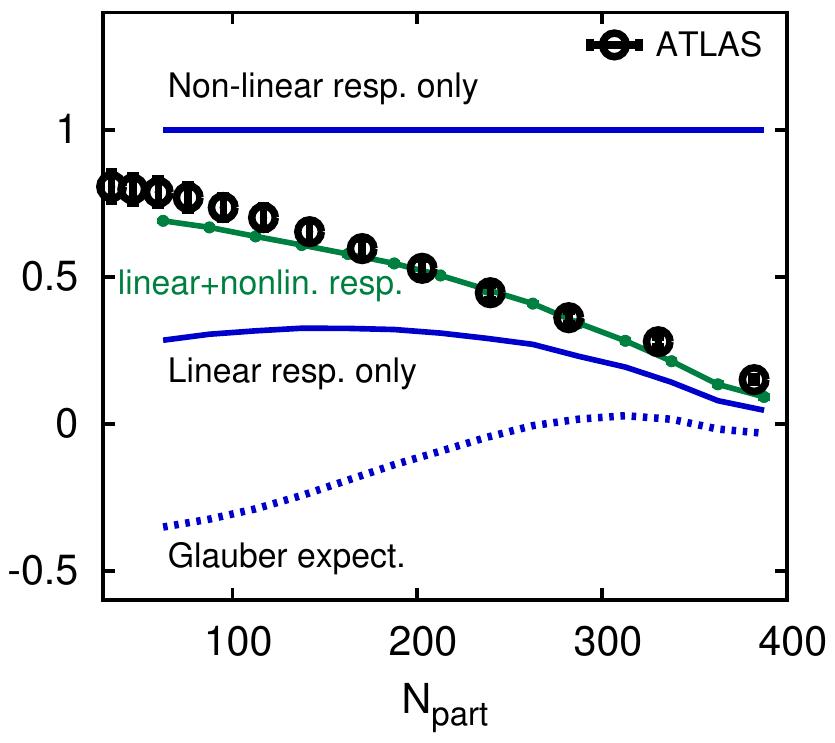}
\figcaption{
\label{fig:pc24_show}
The event-plane correlator $\rho_{24}$ measured by the ATLAS collaboration, with respect to
Pb+Pb at $\sqrt{s_{NN}}=2.76$ TeV. The blue lines are those expected from the medium response scenarios,
when linear or nonlinear part of the quadrangular flow dominates, respectively. Hydro prediction
with $\eta/s=1/4\pi$ interpolating the two scenarios is shown as the green solid line.
Figure reproduced from \Ref{Teaney:2013dta}, with permission.
}
\end{center}

\begin{center}
\includegraphics[width=0.45\textwidth] {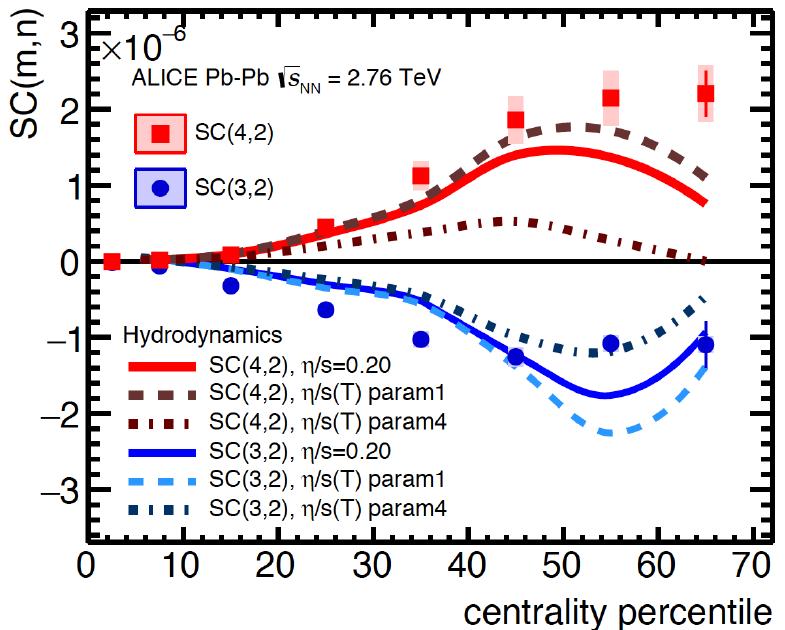}
\figcaption{Symbols correspond to the symmetric cumulants 
$SC(4,2)$ (red) and $SC(3,2)$ (blue) measured by the ALICE
collaboration. Lines of different types are hydrodynamic predictions 
with various parameterizations of $\eta/s$. Figure reproduced from
\Ref{Zhou:2017uuw} (DOI: http://dx.doi.org/10.1016/j.nuclphysa.2017.04.016), under the CC-BY-4.0 license (http://creativecommons.org/licenses/by/4.0/).
\label{fig:sc}
}
\end{center}

It should be noted that the correlation \Eq{eq:ep24} measured in experiments involves
three-particle correlations (in the numerator), and the rms value of $v_4$ and
fourth order moment of elliptic 
flow (in the denominator). The correlation feature of the mixed harmonics is entirely 
captured in the three-particle correlations in the numerator, while the 
denominator in \Eq{eq:ep24} plays a role of normalization. Direct measurements
of the numerator can be carried out in three-particle correlations, which was done
recently by the STAR collaboration~\cite{Adamczyk:2017hdl,Adamczyk:2017byf},
\be
C_{m,n,m+n}\equiv
\Re
\bbra e^{im\phi_p^1+in\phi_p^2-i(m+n)\phi_p^3}\kket
\sim \Re\bbra V_mV_n V_{m+n}^*\kket\,.\cr
\ee
Despite the difference in magnitudes obtained
in the three-particle correlations comparing to the 
event-plane correlators (for obvious reasons), the signs
of the correlations agree in these measurements, as expected.  It is worth mentioning 
that some particular correlations involving dipolar flow $V_1$ get extra contributions 
from the condition of moment conservation, such as $C_{112}$.  It is interesting to notice from
the STAR measurements that a positive $C_{123}$ is observed, which implies that in $V_3$ there is
a nonlinear coupling of modes between $V_1$ and $V_2$.

\subparagraph{Symmetric cumulants $SC(m,n)$}

One difficulty in the measurement of flow is the subtraction of 
non-flow contributions.  
Non-flow effects also present in the event-plane correlators, as expected since there are
 moments involved in the definition of event-plane correlators, e.g. in \Eq{eq:ep24} and
\Eq{eq:ep235}. To subtract non-flow effects, analogous to the cumulants of flow which
suppress non-flow by construction, the correlations
among mixed flow harmonics can be studied in the so-called symmetric cumulants~\cite{Bilandzic:2013kga,ALICE:2016kpq}.
\be
\label{eq:sc}
SC(m,n)=\bbra v_n^2 v_m^2\kket - \bbra v_n^2\kket \bbra v_m^2\kket
\ee
where $m\ne n$ refers to different harmonic orders. As expected in a cumulant, self-correlations are subtracted in the definition, and so are
non-flow contributions.
Apparently, the symmetric
cumulant vanishes once there is no correlation between $v_n$ and $v_m$.
Although the definition of symmetric cumulant relies only on correlations of flow magnitudes, 
symmetric cumulant gets contributions from the correlation of flow phases.
In \Fig{fig:sc}, symmetric cumulants are measured and shown with mixings between
$V_2$ and $V_3$, and $V_2$ and $V_4$. Both of the observed symmetric cumulants
present a trend of 
increasing correlation strength towards large centrality percentiles. 
The correlation between $V_2$ and $V_3$ is negative, while it is positive between $V_2$
and $V_4$, consistent with what was observed in the event-plane correlators.
 The correlation
magnitude and trend are captured by the corresponding
hydrodynamic simulations with specified values of $\eta/s$.

\begin{center}
\includegraphics[width=0.5\textwidth] {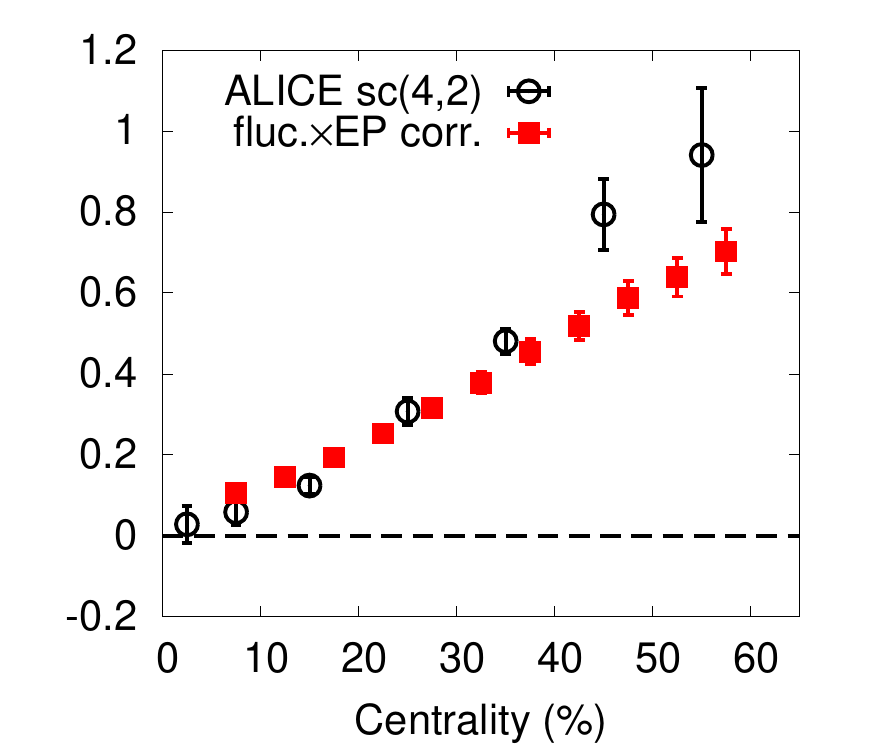}
\figcaption{ A comparison between the normalized symmetric cumulant $sc(4,2)$ 
(open symbols) and the factorization relation \Eq{eq:nsc24} (red points).
Figure reproduced from \Ref{Giacalone:2016afq}, with permission.
\label{fig:sc_factor}
}
\end{center}

Following the same strategy, the observed correlation patterns in 
the symmetric cumulants can be understood in terms of flow response.
Before proceeding, we define the normalized symmetric cumulant~\cite{ALICE:2016kpq},
\be
\label{eq:nsc}
sc(m,n)=\frac{SC(m,n)}{\bbra v_m^2\kket \bbra v_n^2\kket}\,.
\ee
\Eq{eq:nsc} captures the same correlation behavior as in \Eq{eq:sc}, but it is normalized
by taking a ratio with respect to $\bbra v_m^2\kket \bbra v_n^2\kket$. One advantage of 
the normalized symmetric cumulant is, statistical
errors of the measured event-averaged quantities 
cancel out in the ratio to some extent. Again, let us take the correlation between $V_2$ and $V_4$ as
an example. Substituting the response relation of $V_4$ in $sc(2,4)$, one realizes the 
relation
\be
\label{eq:nsc24}
sc(2,4)=\left(\frac{\bbra v_2^6\kket}{\bbra v_2^2\kket\bbra v_2^4\kket}-1\right)\rho_{24}^2\,.
\ee
According to \Eq{eq:nsc24}, the normalized symmetric cumulant is factorized into a factor  that 
records the fluctuation property of $v_2$ (in terms of ratios of moments)
and the event-plane correlator $\rho_{24}$, both of
which are measurables in experiments. Since the underlying assumption behind \Eq{eq:nsc24}
is the flow response relation in the flow paradigm, the validity of \Eq{eq:nsc24} provides an ideal test of
 the flow paradigm we have been discussing so far. 
 
 By taking both the fluctuations of $v_2$ and
 event-plane correlators from experimental results, \Fig{fig:sc_factor} depicts the comparison of 
 the normalized symmetric cumulant $sc(2,4)$ and the factorization relation in \Eq{eq:nsc24}.
 Although the $v_2$ fluctuations and event-plane correlator results 
 are adopted from the ATLAS collaboration, which has different 
 acceptances in transverse momentum $p_T$ and pseudo-rapidity $\eta$ compared to ALICE,
quantitative agreement is achieved in \Fig{fig:sc_factor}. Let us emphasize again that the
comparison made in \Fig{fig:sc_factor} does not involve any hydro simulations with respect
to particular parameterizations, but it manifests the success of the flow paradigm.

\section{Flow paradigm in small colliding systems}
\label{sec:challenge}

In general, small colliding systems, such as p+Pb, are considered in heavy-ion experiments 
as a baseline of the nucleus-nucleus collisions, 
where the generated system is \emph{not} recognized as thermalized. Therefore, techniques of 
perturbative calculations based on QCD dynamics can be applied for theoretical predictions. 
The observed correlation patterns in small colliding systems 
are accordingly expected to be distinguished from those induced by the medium 
collective expansion in nucleus-nucleus collisions.
However, long-range multi-particle correlations were observed in the recent experiments at RHIC 
and the LHC energies, in the very high multiplicity events of the small colliding systems:
p+Pb~\cite{Khachatryan:2015waa,CMS:2012qk,Chatrchyan:2013nka,Sirunyan:2017uyl,Abelev:2014mda,ABELEV:2013wsa,Abelev:2014mda,Aaboud:2017blb,Aaboud:2017acw}, 
d+Au~\cite{Adare:2014keg}, He$^3$+Au~\cite{Adare:2015ctn}
and even p+p~\cite{Khachatryan:2015lva,Khachatryan:2016txc}.
These long-range correlation patterns 
lead to similar measurements of flow harmonics, and fluctuations
and correlations of these flow.
These flow observables
are comparable to the results found in nucleus-nucleus collisions on a general ground. 
Besides, long-range correlations in small
colliding systems can be quantitatively measured.
The measured harmonic flow in small colliding systems 
can be captured by hydrodynamic
simulations~\cite{Bozek:2011if,Bozek:2012gr,Bozek:2015qpa,Bozek:2013uha,Habich:2015rtj,Nagle:2013lja,Bzdak:2013zma,Schenke:2014zha}, 
provided the geometrical information of the initial states 
is incorporated properly.

Based on 
the observed long-range multi-particle correlation patterns and successful applications of hydrodynamics,
it appears tempting to generalize the flow paradigm in the small colliding systems, 
so that the observed flow harmonics can be similarly understood as fluid response to the initial 
state geometrical properties of these systems. 
However, 
there are issues that makes a straightforward generalization questionable. In this section, after 
summarizing some of the experimental observations related to the phenomenon of medium collective
expansion in small colliding systems, in comparison with the results from hydrodynamic simulations and
qualitative (or quantitative) estimate from the idea of the flow paradigm, we discuss some challenges in the theoretical
aspect of applying hydrodynamics in small
colliding systems.

\subsection{Collectivity in small colliding systems}

Application of the flow paradigm to small colliding systems requires 
 medium collective expansion. There has been evidence collected from various measurements
of the flow harmonics in experiments at RHIC and the LHC energies that supports
a medium collective expansion scenario in the small colliding systems. Apart from
experiments, theoretical calculations with viscous hydrodynamics with a proper initialization,
or AMPT model~\cite{Bzdak:2014dia}, also reproduce to a quantitative level 
the flow observables, emphasizing the role of 
late stage evolution in the formation of the observed long-range correlations. With the help
of the medium response relations established in the flow paradigm based on previous
studies of nucleus-nucleus collisions, these observed correlation signatures in small
colliding systems can be understood consistently.

\subsubsection{Experimental evidence}

As we have learned from analyses in nucleus-nucleus collisions, the medium collective 
expansion of a colliding system results in long-range 
correlation in rapidity, which is recognized as the ``ridge'' at $\Delta \phi=0$ and at 
large relative
rapidity in the two-particle correlation function. The non-zero ``ridge'' leads to the measured 
harmonic flow in the Fourier decomposition of  the two-particle correlation function.
\Fig{fig:2pc_pPb} displays the two-particle correlation function measured in p+Pb 
collisions, at $\sqrt{s_{NN}}=5.02$ TeV, by the CMS collaboration~\cite{CMS:2012qk}.
In collision events with low multiplicity productions, as shown in \Fig{fig:2pc_pPb} (a),
there is no such long-range structure observed, consistent with
the traditional understanding of p+Pb collisions. Quite differently,
 the ``ridge'' becomes obvious in the two-particle correlation function in high 
 multiplicity events, shown in \Fig{fig:2pc_pPb} (b).  
 
\begin{center}
\includegraphics[width=0.5\textwidth] {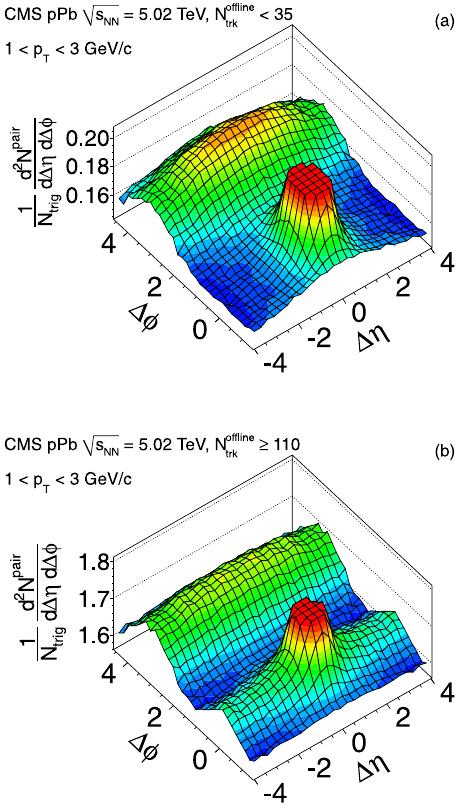}
\figcaption{Two-particle correlation function of the p+Pb colliding system at $\sqrt{s_{NN}}=5.02$ TeV, 
for the (a) low and (b) high multiplicity production events, measured 
by the CMS collaboration. Figure reproduced from \Ref{CMS:2012qk} (DOI: https://doi.org/10.1016/j.physletb.2012.11.025), under the CC-BY-NC-ND license. 
\label{fig:2pc_pPb}
}
\end{center}

Actually, \Fig{fig:2pc_pPb} 
reveals an interesting trend that the observed long-range correlation patterns 
exhibit only in events with sufficiently high multiplicity productions, in the small colliding systems.
Indeed, similar behavior of the long-range azimuthal correlations is 
confirmed in measurements with respect to d+Au~\cite{Adare:2014keg}, He$^3$+Au~\cite{Adare:2015ctn} 
at RHIC, and also p+p collisions
at the LHC energies~\cite{Khachatryan:2016txc,Sirunyan:2017uyl}. 
It is also realized that the fourth power of the four-particle cumulant of elliptic flow  $v_2\{4\}^4$ 
from p+Pb collisions
changes sign as multiplicity 
decreases~\cite{Chatrchyan:2013nka,Khachatryan:2016txc,Sirunyan:2017uyl}.
The role of high multiplicity productions
in the flow paradigm implies high energy/entropy 
density of the created medium, which is necessary for 
the collective evolution according to fluid dynamics. 
Observing the long-range
correlation patterns in high multiplicity events is already an indication that
the system evolution in small colliding systems is compatible to what one would expect 
in a flow paradigm. 

Along the same line of analysis, two-particle correlation functions can be decomposed into
Fourier harmonics to extract harmonic flow, which gives the flow of two-particle
cumulant $v_n\{2\}$. To further investigate and identify the effect of medium collective expansion, flow
harmonics are expected as well from multi-particle correlations.  The measured cumulant of
elliptic flow in the p+Pb system can be found in the right-hand panel of \Fig{fig:v2_cumulant},
by the CMS collaboration~\cite{Khachatryan:2015waa}. 
It is evident from the flow cumulants that the measured elliptic anisotropy
presents, in up to eight-particle correlations, 
which strongly supports the idea of medium collectivity. 

In addition to these \emph{qualitative} agreements with the medium collective expansion, 
there is also evidence that is \emph{quantitatively} consistent with the flow paradigm estimate. 
Compared to the flow multi-particle cumulants in Pb+Pb collisions,
one observes similar hierarchy ordering from $v_2\{2\}$ to higher order 
cumulants as in \Eq{eq:vnorder}, $v_2\{2\}>v_2\{4\}\gtrsim v_2\{6\}\gtrsim v_2\{8\}$,
which implies similar non-Gaussian properties of flow fluctuations.
Following the analysis in the flow paradigm, we understand 
the ordering in flow cumulants as an indication of non-Gaussianity, especially, the 
non-Gaussianity in the initial state ellipticity fluctuations, upon a linear fluid response.
In a p+Pb colliding system, the background of the initial state geometry is dominated by the
shape of the proton, which is approximately spherical. The
elliptic eccentricity in p+Pb is purely fluctuation-driven, and its 
event-by-event fluctuations can be described by a power 
distribution function, \Eq{eq:2d_power}. 

The power distribution allows one to relate the ratio between
flow cumulants analytically. \Fig{fig:pPb_cumu} shows the measured ratios of the cumulants
of the elliptic flow in p+Pb collisions and Pb+Pb collisions 
by the CMS collaboration~\cite{Khachatryan:2015waa}. 
Solid lines correspond to the expectations from the power 
distribution, which are compatible with the measured data of p+Pb collisions within
systematic and statistical uncertainties.
This comparison provides quantitative evidence for the collectivity expansion
in the high multiplicity events of p+Pb collisions.

Recently, similar analyses with respect to the flow cumulant 
were carried out by the ATLAS collaboration~\cite{Aaboud:2017blb,Aaboud:2017acw},
and extended to the high multiplicity events in p+p collisions. 
Given the observed hierarchy of the flow
cumulants, the analytical description using the power distribution function determines an
effective number of independent sources. It is interesting to notice that
the extracted number of independent sources by the ATLAS collaboration are compatible for 
p+p and p+Pb collisions, provided the multiplicity productions in these systems are comparable.
It is worth mentioning that this observation
is consistent with some theoretical analyses based on hydrodynamics, which suggest that 
the system multiplicity plays a dominant role in collective evolution, 
rather than the system size~\cite{Chesler:2016ceu,Basar:2013hea,Yan:2015lfa}.

\begin{center}
\includegraphics[width=0.45\textwidth] {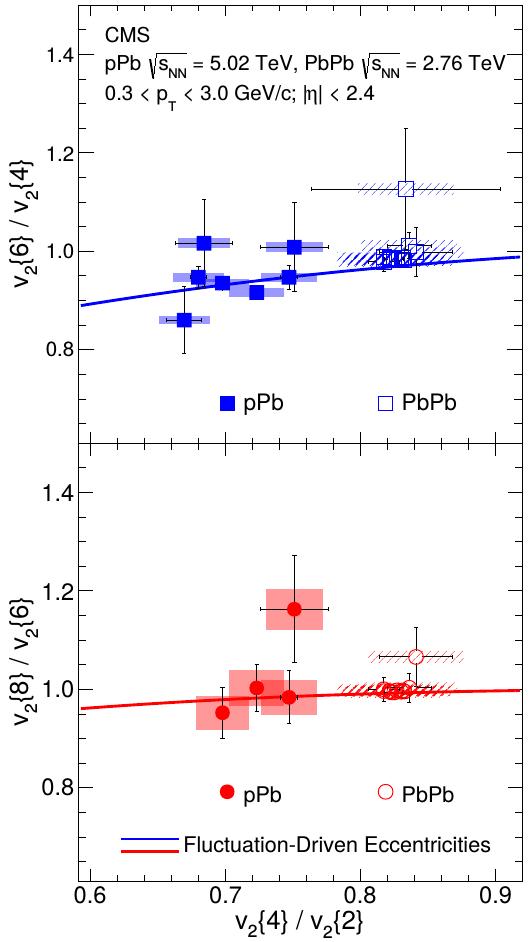}
\figcaption{
\label{fig:pPb_cumu}
Relations betweeen cumulant ratios measured by the CMS collaboration, for the
p+Pb collisions at $\sqrt{s_{NN}}=5.02$ TeV and Pb+Pb collisions at
$\sqrt{s_{NN}}=2.76$ TeV. Solid lines are expected analytical relations 
in the flow paradigm, in terms of the power distribution. Figure reproduced 
from \Ref{Khachatryan:2015waa} (DOI: https://doi.org/10.1103/PhysRevLett.115.012301, ``Evidence for Collective Multiparticle Correlations in p-Pb Collisions''), under the CC-BY 3.0 license.
}
\end{center}

In addition to the flow cumulants of $v_2$, there are other signatures related to flow harmonics that
are captured in the flow paradigm. For instance, the differential spectrum of elliptic flow of
identified particles in the small colliding systems present mass 
ordering~\cite{ABELEV:2013wsa,Adare:2014keg}, 
i.e., the anisotropic momentum spectra of heavier particles get pushed towards larger 
transverse momentum owing to the radial flow~\cite{Shen:2011eg},
an interesting feature anticipated 
due to the conversion to particles from fluid excitations,
although it should be noted that mass ordering is \emph{not} an ad hoc
feature characterized in hydro modelings~\cite{Schenke:2016lrs}.
Besides, the measured flow observables
in small systems also include triangular flow $v_3$~\cite{Adare:2015ctn,Aaboud:2017acw}, quadrangular
flow $v_4$~\cite{Aaboud:2017acw}, and correlations of these flow observables in terms of 
symmetric cumulants~\cite{Sirunyan:2017uyl}.
These observables are compatible with the flow paradigm expectations as well.
For instance, it is noticeable that 
in the small systems, that
the hierarchy of anisotropic flow exists from lower to higher order harmonic 
orders. The sign of the measured symmetric cumulant, e.g.,
$SC(2,3)<0$, agrees with the expected correlation pattern induced by
initial state geometry~\cite{Yan:2015fva}.

\subsubsection{Results from hydrodynamic simulations}

When solving viscous hydrodynamics with respect to small colliding systems, the essential 
adjustments 
are in
the effective descriptions of initial state, compared to the hydro simulations with
respect to nucleus-nucleus collisions. Especially, one should be cautious about potential
contributions from effects of small scales, e.g., sub-nucleon structures.
For high multiplicity events in the light-heavy colliding systems, the background
geometry of the created system is to a large extent 
determined by the configuration of the light nucleus, such as 
a proton, deuteron, or He$^3$. Additionally, the resulting initial state density profile
after nucleon-nucleon collisions is affected 
by details of energy deposition. 
Taking these factors into account on various grounds makes hydro
predictions strongly model-dependent. For instance, it was realized that 
the IP-Glasma realization without intrinsic proton shape
deformation under-predicts the observed elliptic flow in p+Pb collisions~\cite{Schenke:2014zha}.

\begin{center}
\includegraphics[width=0.45\textwidth] {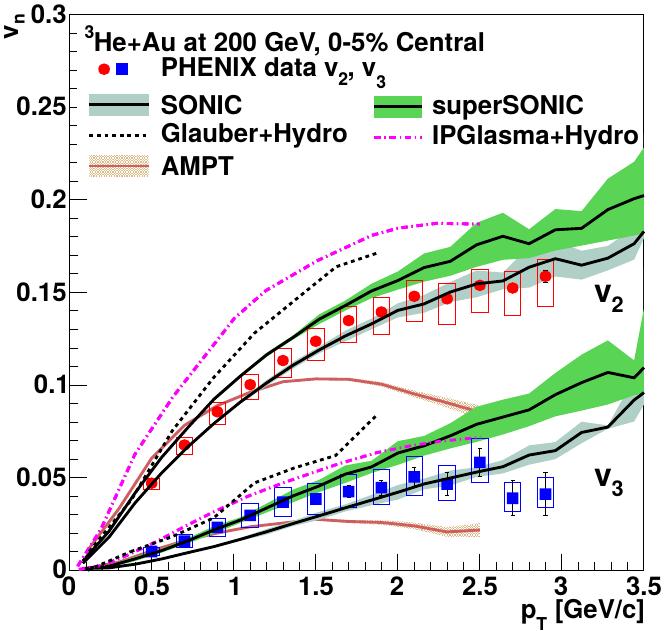}
\figcaption{Symbols: Elliptic flow $v_2$ and triangular flow $v_3$ measured
in He$^3$+Au collisions at $\sqrt{s_{NN}}=200$ MeV, by the PHENIX 
collaboration. Lines: Results from hydrodynamic simulations and AMPT. 
Figure reproduced from \Ref{Adare:2015ctn}, with permission.
\label{fig:vnhe}
}
\end{center}

Nevertheless, an overall agreement of predictions from hydro modelings of small colliding systems
and experimentally measured flow harmonics is achieved.

As an example, 
\Fig{fig:vnhe} presents the 
measured elliptic flow and triangular flow in the He$^3$+Au collisions, at the RHIC energy
$\sqrt{s_{NN}}=200$ MeV~\cite{Adare:2015ctn}. Note that there is an intrinsic triangular asymmetry of the colliding 
system, determined by the configuration of He$^3$. 
Model calculations based on viscous hydrodynamics, or transport model~\cite{Bzdak:2014dia},
are shown as lines of different types. For the He$^3$ system, intrinsic geometry is dominated by the
the three nucleons, not the detailed geometrical configuration of the proton. 
Thereby, with the effective modelings of initial state by the MC-Glauber model~\cite{Bozek:2015qpa}
or IP-Glasma~\cite{Schenke:2014gaa}, hydrodynamic simulations with $\eta/s$ equal or close to 
$1/4\pi$ lead to compatible results of $v_2$ and $v_3$, although the results over-predict 
in comparison with experiment. The over-predictions can be remedied in the hydro 
calculations with larger dissipative corrections. 
In an alternative approach~\cite{Nagle:2013lja}, hydro results from Glauber+SONIC 
 coupled to hadron transport
calculations, which effectively introduces extra sources of medium dissipations,
do lead to reasonable predictions, in spite of the fact that the initial state is modeled
with a similar MC Glauber. When fluid velocity in the initial state is assumed, 
the flow harmonics from hydro predictions are enhanced,
as seen in the results from superSONIC simulations~\cite{Romatschke:2015gxa} in \Fig{fig:vnhe}.

To further examine the response relations in a flow paradigm,  we look at the 
simulated results from Glauber+SONIC simulations~\cite{Nagle:2013lja} for different
colliding systems. Shown in \Fig{fig:pA_v2e2} is a scatter plot of the ratio of elliptic flow magnitude
at $p_T=1$ GeV
to the initial eccentricity, $v_2/\ve_2$, as a function of $\ve_2$, from p+Au (black),
d+Au (red) and He$^3$+Au (blue) collision events. Freeze-out temperatures are used to
control the life-time of hydro evolution during system expansion. The left-hand figure in \Fig{fig:pA_v2e2}
has a $T_{fo}=170$ MeV, while the right-hand one is obtained with $T_{fo}=150$ MeV.
Each point in the figure corresponds to one single event. A constant value in the plots indicates  a
linear flow response relation between $v_2$ and $\ve_2$, which is observed in the three colliding
systems, except in d+Au with a sufficiently large $\ve_2$. The breaking of linear response relation in d+Au
is understood as follows: at very large $\ve_2$, the initial density profile is composited by two
separated hotspots, which hardly merge during expansion at later stages of the medium evolution,
leading to a very small value of $v_2$. Apart from the discrepancy in d+Au of large $\ve_2$, the 
linear response relations of elliptic flow are valid in the hydro simulations of small colliding systems.

\subsection{Challenges of the flow paradigm}

This section is devoted to a 
discussion on challenges in the flow paradigm, from the \emph{theoretical} aspect.
The discussion is on
a more general ground with respect to all the colliding systems 
in heavy-ion experiments, albeit the situation is obviously more serious 
in small colliding systems. 
It should be emphasized that, so far, there have not been strong violations of the
expected features from the flow paradigm observed in \emph{experiments}, 
from large colliding systems
of Au+Au, Pb+Pb, to the recent measurements in small systems, p+Pb, d+Au and He$^3$+Au. 
The discussion will not be conclusive, but give a somewhat brief description of the recent
developments in the theory of hydrodynamics beyond local thermal equilibrium.

The flow paradigm is quite well-established in  large colliding systems, where
the observed long-range multi-particle correlation patterns are understandable as 
a consequence of medium response to initial state geometry. 
In small colliding systems, although the system size decreases dramatically (roughly by a factor
of 10),
hydro modelings provide reasonable characterizations
of the system collective expansion, with predictions capturing a wide spectrum of the observed flow harmonics.
The success of the flow paradigm based on hydrodynamics 
raises a question : what is the limit of applying 
fluid dynamics in heavy-ion collisions? Or alternatively, what is the smallest
droplet of fluid system generated in heavy-ion collisions?

Although it is natural to implement viscous hydrodynamics
for the description of medium collective expansion in the flow paradigm, 
the picture of 
medium response relies on the dominance of hydrodynamic modes, over non-hydro
modes from many origins. The applicability condition of
hydrodynamics must be satisfied to ensure hydro mode evolution. 
There are two crucial issues one needs to take into
account regarding the applicability of viscous hydrodynamics in heavy-ion collisions:
1) the thermalization of the created quark-gluon system; and 2) the validity of 
truncation of the gradient
expansion in viscous hydrodynamics. 

\end{multicols}
\begin{center}
\includegraphics[width=0.95\textwidth] {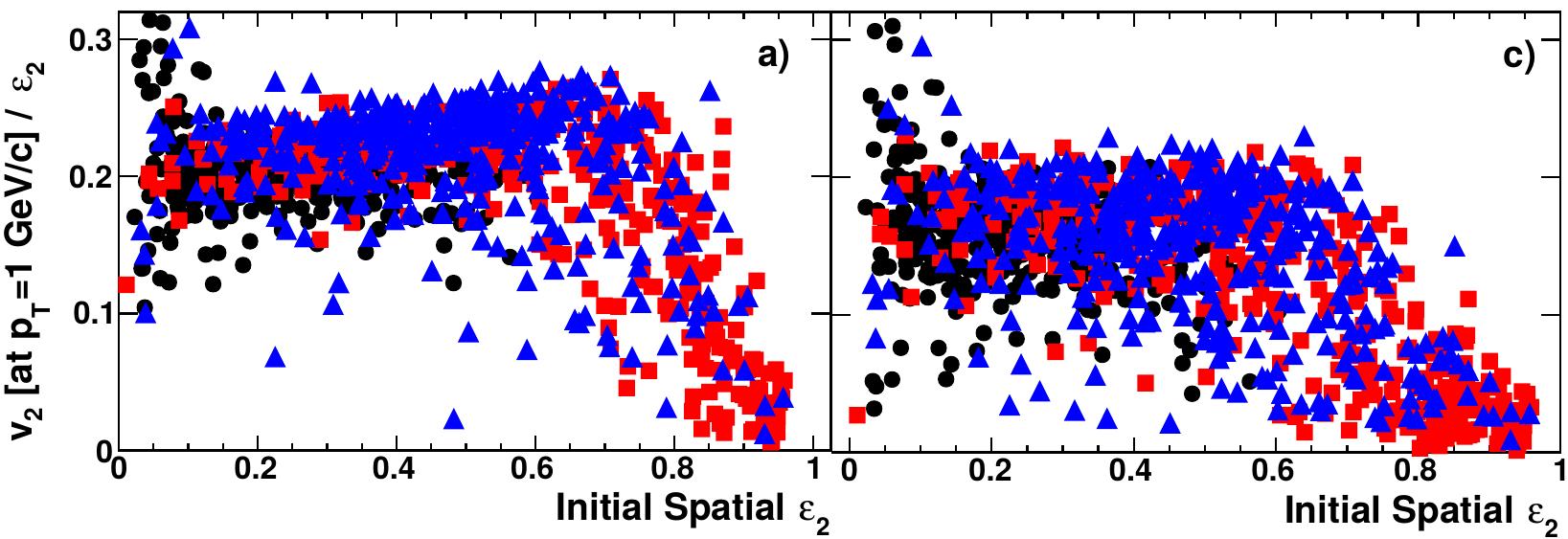}
\figcaption{Scatter plot of $v_2/\ve_2$ at $p_T=1$ GeV from Glauber+SONIC hydro simulations 
with respect to p+Au (black symbols), d+Au (red symbols) and He$^3$+Au (blue symbols)
collisions. Different freeze-out temperatures, $T_{fo}=150$ MeV (left panel) and
$T_{fo}=170$ MeV (right panel), are applied to control the life-time of hydro
evolution of the medium collective expansion. Figure reproduced from \Ref{Nagle:2013lja}, with permission.
\label{fig:pA_v2e2}
}
\end{center}
\begin{multicols}{2}

\subsubsection{Thermalization of QGP}
\label{sec:thermalization}

Thermalization is a necessary condition for viscous hydrodynamics, so that hydro variables
are well-defined in theory. However, to realize a thermalized medium in heavy-ion collisions
in a short time scale
is challenging for the weakly-coupled QCD dynamics, 
even in large colliding systems (cf \Ref{Fukushima:2016xgg} for
a recent review).
In particular, considering
the required starting time for hydro modelings determined by phenomenological analyses, 
namely, the required thermalization time, being around $O(1)$ fm/c,
perturbative QCD gives a much longer
estimate of time scale
$\tau\gtrsim 1.5\alpha_s^{-13/5}Q_s^{-1}$~\cite{Baier:2000sb}
\footnote{Even with the strong coupling constant taken at a relatively large value, 
$\alpha_s\sim 0.3$, the saturation scale $Q_s$ 
expected in nucleus-nucleus collisions cannot result in a 
thermalization time scale comparable to $O(1)$ fm/c. The expected time scale from perturbative
QCD  is
even longer in small systems, as $Q_s$ is smaller.
}.

Using the condition of the onset of hydrodynamics rather than local thermal equilibrium,
the discrepancy in the time scale of thermalization can be partly remedied. 
The onset of hydrodynamics, sometime known 
as \emph{hydrodynamization}, refers to a state of the medium at which a system starts to evolve 
hydrodynamically. By solving kinetic theory for a weakly-coupled medium, for the pre-equilibrium 
evolution in heavy-ion
experiments, it is indeed confirmed that \emph{hydrodynamization} does not require
local thermal equilibrium~\cite{Kurkela:2015qoa}, with deviations from thermal equilibrium correspond to viscous
corrections~\cite{Blaizot:2017lht}. Actually, in the result of the Boltzmann equation, 
\emph{hydrodynamization} is achieved earlier than isotropization~\cite{Arnold:2004ti},
at which the longitudinal pressure $\P_L$ and transverse pressure $\P_T$,
\begin{align}
\label{eq:plpt}
\P_L=&\int \frac{d^3p}{(2\pi)^3p^0} p_z^2 f(t, \vec x,\vec p)\,,\cr
\P_T=&\int \frac{d^3p}{2(2\pi)^3p^0} (p_x^2+p_y^2) f(t, \vec x,\vec p)\,,
\end{align}
become comparable. The condition of $\P_L\approx \P_T$ is known
as \emph{isotropization}. In \Eq{eq:plpt}, $f(t,\vec x, \vec p)$ is the phase space distribution
of the out-of-equilibrium system. One may check that, for the Bjorken flow, the
pressure difference is related to viscous corrections in hydrodynamics, 
\be
\P_L-\P_T=-2\eta/\tau+O(1/\tau^2)\,,
\ee
which gives an explicit example showing the relation between viscous corrections in hydrodynamics
and effects of out-of-equilibrium dynamics.
For strongly coupled systems, numerical simulations have been developed to mimic 
colliding systems of different system sizes, 
based on gauge/gravity duality techniques (cf. \Ref{Chesler:2008hg,Chesler:2015wra}).
A short time scale of \emph{hydrodynamization} that is comparable to the estimate in the 
flow paradigm is realized. 

\begin{center}
\includegraphics[width=0.45\textwidth] {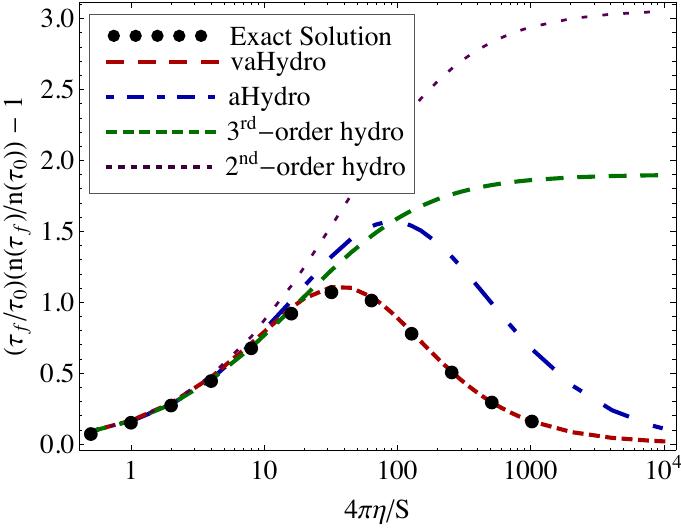}
\figcaption{ Convergence of ahydro and vahydro in comparison with the exact solution 
of Boltznmann equation with relaxation time approximation. The parameter $\eta/s$ can
be understood as a quantity 
controlling deviations of the system from local thermal equilibrium.
Figure from
\Ref{Bazow:2013ifa}, with permission.
\label{fig:ahydro}
}
\end{center}

Despite the progress made in the analyses of thermalization or  the
onset of hydrodynamics, by solving out-of-equilibrium dynamical evolutions, an alternative 
is to amend the hydrodynamic framework to include out-of-equilibrium effects.
Such a strategy leads to the development of anisotropic hydrodynamics (ahydro)~\cite{Martinez:2010sc,Florkowski:2010cf}.

In anisotropic hydrodynamics, 
following the standard derivation of 
hydrodynamics from kinetic theory, the gradient expansion is made  on top of an anisotropic background.
For instance, for a system with Bjorken boost invariance, 
the Romatschke-Strickland distribution function~\cite{Romatschke:2003ms}
can be used to effectively describe an anisotropic background,
\be
\label{eq:ahydro_dis}
f(t,\vec x, \vec p) = f_{\mbox{\tiny iso}}
\left(\sqrt{\vec p^2+\xi(\tau)p_z^2}/\Lambda(\tau)\right)\,,
\ee
with $\xi$ characterizing momentum anisotropy, and $\Lambda(\tau)$ the energy scale.
By doing so, out-of-equilibrium effects of the system are to some extent absorbed into the  
anisotropic background. Similar gradient expansion including higher viscous corrections can be 
applied iteratively, leading to 
the viscous version of anisotropic hydrodynamics (vahydro)~\cite{Bazow:2013ifa}.
Therefore, one expects improvements in the solution of out-of-equilibrium
evolution, compared to the traditionally derived viscous hydrodynamics. This is demonstrated in
\Fig{fig:ahydro}, where the produced entropy density of a Bjorken boost invariant system
is calculated in viscous hydrodynamics, ahydro, vahydro and the corresponding 
Boltzmann equation with relaxation time approximation.
The relaxation time, which characterizes how fast the system relaxes towards equilibrium,
 is taken inversely proportional to the local temperature, $\tau_{rel}\propto \eta/sT$. As 
 a result, $\eta/s$ effectively controls the deviations of the system from local equilibrium.
Ahydro and vahydro do have a better convergence behavior towards the exact solution from the 
Boltzmann equation, even though the system is sufficiently far from local thermal
equilibrium, with $\eta/s\sim 10^4$. One may refer to \Ref{Strickland:2014pga}
for a detailed review of ahydro.

\subsubsection{Gradient expansion in viscous hydrodynamics}
Although ahydro (or vahydro) extends the framework of hydrodynamics to
out-of-equilibrium systems, which accordingly helps to relieve the tension between
out-of-equilibrium system evolution and  the flow paradigm, 
 it does not conceptually provide
an answer to the question raised in this section. Especially, 
the present success of the flow paradigm
relies on the application of second order viscous hydrodynamics, as a truncated gradient expansion
at the second order 
with respect to an isotropic background. 

In the canonical formulation of viscous hydrodynamics, 
the gradient expansion corresponds to viscous corrections to the system evolution order-by-order.
It requires a convergence condition of the gradient expansion, 
so that viscous hydrodynamics can be applied to a system in when higher order viscous corrections
are subdominant. For the practical formulation of hydrodynamics, truncation of the gradient expansion
is commonly taken at second order to avoid acausal mode evolution. 
For a relativistic fluid, the convergence condition of the gradient expansion is reflected by the smallness of the
Knudsen number Kn, (recall the definition of the Knudsen number as 
a ratio between a microscopic scale, e.g., mean-free path
$l_{\mbox{\tiny mft}}$,
and a macroscopic scale, e.g., system size $L$, Kn$\sim l_{\mbox{\tiny mft}}/L$). As a result, when the 
system size
gets smaller and smaller, as in the cases of small colliding systems in heavy-ion collisions,
 the convergence of gradient expansion is more likely to be violated. It is indeed found in 
 realistic hydro simulations, that the Knudsen number in small colliding systems is
 larger than that in nucleus-nucleus collisions~\cite{Niemi:2014wta}.

A large Knudsen number in small colliding systems implies the significance of higher order terms in the gradient 
expansion or higher order viscous corrections, in  the application of hydrodynamics. 
Besides, one also notices that the inclusion of higher order viscous corrections 
extends the applicability of hydrodynamics in an out-of-equilibrium system, owing to
the correspondence between effects of out-of-equilibrium dynamics and viscous corrections.
However, to include higher order viscous corrections in a theoretical framework of
hydrodynamics is complicated, due to fact that the gradient expansion leading to viscous 
hydrodynamics is asymptotic (zero radius of convergence) rather 
than convergent~\cite{groot1980relativistic}. 

The divergence of the gradient expansion was recently explored in the context of 
Bjorken flow~\cite{Heller:2016rtz,Denicol:2016bjh}, where analysis is simplified as
a consequence of symmetry conditions. For instance, the expansion rate of the Bjorken
flow is solely determined by the proper time, $\nabla\cdot u=1/\tau$. Accordingly
the measure of out-of-equilibrium 
can be chosen to be the dimensionless
quantity, 
\be
w=\tau T.
\ee
so that the gradient expansion of hydrodynamic variables in hydrodynamics 
is written in terms of $1/w$. One may check that in the Bjorken flow, the Knudsen number
Kn$\sim 1/w$.
In particular, the function $f(w)\equiv  \tau \partial_\tau \ln w=1+\frac{\tau}{4}\partial_\tau\ln \epsilon$, that 
characterizes the decay rate of the local energy density in a system experiencing Bjorken expansion, 
is expanded as~\cite{Heller:2015dha}
\be
\label{eq:gra_f}
f(w)
=\sum_n f_n w^{-n}\,.
\ee
For the function $f(w)$, it is clear that the gradient expansion is divergent, since the 
constant coefficients 
$f_n$ exhibits a facotrial growth at large $n$. Nevertheless, the expansion is Borel resummable. 
To do so, the first step is to take a Borel transformation,
leading to the Borel transform of $f(w)$,
\be
f_B(\xi) = \sum_n \frac{f_n}{n!} \xi^{n}\,.
\ee
An analytical continuation should be applied to $f_B(\xi)\rightarrow \tilde f_B(\xi)$ to help locate the
singularities of the function, before the inverse Borel transformation giving rise to the resum of
a generalized gradient expansion,
\be
f_R(w) = w\int_C d\xi e^{-w\xi} \tilde f_B(\xi)\,.
\ee
The integration follows a contour $C$ between the origin and infinity on the complex plane of $\xi$.
Corresponding to the fluid dynamics derived using gauge-fluid duality, a set of singular poles can be 
identified~\cite{Heller:2013fn}. 
In addition to the part that is analytic, the singular poles result in exponential decay modes 
in the inverse transform $f_R$. The lowest order one is 
\be
\label{eq:non-hydro}
\delta f_R \sim w^{-\gamma} e^{-w\xi_0}\,,
\ee 
where $\gamma$ and $\xi_0$ are constant obtained with respect to the singularities
of $\tilde f_B(\xi)$.

\begin{center}
\includegraphics[width=0.45\textwidth] {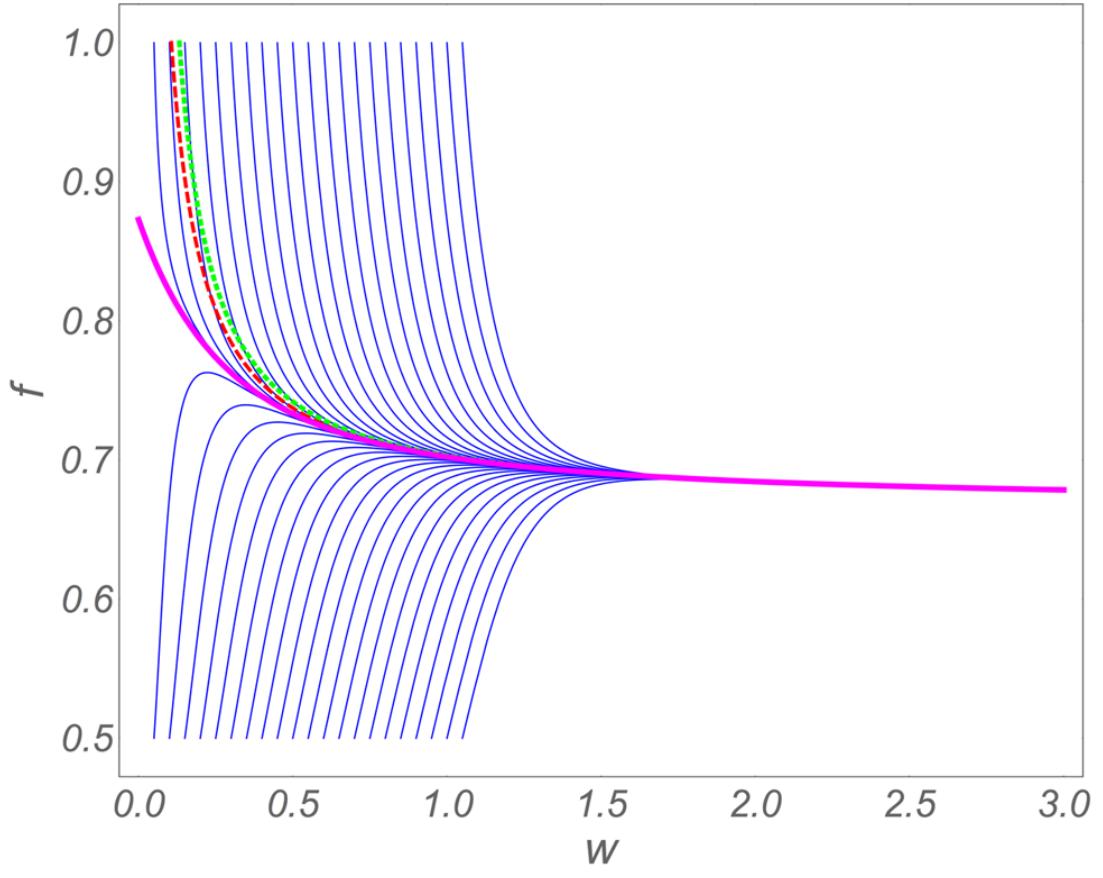}
\figcaption{Solutions of MIS hydro in the Bjorken expansion, with respect to various initial
conditions (blue lines). The thick magenta line is the numerically determined hydro attractor.
Hydro expectations from first and second viscous hydro are shown as the red dashed and green 
dotted lines. Figure reproduced from \Ref{Heller:2015dha}, with permission.
\label{fig:hydro_att}
}
\end{center}
 
In total, the resummation procedure for $f(w)$ leads to a trans-series consisting of
polynomials of $1/w$ and the exponential decay parts, instead of the normal gradient
expansion.  The physical interpretation of the trans-series is to identify the polynomial of $1/w$ as
the corresponding resummation of hydro modes, while non-hydro modes are 
recognized as short-lived decay modes~\cite{Heller:2013fn}.
With respect to the fluid dynamics derived from gauge-gravity duality, non-hydro modes are found
to coincide with quasi-normal modes in gravitational fluctuations~\cite{Heller:2013fn,Kovtun:2005ev}. 
The appearance of the exponential decay parts 
in the Borel resummation is contour dependent, but there exists an unambiguous and physically sensible 
result in the trans-series. It is a resumed result where the exponential decay parts cancel consistently, which, as will become
clear later, corresponds to the hydrodynamic attractor solution, a property 
related to the theory of \emph{resurgence} in quantum theories 
(cf. \Ref{Dunne:2012ae,Dunne:2015eaa,Dorigoni:2014hea,Buchel:2016cbj}).

The above analysis employs only the formal behavior of the gradient expansion of the hydro function 
$f(w)$. The obtained structure can be validated in practical hydro calculations. For instance, the 
Muller-Israel-Stewart (MIS) hydrodynamics has the following equations of motion regarding Bjorken 
flow,
\begin{align}
&\tau \partial\epsilon=-\frac{4}{3}\epsilon + \Phi\,,\cr
&\tau_\pi \partial\Phi=\frac{4\eta}{3\tau}-\frac{4\tau_\pi\Phi}{3\tau} - \frac{\lambda_1\Phi^2}{2\eta}-\Phi\,,
\end{align}
where $\Phi=-\pi^\xi_\xi$,  and $\tau_\pi$, $\lambda$ are second order transport 
coefficients. The MIS hydro equations can be written as a nonlinear differential equation in terms of $f(w)$,
with solutions obtained accordingly. 

In \Fig{fig:hydro_att}, the numerical solutions to MIS hydro, with respect to various initial conditions, 
are displayed in thin blue lines. As time evolves ($w$ increases), the hydro solutions tend to collapse
towards an attractor solution.
The attractor solution associated with the nonlinear differential equation can be identified
numerically, or analytically through the slow-roll approximation. The attractor solution is numerically solved
and shown as the magenta line in \Fig{fig:hydro_att}. Starting from different initial conditions, one indeed 
observe a trend of exponential decay from the random 
hydro solutions towards the hydro attractor. Actually,
one can even determine that to the linearized order, the decay mode of the MIS hydrodynamics is consistent with
that expected in the generalized Borel resummation, \Eq{eq:non-hydro}.
Note that a large $w$ indicates the system is close to local thermal equilibrium, which explains the agreements
of the numerical solution with the analytical expectations from first and second order viscous hydro, in
\Fig{fig:hydro_att}.

\Fig{fig:hydro_att} is a good demonstration of the existence of the hydro attractor, and the non-hydro mode
decay. The hydro attractor effectively extends  the description of hydrodynamics for out-of-equilibrium systems, 
regarding
the small $w$ region in \Fig{fig:hydro_att}. Note that at the initial state $\tau_0=0.5$ fm/c 
in the Au+Au collisions at RHIC, 
$w$ can be as small as $0.5$. For a system evolution even out-of-equilibrium, the hydro attractor 
captures contributions from the summed hydro modes, while
non-hydro modes are short-lived, and decay exponentially towards the hydro attractor.
The existence of attractor solutions can be used to
explain why (second order viscous) hydrodynamics provides such a remarkable 
description in heavy-ion collisions, from large to small colliding systems, despite the apparent
challenges from thermalization and out-of-equilibrium influences.

The existence of hydro attractors appears promising regarding the extension of
hydro modeling in heavy-ion collisions, and hence for a generalized application of
the flow paradigm, to an out-of-equilibrium system. 
It offers a possible answer to the questions raised at the beginning of this 
section, that the application of hydrodynamics is extended to cases when
hydrodynamic attractor solution dominates.
Nonetheless, so far, 
the analyses of out-of-equilibrium system evolution, and the incorporation
of hydrodynamics in the out-of-equilibrium system with an attractor solution
are carried out in the context of Bjorken expansion with respect to conformal
symmetry~\cite{Strickland:2017kux,Denicol:2017lxn}. 
Realistic systems in heavy-ion collisions are more complicated with
much weaker symmetry constraints~\cite{Romatschke:2017acs}, especially in the case of
QCD dynamics, in which the role of hydrodynamic attractors has not yet been clarified. 

\section{Summary } 
\label{sec:sum}

Based on the success of hydro modelings of heavy-ion collisions, the flow paradigm
is established to analyze the observed of harmonic flow $V_n$. In the flow paradigm,
various properties of the harmonic flow can be interpreted as a consequence 
of the medium collective expansion regarding the initial state geometrical information. 

Although
the flow paradigm relies heavily on a hydro description of the medium collective expansion,
it captures the conceptual ideas of medium collective evolution. Especially,
the physics picture constructed in the flow paradigm helps to reduce the influence
of effective parameterizations
in model simulations. 
In the flow paradigm, fluid dynamics is often employed to take into account the 
effect of medium collective expansion, 
and particularly fluid dynamics captures the 
hydro mode mode evolution associated with the decomposed modes according to
the azimuthal symmetry. The decomposed modes in harmonics, 
with respect to the initial state geometrical fluctuations, are characterized as 
the initial state eccentricities $\E_n$'s. These eccentricities $\E_n$'s are responsible
for the observed anisotropic 
momentum spectrum in experiments, and hence the harmonic flow, upon medium response
relations. These response relations  (\Eqs{eq:vn_form})
proposed in the flow paradigm, have been examined to a quantitative level,
in event-by-event hydrodynamic simulations. Some of the experimentally measured 
signatures, such as the factorization relations among different types of symmetric cumulants,
and the relative scales of the measured nonlinear medium response coefficients $\chi$'s,
are found to be consistent within the medium response framework to a quantitative level. 

In the flow paradigm, given these medium response relations,
one is allowed to disentangle the physics information of initial state fluctuations,
from the dynamics of medium evolution, from the measured flow harmonics. 
This is of great significance because it provides potentially model-independent analyses
in heavy-ion collisions, with emphases laid directly on the properties of the
initial state and medium transport.
We have presented in this review, as an example, that
the event-by-event fluctuations of elliptic 
flow $v_2$ reveal the fluctuation behavior of initial ellipticity $\ve_2$. Through the
fit of the probability distribution function via an elliptic-power function, 
the linear medium response coefficient $\kappa_2$
is approachable, together with the parameters associatd with initial state. 
In addition to the flow fluctuations,  another way of extracting the medium 
dynamical properties comes from the analyses of flow correlations. This is again, a model-independent procedure. 
In the flow 
correlations involving higher order harmonics, proper cancellations in the ratio of different types 
of the harmonic flow lead to the measurements of medium response 
coefficients, $\chi$'s.

Although the flow paradigm is mostly established in 
 high energy nucleus-nucleus collisions, its generalization to small 
colliding systems appears straightforward. 
In the recent experiments involving small colliding systems, the scenario of 
medium collective expansion gets support from the similar results of long-range multi-particle
correlations. Quantitatively, these correlations are compatible with hydrodynamics
in terms of the predicted harmonic flow, of various types. For instance, the elliptic and triangular
flow from two-particle correlations in He$^3$Au, dAu and even proton+proton,
can be well reproduced by viscous hydrodynamics. 
In particular, the fluctuations of $v_2$ in proton-lead collisions, 
which exhibit a consistent pattern with the power distribution function, 
have not only  been used as the most convincing evidence of the medium collective expansion,
but also provide information of the fluctuating initial state.
 
The success of the flow paradigm depends on the applications of viscous
hydrodynamics in the quark-gluon systems created in heavy-ion collisions. 
However, the application hydrodynamics in small colliding systems, where one requires 
the thermalization of quarks and gluons to be approached in an extremely short time
scale, is questionable. It then motivates the extension of the 
flow paradigm, or to say, the extension of the theoretical
framework of viscous hydrodynamics, to systems which are out of local thermal
equilibrium. In addition to the theoretical studies of the thermalization process
in weakly-coupled or strongly-coupled systems, one promising progress is 
the discovery of hydrodynamic attractors, although the present
investigations of hydrodynamic attractors are mostly limited to cases with
stong symmetry conditions.

\acknowledgments
{
The author is grateful to Jean-Yves Ollitrault for carefully reading the manuscript and 
very valuable comments. 
This work is supported in part by the Natural Sciences and Engineering Research Council of Canada.}

\end{multicols}

\vspace{10mm}

\begin{multicols}{2}

\end{multicols}

\vspace{-1mm}
\centerline{\rule{80mm}{0.1pt}}
\vspace{2mm}

\begin{multicols}{2}
\bibliographystyle{unsrt}    
\bibliography{refsRev}

\end{multicols}

\clearpage

\end{document}